\documentclass[twocolumn]{aastex7}

\usepackage{fp}
\usepackage{amsmath}
\usepackage{soul}

\newcommand{\muONE}{29.189}

\newcommand{\muFOU}{31.732}

\FPeval{\distONEtoFOU}{\muONE-\muFOU}

\FPeval{\muONEround}{round(\muONE,3)}
\FPeval{\muFOUround}{round(\muFOU,3)}
\FPeval{\distONEtoFOUround}{round(\distONEtoFOU,3)}

\defcitealias{Hoyt_2021}{H21}
\defcitealias{Jang_2017_h0}{JL17}
\defcitealias{Freedman_2019}{F19}
\defcitealias{Freedman2021ApJ...919...16F}{F21}
\defcitealias{Hoyt_2023}{H23}
\defcitealias{Uddin2023arXiv230801875U}{U23}
\defcitealias{Riess_2022ApJ...934L...7R}{R22}
\defcitealias{Riess_2016ApJ...826...56R}{R16}

\newcommand{\colorTHREEFIVE}{$(\mathrm{F115W}-\mathrm{F356W})$}
\newcommand{\colorFOURFOUR}{$(\mathrm{F115W}-\mathrm{F444W})$}
\newcommand{\magSlopeErr}{$\mathrm{mag}~\mathrm{mag}^{-1}$}
\received{\today}

\begin{document}

\title{The Chicago Carnegie Hubble Program: Improving the Calibration of SNe~Ia with JWST Measurements of the Tip of the Red Giant Branch}

\correspondingauthor{Taylor J. Hoyt}
\email{taylorjhoyt@gmail.com}

\author[0000-0001-9664-0560]{Taylor J. Hoyt}
\affiliation{Physics Division, Lawrence Berkeley National Lab, 1 Cyclotron Road, Berkeley, CA 94720, USA}
\email{taylorjhoyt@gmail.com}

\author{In Sung Jang}
\affiliation{Department of Astronomy \& Astrophysics, University of Chicago, 5640 South Ellis Avenue, Chicago, IL 60637}
\email{hanlbomi@gmail.com}

\author{Wendy L. Freedman}
\affiliation{Department of Astronomy \& Astrophysics and Kavli Institute for Cosmological Physics, University of Chicago, 5640 South Ellis Avenue, Chicago, IL 60637}
\email{wfreedman@uchicago.edu}

\author{Barry F. Madore}
\affiliation{The Observatories of the Carnegie Institute of Washington, 813 Santa Barbara St, Pasadena, CA 91101}
\affiliation{Department of Astronomy \& Astrophysics, University of Chicago, 5640 South Ellis Avenue, Chicago, IL 60637}
\email{barry.f.madore@gmail.com}

\author{Kayla A. Owens}
\affiliation{Department of Astronomy \& Astrophysics, University of Chicago, 5640 South Ellis Avenue, Chicago, IL 60637}
\email{kaowens@uchicago.edu}

\author{Abigail J. Lee}
\affiliation{Department of Astronomy \& Astrophysics, University of Chicago, 5640 South Ellis Avenue, Chicago, IL 60637}
\email{abbyl@uchicago.edu}

\begin{abstract}
We present distances to ten supernova (SN) host galaxies determined via the red giant branch tip (TRGB) using JWST/NIRCAM and the F115W, F356W, and F444W bandpasses. Our analysis, including photometric catalog cleaning, adoption of disk light profiles, TRGB color slope estimation, and a novel technique for identifying the infrared TRGB, was conducted blinded. The new F115W TRGB distances agree well with our previously derived HST TRGB distances, differing by only 1\% on average and 4\% on a per-galaxy basis. 
The color-corrected F115W TRGB is therefore equally precise a method of distance measurement as, and offers unique advantages over, its color-insensitive, $I$-band counterpart.
Using these distances, we update the absolute calibrations of eleven calibrator SNe, yielding $68.4 \leq H_0 \leq 69.6$ km/s/Mpc depending on which of four sets of SN magnitudes are used. We expand the sample of calibrator SNe to 24 by combining with HST TRGB distances. Doing so increases our $H_0$ estimate based on the Carnegie Supernova Project II (CSP-II) by $0.8$~km/s/Mpc ($1.4 \sigma$) demonstrating that our JWST $H_0$ based on 11 SNe is not significantly biased toward lower values.
In contrast, the Pantheon+ calibration shifts higher by $+2$~km/s/Mpc ($3.1 \sigma$ significance), a significantly larger increase than seen in both the CSP and the Pantheon team's own SuperCal analysis. More JWST observations of the TRGB as well as independent analyses of low-redshift SNe are needed to continue unraveling the true nature of the Hubble Tension.
\end{abstract}


\section{Introduction} \label{sec:intro}

\subsection{Measuring the Hubble constant via the Astrophysical Distance Scale}

A redshift-independent, absolute calibration of the universe's current expansion rate, $H(z=0)=H_0$, referred to as Hubble's constant, is determined by chaining together a ``ladder'' of independent extragalactic distance indicators that can be observed at the same distances with sufficient frequency so as to enable a cross-calibration between them. These nodes of overlap, or ``rungs'', in concert constrain the physical scale of the universe. In the modern day, the $H_0$ distance ladder is composed of just three rungs, beginning with trigonometric measurements in the kpc to Mpc universe (e.g., parallaxes or orbiting masers) and ending with probes of the smooth Hubble expansion observed at Gpc distances. Type Ia supernovae (SNe~Ia) provide the most precise and accurate option for this. Less reliable methods such as surface brightness fluctuations (SBFs) or the Tully-Fisher relation can be used as interesting alternative checks, but ultimately cannot reach the statistical constraining power of SNe~Ia. For a review, see \citet{Freedman_2010_annrev}.

Reducing the uncertainties associated with each cross-calibration rung is key to improving the precision and \textit{accuracy} of $H_0$ measurements. And due to methods of trigonometric distance measurement recently achieving percent-level precision \citep{Pietrzynski_2019, Reid_2019, Lindegren_2021}, the proliferation of low-$z$ SN surveys \citep{Hamuy_1996AJ....112.2438H, Riess_1999AJ....117..707R, Aldering_2002SPIE.4836...61A, Krisciunas_2004AJ....128.3034K, Hamuy_2006, Jha_2006AJ....131..527J, Hicken_2009, Burns_2018, Phillips_2019}, as well as increasingly comprehensive compilations of such surveys \citep{Kowalski_2008, Suzuki_2012ApJ...746...85S, Scolnic_2018ApJ...859..101S, Scolnic_2022ApJ...938..113S, Rubin_2024}, the contemporary focus is on the middle, or second, rung. The second rung is based on the use of a primary distance indicator, such as the Cepheids or the Tip of the Red Giant Branch (TRGB), to provide the link between trigonometric measurements in the first rung with probes of cosmological distance in the third rung.

\subsection{Cepheids and the Hubble Tension}
Cepheids were for many decades the most precise tool for calibrating the Hubble expansion \citep{Hubble1929PNAS...15..168H, Sandage_1970, deVac_1978}. Their accuracy was improved via advancements in the use of multi-wavelength observations \citep{Madore_1976RGOB..182..153M, Madore_1982ApJ...253..575M, Freedman_1988ApJ...326..691F}, the launch of the Hubble Space Telescope \citep[HST][]{Freedman_1994Natur.371..757F, Mould_2000ApJ...529..786M, Freedman_2001ApJ...553...47F}, and the incorporation of infrared (IR) observations \citep{McGonegal_1982ApJ...257L..33M, Macri2001ApJ...549..721M, Freedman2008ApJ...679...71F,Riess_2009ApJ...699..539R,Riess_2011ApJ...730..119R,Freedman2012ApJ...758...24F,Riess_2016ApJ...826...56R}. Today, the Cepheid-SN distance ladder value of the Hubble constant is reported by the SH0ES team to be $H_0 = 73.04 \pm 1.04$~km/s/Mpc, \citep{Riess_2022ApJ...934L...7R}, indicating a $5\sigma$ tension with the standard $\Lambda$CDM model of the universe, which has been found to be consistent with $H_0$ values closer to 68 km/s/Mpc \citep{Planck2020A&A...641A...6P, DESI2024arXiv240403002D}. However, the quoted precision of that Cepheid-based measurement is still under debate \citep{Freedman2021ApJ...919...16F, Owens2022ApJ...927....8O, Freedman_2023}. The values of $H_0$ published by SH0ES have remained stable to $<1.0$~km/s/Mpc for over a decade despite many constituent pieces of their analysis fluctuating at levels in significant excess of that amount.
JWST observations are being used to stress-test aspects of the Cepheid error budget, both in a different branch of our program (Owens et al., in prep.) as well as in others \citep{Riess2024ApJ...962L..17R}. It should be noted, however, that the improved IR resolution and signal-to-noise brought by JWST cannot address concerns that are inherent to the method itself (e.g., metallicity).

\subsection{The Tip of the Red Giant Branch: A Vital Check on the Cepheid-driven $H_0$ tension}

Contrary to the high value of $H_0$ from Cepheids, the TRGB calibration of SNe~Ia has yielded values more consistent with $\Lambda$CDM predictions \citep{Jang_2017_h0, Freedman_2019}.\footnote{Instances where the TRGB resulted in larger values of $H_0$ occurred in reanalyses of the CCHP data that either used lower signal-to-noise data to re-calibrate the TRGB \citep{Anand2022ApJ...932...15A}, or implemented a new ``unsupervised'' methodology that left the TRGB measurements uncorrectably contaminated by intermediate-age stellar populations \citep{Scolnic_2023}. It turns out that these attempts to revise the CCHP TRGB distances into agreement with the pre-2022 SH0ES distance scale were not warranted. 
On the contrary, in 2022 SH0ES would later revise their distances by up to 15\% for some SN host galaxies, bringing them into significantly better agreement the with CCHP \citep{Riess_2022ApJ...934L...7R}.} \citet{Freedman2021ApJ...919...16F}, which we will refer to as CCHP-21, reported a value of $H_0 = 69.8 \pm 1.8$~km/s/Mpc based on a TRGB calibration of 20 SNe~Ia and four geometric anchors (LMC, SMC, NGC~4258, Milky Way)---this provided a culmination of what we will refer as the first phase of the Carnegie Chicago Hubble Program (CCHP). 

In addition to its new and independent estimate of $H_0$, the CCHP TRGB distances to SN host galaxies provided a crucial check on the SH0ES Cepheid-based result that has dominated evidence of a Hubble Tension \citep{Riess_2016ApJ...826...56R, Riess_2019}. For the same ten SNe~Ia in common between CCHP and SH0ES-19, the CCHP distances were $+$0.06~mag farther (equivalent to 2~km/s/Mpc lower in $H_0$). Furthermore, the distribution of SN magnitudes in the CCHP calibration had a dispersion of $0.11$~mag, while the same quantity was equal to $0.15$~mag in the SH0ES-16 calibration of those same SNe \citep{Freedman_2019}. The CCHP calibration of SNe was therefore significantly and unambiguously more precise, despite both studies quoting comparable uncertainties. 

Following the CCHP result, several studies from the SH0ES team argued (in error) that the luminosity of the TRGB that was adopted by the CCHP needed to be revised in order to force agreement with their Cepheid distances \citep{Yuan_2019, Reid_2019, Soltis_2021, Li_2022, Anand2022ApJ...932...15A}. However, those studies have not held up amidst scrutiny \citep{Hoyt_2023, Freedman_2023}, nor in light of new, higher quality calibration data \citep[HST-GO-16743,][]{Hoyt2021hst..prop16743H}.
Indeed, the CCHP calibration of the TRGB zero point was recently confirmed by SH0ES' latest TRGB zero point calibration presented in \citet{Li_2023}. Those authors analyzed yet-to-be-published, high signal-to-noise HST data acquired of RGB stars in the halo of geometric anchor NGC~4258 \citep[HST-GO-16743,][]{Hoyt2021hst..prop16743H} and recovered a TRGB magnitude in 1\% agreement with the CCHP value that has remained stable over the years \citep{Lee_1993, Freedman_2019, Freedman_2020, Freedman2021ApJ...919...16F, Jang_2021, Hoyt_2023}. This recent SH0ES TRGB finding marked a significant departure from their previous claims that TRGB miscalibration could be the cause of the disagreement between (pre-2022) SH0ES and CCHP.

The real resolution to the CCHP-SH0ES distance discrepancy arrived in the latest major iteration of the SH0ES project \citep[][hereafter SH0ES-22]{Riess_2022ApJ...934L...7R}, in which they updated their previous IR flux measurements of Cepheids in nearby, crowded galaxies by as much as $0.3$~mag in the most extreme case and in the average by $0.06$~mag from an updated analysis of the exact same HST imaging data. The revised SH0ES-22 Cepheid distances immediately relaxed the CCHP-SH0ES distance discrepancy after having come into 1\% agreement with the CCHP TRGB distances. The dispersion in the SH0ES calibration to the same 10 SNe also decreased from $0.15$ to $0.10$~mag, which is much more in-line with the $0.11$~mag dispersion that was seen in the CCHP's TRGB-SN calibration. 

Despite this significant $0.06$~mag average shift in the fluxes measured by SH0ES to the same Cepheids from the same data, the $H_0$ value they reported only shifted by $-0.01$~mag, or $-0.3$~km/s/Mpc. Other changes included a 0.3~dex systematic shift applied to the metallicity scale of only Cepheids observed in distant hosts, equivalent to a systematic shift in $H_0$ of order 1~km/s/Mpc, with the exact size depending on the choice of geometric anchor. Similarly, the magnitudes of the SNe shifted fainter in the average by 0.037~mag, equivalent to increasing $H_0$ by 1.28~km/s/Mpc. Some of these analysis-level tensions contributing to the Hubble tension are discussed later in this manuscript and in this program's report on the cosmology results \citep{Freedman2024arXiv240806153F}. Given these ongoing changes, we conclude that more data are needed to better assess the accuracy of HST-based Cepheid distances \citep{Hoyt_2024, Freedman2024arXiv240806153F, Riess2024arXiv240811770R}. Checks on the Pantheon+ analysis of SNe that is part of the SH0ES estimate of $H_0$ are also warranted.

\subsection{The Next Phase of the CCHP with JWST}

The contents of this article constitute one part of JWST program GO-1995 (PI: Freedman, co-PI: Madore), which was designed to search for yet-undiscovered systematics in the distance scale as well as to estimate a new value of $H_0$ using JWST. To do so, three independent distance indicators had to be imaged and accurately photometered from the same set of imaging per target galaxy. The three indicators include: Cepheids, the TRGB, and the recently developed branch of carbon-enhanced asymptotic giant branch (AGB) stars that exhibit a $J$-band magnitude that is insensitive to the observed temperature/color, referred to as JAGB.\footnote{Note that a JAGB star should not be confused with a ``J-star;'' the latter is a subclass of 10-15\% of carbon stars that are identified spectroscopically and likely formed through mass transfer \citep{Abia_2000, Morgan_2003}. However, the JAGB population used as a distance indicator is selected on the infrared H-R diagram and instead are formed through a different mechanism \citep[see][]{Nikolaev_2000, Weinberg_2001}.}

In this manuscript, we focus on the TRGB aspect of the program. As mentioned, the new distances constitute part of the larger CCHP goal to measure $H_0$ \citep{Freedman2024arXiv240806153F}, and will be used also to assess and check the SH0ES Cepheid calibration of SNe. The new TRGB measurements are based on IR data taken at 1.2~\micron{} and near $\sim\!\!4$~\micron. and are thus independent (in wavelength) of previous TRGB programs, which were based on observations acquired in bandpasses centered around 0.8 \micron.

In section 2 we will introduce the TRGB, its applications, and its manifestation at IR wavelengths. In section 3 we present the data, their reduction, and sample selection. In section 4, we describe the TRGB measurement procedure. In Section 5 we compare the new distances with external measurements. In section 6 we present revised calibrations of the SNe based on the new JWST measurements. We discuss the results in Section 7 and conclude in Section 8.

\section{The Tip of the Red Giant Branch} 

\subsection{Background}

The TRGB is observed as a dramatic decrease in the numbers of low-mass, first-ascent red giants after they reach their peak luminosity on the red giant branch (RGB). This steep drop in the RGB luminosity function (LF) is caused by the onset of helium fusion in the star's core, which has been left degenerate after having had $t_{RGB}\sim1$~Gyr worth of helium ash dumped onto it as a byproduct of hydrogen shell burning.
The eventual onset of He burning in the core runs away due to its degenerate equation of state being decoupled from its temperature. This thermonuclear runaway event, or the ``helium flash,'' injects via the triple alpha process a galaxy's worth of luminosity into the surrounding region. 

The helium flash lasts only a few seconds until the intense energy input raises the gaseous pressure in the core enough to lift the electron degeneracy, thereby reintroducing a temperature-dependent equation of state and allowing the core to finally expand. 
This expansion brings the core out of the degenerate regime and a steady state for helium core burning is established. As a result, the star rapidly migrates off the RGB and to the horizontal branch on a timescale $t_{RGB\rightarrow HB}\sim$1~Myr.
Finally, since $t_{RGB} >> t_{RGB\rightarrow HB}$ and $P(observe) \sim t$, a composite stellar population will be left with a sharp decline in the numbers of stars observed just above the TRGB.

\begin{figure}
    \centering
    \includegraphics[width=0.99\linewidth]{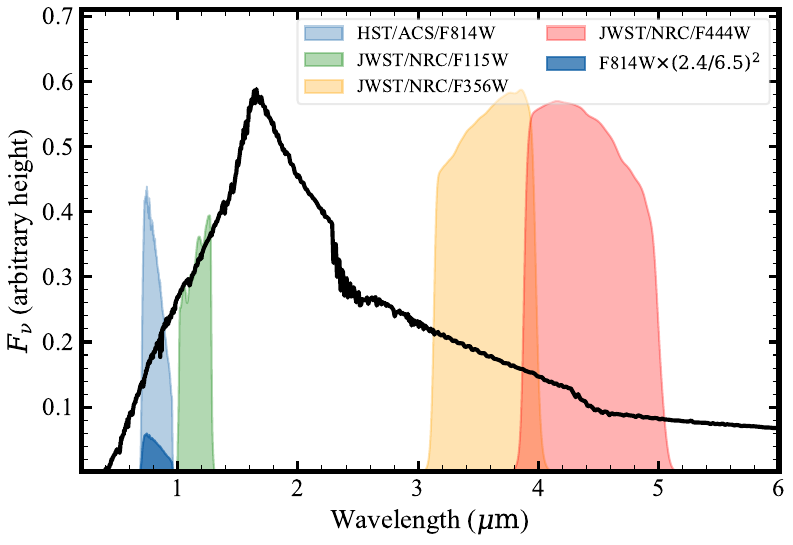}
    \caption{Model atmosphere from \citet{Castelli_2003} for a K5I star with log(g) = 0.00 (black curve) intended to approximate a TRGB star's atmosphere. Four filter transmission curves are plotted (lightly shaded regions). Also shown is an additional F814W transmission curve that has been renormalized by the ratio of HST to JWST's mirror size (dark blue shaded region).}
    \label{fig:trgb_spectrum}
\end{figure}

There are two unique characteristics of a low-mass ($M \lesssim 2M_{\odot}$) RGB star's ascent that then cause the TRGB luminosity to manifest as a nearly constant quantity. First, the steep pressure gradient between an RGB star's degenerate helium core ($0.1 M_{\odot} \lesssim M_{core} \lesssim 0.5 M_{\odot}$) and its convective hydrogen envelope means the core mass dictates the stellar processes and structure, including the rate of hydrogen burning in the thin ($10^{-3} M_{\odot}$) shell. This is the well known core mass luminosity relation in low-mass RGB stars \citep{Kippenhahn1990sse..book.....K, Salaris2002PASP..114..375S, Kippenhahn2013sse..book.....K}, or, 
\begin{equation}
    L \sim M_{core}^{\varphi}
\end{equation}
where typical values of the power law index $\varphi$ are near 7
\citep{Kippenhahn1990sse..book.....K, Salaris_2005}. 
Moreover, in these $M \lesssim 2M_{\odot}$ RGB stars, the mass of the core at the onset of helium burning ($M_{core} \simeq 0.5 M_{\odot}$) is nearly invariant, with a slight dependence on metallicity, regardless of initial mass. These points together ensure that low mass stars reach a near constant bolometric luminosity at the TRGB.

The reader is referred to the extensive literature for more details on the TRGB and its physical underpinnings \citep{Frogel_1981, Kippenhahn1990sse..book.....K, Salaris1998MNRAS.298..166S, Madore_1999, Salaris2002PASP..114..375S, Salaris_2005}.
Indeed, it is a testament to the progression of the field, and the mature understanding of the \textit{core} physics underlying the TRGB, that articles such as this one can confidently lead with what is now considered to be the well established stellar physics.

\begin{figure*}
    \centering
    \includegraphics[width=0.7\linewidth]{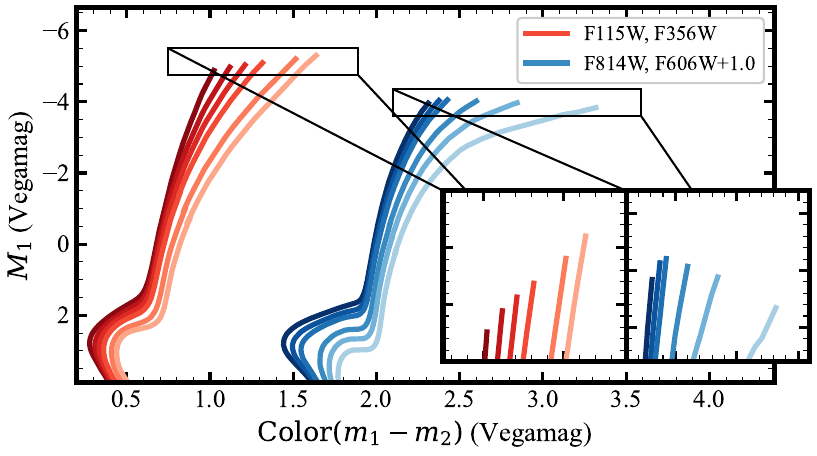}
    \caption{Optical and infrared 10~Gyr stellar evolution tracks, or isochrones, from the BaSTI suite \citep{Hidalgo_2018, Pietrinferni_2021}. The vertical axis (labeled $M_1$) corresponds to either the ACS/WFC/F814W (shades of blue) or JWST/NIRCAM/F115W (shades of red) absolute magnitudes. On the horizontal axis is the color index as conventionally computed with the redder magnitude subtracted from the bluer one, notated as $m_1-m_2$ here. The HST and JWST isochrone magnitudes were shifted by $+0.16$~mag and $-0.10$~mag, respectively, to match empirical zero point measurements. The $\mathrm{F606W} - \mathrm{F814W}$ colors of the isochrones were shifted by 1~mag for visual clarity. The metallicity range plotted is $[Fe/H]=\{-1.90, -1.55, -1.30, -1.05, -0.60, -0.30\}$. Lighter shades represent increasingly metal-rich isochrones. Inset zoom panels are included to highlight the linearity of the IR TRGB and the quadratic, nonlinear form of the optical TRGB.}
    \label{fig:oir_isochrones}
\end{figure*}

\subsection{The TRGB in the Optical and Infrared}
The TRGB can be represented as a luminosity and temperature locus on the H-R diagram, i.e., a magnitude and color coordinate of finite size. The shape and position of this locus depends on the filters through which the stellar fluxes are transmitted. We will in this subsection provide an overview of this wavelength-dependent behavior.

In \autoref{fig:trgb_spectrum}, we show a model spectrum meant to approximate a TRGB star's spectral energy distribution (SED). The displayed SED (black curve) is a K5I type star with $T_{eff} = 3850$~K and $\log g = 0.00$ from the \citet{Castelli_2003} library. Note the units are $F_\nu$ and not $F_{\lambda}$. Plotted are the transmission curves for the three JWST/NIRCAM bandpasses that will be used in this study: F115W, F356W, and F444W as green, yellow, and red shaded regions, respectively. Also plotted as a light blue shaded region is the transmission curve for the HST ACS/WFC/F814W filter, the bandpass most widely used in extragalactic distance measurements made with the TRGB \citep{Sakai_2004, Rejkuba2005ApJ...631..262R, Makarov_2006, Tully_2009, Jacobs_2009}, and that was used in all past TRGB calibrations of the distance ladder $H_0$ \citep{Mould_2009, Jang_2017_h0, Freedman_2019, Anand2022ApJ...932...15A}. A second transmission curve for F814W, rescaled by the square of the ratios of the JWST and HST mirror diameters $(2.4/6.5)^2 \simeq 1/7$, is plotted to provide a rudimentary visualization of the difference in collecting area between HST and JWST.

The brightness of a TRGB star when observed near 800-900~nm is known to be insensitive to its metallicity $[M/H]$ when $[M/H] \leq -0.5$~dex \citep{Lee_1993}, making it a near perfect standard candle.\footnote{Note that we use $[M/H]$ here instead of an empirical metallicity scale such as $[Fe/H]$. The global metallicity provides a more complete description of the TRGB luminosity $M_{bol}$ because it encompasses the effects of $\alpha$ enhancement. See, e.g., \citet{Salaris1998MNRAS.298..166S} and references therein.}; this postulate has stood the test of time \citep{Tammann_2008, Freedman_2019, Freedman2021ApJ...919...16F}. 
The underlying mechanism behind this can be understood by starting with the $I$-band TRGB distance as described in \citet{Lee_1993},
\begin{equation}
  (m-M)_I = I_{TRGB} + BC_I - M_{bol,TRGB}
\end{equation}
where $(m-M)_I$ is the distance modulus, $I_{TRGB}$ the apparent TRGB magnitude, $BC_I$ the bolometric correction for a TRGB star's atmosphere, and $M_{bol,TRGB}$ the bolometric magnitude of the TRGB. It turns out that the $M_{bol,TRGB}$ and $BC_I$ terms have a dependence on metallicity that is similar in size, but opposite in sign \citep{Dacosta_1990, Salaris1998MNRAS.298..166S}. The terms then mostly cancel, making $M_I$ a near exact standard candle in this lower metallicity regime.

Findings based on the magnitudes and colors of TRGB stars in the Magellanic Clouds (MCs) further confirmed the insensitivity of the $I$-band TRGB to metallicity/color. That is, the magnitudes and colors of SMC and LMC TRGB stars are observed to be consistent with the metal-poor, standard candle postulate except for a small fraction in the inner regions of the LMC. These TRGB stars exhibited marginally, but measurably, fainter magnitudes (by $\sim 0.02$~mag) and redder colors than those in the outer regions \citep[see, e.g.,][]{Jang_2017_color, Hoyt_2023, Bellazzini_2024}. APOGEE abundance mapping showed that the distribution of $[Fe/H]$ of color-magnitude selected RGB stars in the LMC prefers $-0.90 < [Fe/H] < -0.60$~dex, with a steep tail up to $-0.35$~dex \citep{Nidever_2020}. This is consistent with the inner TRGB stars being just metal-rich enough to lie outside of the $I$-band standard candle regime.

The magnitude of the infrared (IR) TRGB, however, exhibits a strong dependence on metallicity, and there is no regime where this dependence is reduced \citep{Bellazzini_2004, Valenti_2004}. At the same time, the relationship between metallicity and IR color for (T)RGB stars is tight and linear in form \citep{Frogel_1983, Valenti_2004, Bellazzini2008MmSAI..79..440B}. The pairing of these two well-defined relations forms the basis for using the observed color of the IR TRGB to standardize its apparent magnitude, making it possible to use it as a precision distance indicator.

The reader is referred to \citet{Streich_2014} for a thorough examination of the TRGB color-metallicity relation using ACS/WFC imaging of 71 globular clusters. Their Fig. 2 demonstrates the nonlinear relation between the metallicity and optical $(V-I)$ colors of upper RGB stars. \citet{Valenti_2004} present a similar analysis of 24 globular clusters in the NIR $JHK$ bandpasses.

We contrast the behavior of the $I$-band TRGB with that of the NIR TRGB in \autoref{fig:oir_isochrones}. Plotted are 10~Gyr BaSTI RGB isochrones computed for either the HST/ACS/WFC F606W and F814W (equivalent to $I$-band ) bands or the JWST/NIRCAM F115W (similar to $J$-band) and F356W bands. The HST isochrones are plotted as blue curves shaded by iron abundances that range from an $[Fe/H]$ of $-1.9$ to $-0.3$~dex \citep{Hidalgo_2018, Pietrinferni_2021}. The peak-to-peak variation of the TRGB F814W magnitude when $[Fe/H] < -0.6$~dex is estimated to be 0.1~mag or less. At the same time, the isochrones can be seen to cluster tightly in the very metal-poor regime, then fan out as the metallicity increases. This is the non-linear color-metallicity relation of the $I$-band TRGB in action. 
The NIR isochrones are plotted as red curves that are similarly shaded according to iron abundance. Different from the $I$-band ones, these are evenly spaced on the CMD near the TRGB. This reflects the TRGB's smooth, linear dependence on metallicity/color in the NIR.\footnote{\citet{Riess2024arXiv240811770R} claim that ``in the NIR, theory shows that the TRGB is tilted and non-linear.'' However, there is consensus in both the empirical and theoretical literature that it is in fact definitively linear \citep{Valenti_2004, Bellazzini2008MmSAI..79..440B, Serenelli_2017, Madore_2018, Hoyt_2024, Newman_2024}. See \autoref{fig:oir_isochrones}.} 

These differences suggest that, for the NIR TRGB, while the size of a correction for metallicity is indeed larger than in the $I$-band, the effect is also significantly easier to measure and calibrate out. This may enable the NIR TRGB to ultimately reach a comparable precision to the $I$-band. That is, the color dependence of the IR TRGB can be measured above the noise even at the distances and reasonable exposure time constraints characteristic of SN host galaxy observations ($SNR \sim 10$ per TRGB star). In contrast, extremely high precision photometry ($SNR \sim 100$ per TRGB star) is required in order to measure above the noise the fine-scale structure of the metal-poor, optical TRGB, such as the OGLE photometry of the MCs discussed above. The problem is complicated further by small age effects in $I$-band that will blur the highly compressed, non-linear magnitude-metallicity relation in the $I$-band.
On the other hand, part of the IR TRGB age/dependence is self calibrating, due to the age and metallicity slopes being at least partially parallel. We will explore this further in the following subsection.

\subsection{Standardizing an IR TRGB Distance Scale}

We now narrow our focus to the determination of extragalactic distances using the IR TRGB. In order to do so, there needs to exist a relation that can predict the intrinsic brightness of the TRGB given some additional observable constraint.

As discussed in the previous section, the flat $I$-band TRGB can be described simply by,
\begin{equation}
    M_I = constant
\end{equation}
The IR TRGB, on the other hand, is not so lucky and, for illustrative purposes, can be naively described by,
\begin{equation}
    M_{IR} = constant + \beta_1(metallicity) + \beta_2(mass/age)
\end{equation}
where $\beta_{1/2}$ represent the different relations between the magnitude of the TRGB and its metallicity or age. However, we cannot in practice measure either the average metallicity and/or age of stars at the TRGB (without exorbitant modeling assumptions and associated uncertainties), so we must find some way to simplify the standardization relation to be based on an observable that is more feasible to measure. 

Fortunately, the dependence of the TRGB \textit{color} with either metallicity and age is not perfectly degenerate, allowing us some room for making motivated assumptions for standardization. Indeed, the typical variation of TRGB color and magnitude due to metallicity is significantly larger than it is due to age, i.e., $\beta_{1} >> \beta_{2}$. For example \citet{McQuinn_2019} in their isochrone study report that the TRGB magnitude is expected to vary in the F115W band by 0.3~mag due to metallicity and 0.09~mag due to age.

This coupled with the tight, linear NIR TRGB color-metallicity relation forms the basis of a \textit{standardized} IR TRGB distance scale, or,
\begin{equation}
    M_{IR} \simeq constant + \beta(color)
\end{equation}

Encouragingly, the likelihood of age impacting this postulate is significantly reduced in older stellar populations (about $t > 4-6$~Gyr). We demonstrated this in the Appendix of \citet{Hoyt_2024}, which plotted the BaSTI predictions for the F115W TRGB metallicity slope (mapped into color), as a function of age. We saw that the TRGB magnitude-metallicity-color relation becomes more stable in older populations, indicating that the approximation $color \sim [Fe/H]$ becomes increasingly exact in older stellar populations. We will see this borne out further throughout this study. It also leads us to defining an additional requirement for measuring accurate IR TRGB distances. That is, the TRGB should be measured from an older stellar population to reduce systematics due to age. This is, in fact, not a novel postulate and is shared with the requirements for measuring $I$-band TRGB distances. 

Finally, we can convince ourselves that contributions from $\beta_{age}$ vanish further if the IR TRGB distance scale is built from a sufficiently large sample of galaxies. We would, as a result, be averaging over variations in the distributions of age per each observed composite population, which, in turn, are functions of variations in star formation history.

\begin{figure}
    \centering
    \includegraphics[width=0.95\linewidth]{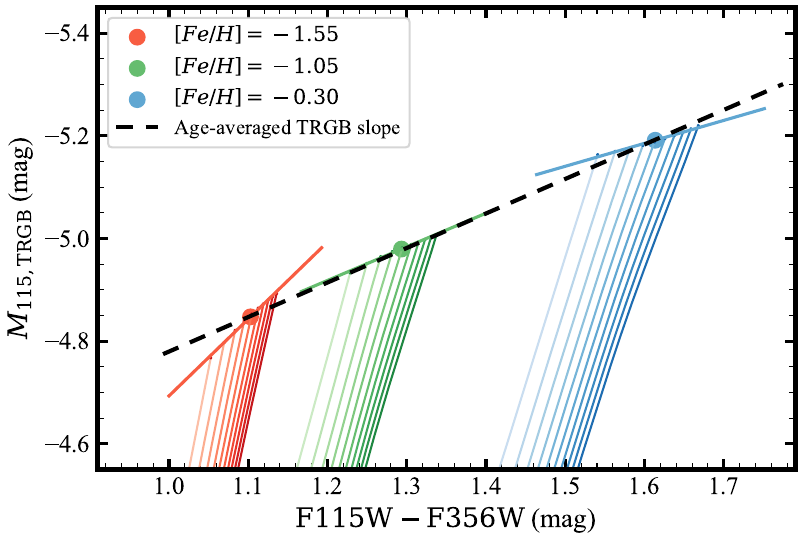}
    \caption{Conception of a standardized IR TRGB distance scale. A set of BaSTI isochrones that vary in age between 4 and 13~Gyr in 1~Gyr steps is plotted for each of three discrete bins in $[Fe/H]$: $-1.55$~dex (red, left) $-1.05$~dex (green, middle), and $-0.30$~dex (blue, right). Younger isochrones are shaded more faintly than older ones. Plotted at the tip of each set of isochrones is the apparent slope of the TRGB (solid colored lines) due to varying the age. Within each metallicity bin, an age-averaged color-magnitude coordinate is computed (filled circles colored according to metallicity).
    The color-standardized IR TRGB distance scale is founded based on the line that connects each of the age-averaged color-magnitude coordinates (dashed black line).}
    \label{fig:age_averaged_slope}
\end{figure}

We illustrate this in \autoref{fig:age_averaged_slope}, in which BaSTI isochrones are plotted in the F115W and F356W color magnitude diagram (CMD) for a range of ages 4 to 13~Gyr in 1~Gyr steps for three discrete metallicity bins $[M/H] = \{-1.55, -1.05, -0.30\}$~dex. We chose these metallicity bins because they contain the full range of TRGB colors observed in our actual JWST data (see Section 4). It can be seen that an age spread in a metal-poor population is expected to yield a steeper observed TRGB color slope, and vice versa for a more metal-rich population. Of course, there is a correlation between age and metallicity, so this illustration cannot be exact. However, because the TRGB's metallicity-dependent variation in magnitude is over a factor of three larger than age-dependent variations, the analogy is a good approximation to reality.
The underlying principle of our proposed universal IR TRGB distance scale then becomes readily apparent. Despite the age dependence skewing the TRGB \textit{slope} that is observed in any one population, the age-averaged magnitude and color provides a stable quantity upon which a universal IR TRGB color correction can be built. This is represented by the black dashed line, which connects the age-averaged midpoints of all isochrone tips contained within the metallicity bins. This then brings to the fore the final requirement needed to establish a self-consistent, universal IR TRGB distance scale: that the TRGB slope needs to be derived from a large enough sample of stellar populations to average over age effects. 

A corollary of this postulate is that, if we measure the TRGB slope from several composite stellar populations, we can expect a wide range in tip slopes similar to those pictured in \autoref{fig:age_averaged_slope}. We will later see that this is exactly what we find in the data. 

\begin{deluxetable*}{lccrrrccrr}
\tablecaption{Target List, Adopted Disk Profiles, and TRGB Selections\label{tab:host_profiles}} 
\tablehead{ 
\colhead{Target Name} &
\colhead{$\alpha_0$} &
\colhead{$\delta_0$} &
\colhead{$a$} &
\colhead{$b$} & 
\colhead{$\phi$} &
\colhead{$r^{10,50,90\%}$} &
\colhead{``Halo'' SMA} &
\multicolumn{2}{c}{SNR@TRGB} \\
\colhead{} &
\colhead{} &
\colhead{} &
\colhead{} &
\colhead{} &
\colhead{} &
\colhead{} &
\colhead{} &
\colhead{F115W} &
\colhead{Color}
\\
\colhead{} &
\colhead{[hh:mm:ss]} &
\colhead{[dd:mm:ss]} &
\colhead{[arcmin]} &
\colhead{[arcmin]} & 
\colhead{[deg]} &
\colhead{[arcmin]} &
\colhead{[arcmin]} &
\colhead{[mag]} &
\colhead{[mag]}
}
\startdata
M101                   & 14:03:12.583 & $+$54:20:55.50 & 11.99  & 11.53  & 0      & $[ 0.98, 2.95, 5.28]$ & 6.35 & 62.5 & 29.4 \\
N1365\tablenotemark{a} & 03:33:36.458 & $-$36:08:26.37 & N/A    & N/A    & N/A    & $[ 2.04, 2.85, 3.68]$ & N/A  &  26.3 & 20.8\\
N2442\tablenotemark{a} & 07:36:23.770 & $-$69:31:51.00 & N/A    & N/A    & N/A    & $[ 1.33, 2.48, 3.59]$ & N/A  &  17.9 & 13.3\\
N3972                  & 11:55:45.090 & $+$55:19:14.65 & 1.79   & 0.49   & 117.5  & $[ 1.18, 1.75, 2.62]$ & 2.42 &  17.4 & 12.2 \\
N4038/9\tablenotemark{b} & 12:01:53.800 & $-$18:53:06.00 & 1.07 & 0.48   & 75.0  & $[ 0.93, 1.43, 2.22]$ & 1.60 &  18.9 & 14.5\\
N4258-Halo             & 12:18:57.620 & $+$47:18:13.39 & 8.49   & 3.62   & 150.0  & $[14.9 , 17.2, 20.0]$ & \nodata &  90.9 & 45.5\\
N4258-Inner            & \nodata                    & \nodata & \nodata & \nodata & \nodata & $[ 2.94, 6.39, 8.56]$ & 8.07 &  90.9 & 43.5\\
N4258-Outer            & \nodata                    & \nodata & \nodata & \nodata & \nodata & $[ 8.34, 9.61, 12.2]$ & 9.34 & 100.0 & 90.9\\
N4424\tablenotemark{c} & 12:27:11.574 & $+$09:25:14.33 & 1.52   & 0.79   & 92.9   & $[ 0.56, 0.96, 2.51]$ & 1.52 &  35.7 & 28.6\\
N4536\tablenotemark{c} & 12:34:27.100 & $+$02:11:18.00 & 3.54   & 1.27   & 119.9  & $[ 1.59, 2.50, 3.95]$ & 3.26 &  31.2 & 10.6\\
N4639\tablenotemark{c} & 12:42:52.379 & $+$13:15:26.71 & 1.43   & 0.94   & 133.2  & $[ 0.57, 0.89, 1.19]$ & 1.26 &  13.5 &  9.6\\
N5643                  & 14:32:40.778 & $-$44:10:28.60 & 2.62   & 2.29   & 98.0   & $[ 1.54, 2.60, 3.63]$ & 2.62 &  40.0 & 31.2\\
N7250                  & 22:18:17.776 & $+$40:33:44.66 & 1.07   & 0.48   & 159.7  & $[ 1.56, 3.13, 5.24]$ & 4.28 &  20.0 &  6.1
\enddata
\tablecomments{Unless otherwise specified, all adopted profile shapes are the $B$-band $D_{25}$ ellipses computed by and compiled in HyperLEDA \citep[][\url{http://leda.univ-lyon1.fr/}]{Makarov_2014}. $\alpha_0$ and $\delta_0$ is the central coordinate of the adopted disk light profile, not the center of the JWST imaging. $r^{X\%}$ is the size of the ellipse which contains $X\%$ of sources in the cleaned stellar catalog, i.e., 10\% of all well-measured point sources in an imaged field are located within $r^{10\%}$ and, similarly, 10\% lie outside $r^{90\%}$. Halo SMA is the boundary outside of which the TRGB is measured. SNR@TRGB is the photometric typical (median) photometric signal-to-noise for a TRGB star in the F115W and either the F356W or F444W bands.}
\tablenotetext{a}{Manual, non-elliptical spatial selection to account for spiral arm dominating field of view (see Appendix). $r$ percentiles quoted are based on $R_{GC}$.}
\tablenotetext{b}{Profile based on NGC~4039, the southern of the Antennae pair (see text).}
\tablenotetext{c}{D26 $r$-band isophote from the SGA-2020 adopted \citep[][\url{https://sga.legacysurvey.org/}]{Moustakas_2023}.}
\end{deluxetable*}

\begin{figure*}
    \centering
    \includegraphics[width=0.99\linewidth]{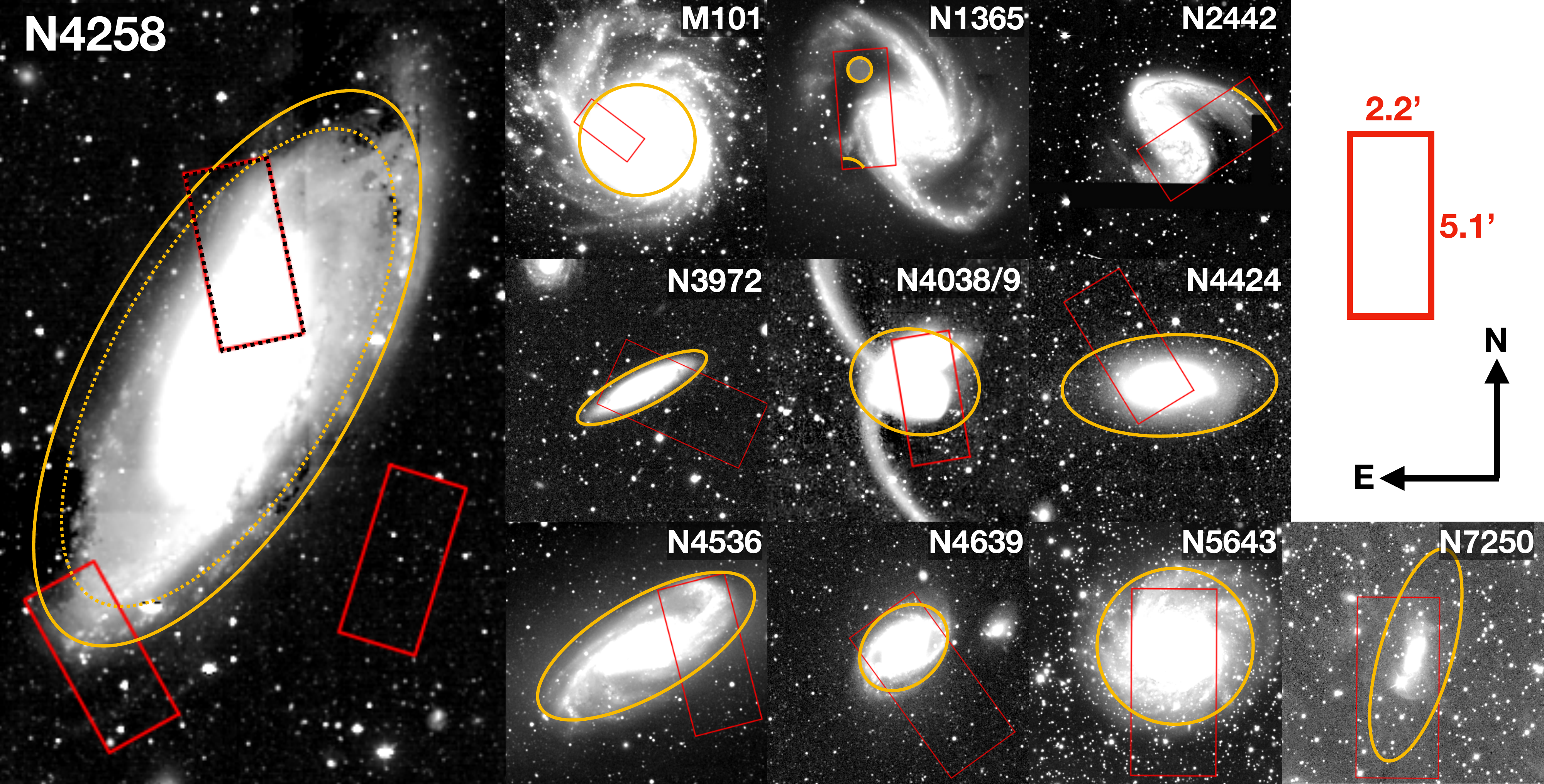}
    \caption{Spatial selection of TRGB fields plotted over wide-area ground-based $g$-band imaging. In all cases DeCALS DR9 or 10 data are plotted, except for N5643 and N7250 for which dedicated DECam imaging and SDSS are used, respectively. The central regions of each galaxy are deliberately saturated to emphasize instead the outermost reaches of their disk light. JWST imaging footprints (red rectangles) are plotted along with the adopted disk boundaries (gold ellipses). For N1365, there is an additional gold shaded circle that encloses (rather than excludes) the adopted selection in the gap of the galaxy's spiral arms. And due to the warping of the N4258 disk at the edges of its major axis, two separate, but still similar, selections were adopted. The inner (gold dotted) and outer (gold solid) ellipses were adopted for the North and South fields, respectively.
    }
    \label{fig:spat_select_comp}
\end{figure*}

\section{Data}
The imaging data used in this study were acquired as part of JWST program GO-1995 (PI: Freedman, Co-PI: Madore). The program was designed to at once measure three different stellar distance indicators from one set of imaging per target \citep{Freedman_2023, Hoyt_2024}. The targets were imaged in the NIRCAM short wavelength (SW) bandpass F115W. Taking advantage of NIRCAM's ability to simultaneously acquire cospatial imaging in its SW and long wavelength (LW) channels we also observed each target in either the F356W or F444W bandpasses. We use these LW data to provide the color baseline for the F115W magnitudes.

\subsection{Observations}
All observations were positioned with the pivot axis pinned to the main body of each galaxy where Cepheids had been previously discovered at optical wavelengths \citep{Hoffmann_2016, Riess_2022ApJ...934L...7R}. The telescope rotation angles were then constrained to sample the thick/outer disk components of the same target galaxies. These galactic structures are thought to be ideal for measurement of JAGB stars. Finally, the outermost portions of each imaging footprint had to be utilized for TRGB measurement. Fortunately, the rectangular footprint of the NIRCAM aperture allowed us to probe radial distances sufficiently outside the target galaxies' disks so as to enable a TRGB measurement from all target fields. Most target fields covered significant portions of each host galaxy's outer regions, except in the fields pointed at host galaxies with the greatest angular extent (e.g., M101) and/or that contain very large, nonlinear structures (e.g., NGC~1365). See \citet{Hoyt_2024} for more discussion of the observation planning.

We targeted ten galaxies that have hosted SNe~Ia and which have been included in the SH0ES Cepheid-SN estimates of $H_0$ \citep{Riess_2016ApJ...826...56R}. We also included in the target list three distinct fields in megamaser host galaxy NGC~4258 to both test the repeatability of the F115W TRGB as well as to set the zero point of our program distance scale. This galaxy's 1.5\% geometric distance is derived by tracking the orbital motions of masing $H_2O$ clouds that are kept in an inverted state by the high energy photons emitted from the galaxy's central black hole \citep{Herrnstein_1999, Humphreys_2008, Humphreys_2013, Reid_2019}. Nine to twelve dithers of six groups and one integration were acquired of each field in the target list, resulting in effective exposure times of 2802~sec or 3607~sec for the nine and twelve dither sequences, respectively.

In \autoref{fig:spat_select_comp}, deep DECam \textit{g}-band imaging \citep{Dey_2019} is shown for all but two target galaxies; the image of M101 is from the Dragonfly Array \citep{Abraham_2014} and the image of NGC~7250 was queried from the (now-defunct) Mosaic tool that was managed as part of the Sloan Digital Sky Survey (SDSS) Science Archive Server (SAS). Outlined in red are the footprint(s) of the JWST/NIRCAM data used in this study. The stretch of the images is set to saturate the bright disk light and emphasize regions of lower surface brightness. These outermost regions of a galaxy's disk best trace the transition from a disk- to a halo-dominated population, the most reliable tracer when making a spatial selection for precision TRGB distance measurement \citep{Beaton_2019, Jang_2021, Hoyt_2021, Hoyt_2024}. Each adopted spatial selection is shown as a gold ellipse. 

In \autoref{tab:host_profiles}, Column (1) lists the unique name associated with each target field. The remaining columns pertain to the spatial selection and signal-to-noise estimates at the magnitude of the TRGB and will be introduced in subsequent sections.

\subsection{Photometry}

\begin{figure}
    \centering
    \includegraphics[width=0.95\linewidth]{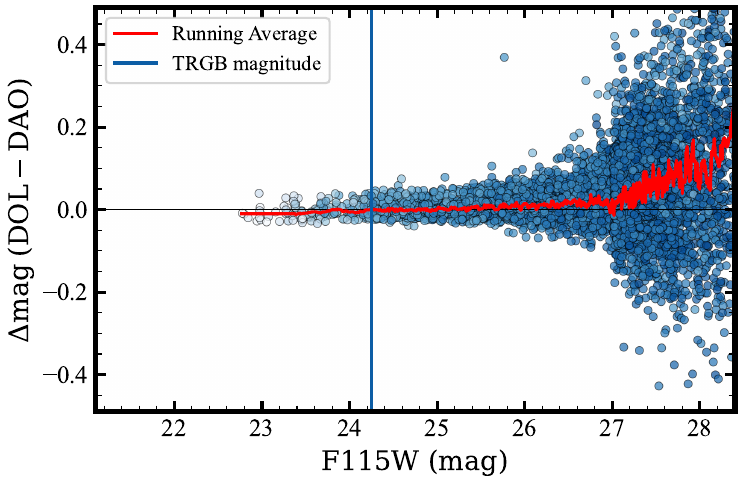}
    \caption{Comparison of independent magnitude estimates for sources in the NGC~4258-Outer field. One set of photometry, plotted on the $x-axis$, is based on DAOPHOT \citep{Stetson_1987, Stetson_1994}, and the other on DOLPHOT \citep{Dolphin_2000, Dolphin_2016, Weisz_2024}. Star-by-star magnitude differences are shown (blue dots) along with the running average (red curve). Lighter shades of blue indicate higher \texttt{chi} values reported by DAOPHOT, an estimate of the goodness-of-fit of the PSF photometry.
    The RMS scatter for all sources brighter than 26th mag is equal to 0.02~mag. For reference, the magnitude of the TRGB in NGC~4258 ($\simeq 25.2$~mag) is plotted (blue vertical line).}
    \label{fig:dol_dao_compare}
\end{figure}

The \textit{cal} images were downloaded from MAST at the same time (2024FEB12) to ensure all images were processed with the same version of the \textit{jwst} calibration pipeline. The images were then aligned to archival HST observations via \texttt{tweakreg} and \texttt{astrodrizzle}. The images were photometered using the \texttt{DOLPHOT} NIRCAM module \citep{Weisz_2024}. The photometry was performed in warm-start mode with the F115W imaging used as the reference image set. A detailed description of the photometric reductions, including artificial star experiments and tests of DOLPHOT's automated aperture corrections, will be presented in a dedicated paper on the photometry pipeline (Jang et al., in prep.). Though it is more than sufficient to note that our procedure closely follows that outlined by the resolved stellar populations ERS program \citep{Weisz_2023, Weisz_2024}. Conservative, uniform criteria were imposed on the photometric catalogs to select for well-measured point sources. We describe this in more detail in Appendix A, including figures that present the impact of the data quality selections on the noisiest (NGC~4038/9) and cleanest catalogs (NGC~4258-Halo).

In the last two columns of \autoref{tab:host_profiles}, we list the SNR of flux measurements for TRGB stars in the F115W band and of the color it forms with either F356W or F444W. For F115W, the typical SNR of a single star at the TRGB ranged from 13.5 to over 100. For the TRGB color measurements, the SNR of a typical TRGB star ranged from 6.1 to 90.9. Note the minimum in the color SNR for NGC~7250 is anomalously low for the sample. This was the first galaxy imaged in our program sample. We found that the image quality degradation in F444W was too severe for efficient application at distances beyond $\sim 15$~Mpc. This prompted us to switch from F444W to F356W as the bandpass providing the color baseline on the CMD, except for the most nearby targets (M101, NGC~4258, and NGC~4536). The SNR of the TRGB color measurements for all other targets is equal to 10 or higher. Recall these are per-star quantities, not uncertainties on the full TRGB measurements, which will be made up of may tip stars.

We evaluated the photometric integrity of the DOLPHOT catalogs by comparing in one of our target fields with photometry estimated from a custom pipeline built around the DAOPHOT software suite for doing PSF photometry \citep{Stetson_1987, Stetson_1994}. We show the results of the comparison in \autoref{fig:dol_dao_compare} which plots the per-source difference in F115W magnitudes between DOLPHOT and the custom DAOPHOT reductions. There is excellent and exact agreement in the average across a 4 magnitude dynamic range (23 to the 27th mag). The per-star RMS of the comparison is equal to $0.02$~mag in this magnitude range. Notably, this region of excellent photometric agreement range encloses all of the TRGB magnitudes observed in our sample, which range from 24.0 to 26.7~mag. This suggests our photometry is robust at this level across the entire program.

Note that for the development and implementation of all methods used in this study, the photometric catalogs were ``blinded.'' Typically, blinding refers to the process of masking the true parameter estimates of one's analysis and relies on the complexity of the dataset to provide a sufficient buffer against potential investigator-dependent bias. In other cases blinding can refer to the use of simulated mock data for the entirety of one's analysis, only swapping in the real data at the end. In our case, we introduced an effective blinding by injecting into each of our 13 photometric catalogs a different, random offset drawn from the uniform distribution $[-0.2,+0.2]$~mag. This is similar in nature to the ``salting'' approach wherein artificial events are secretly injected into an experiment's detection pipeline, used in physics experiments such as LIGO and Lux-Zeplin \citep[LZ,][]{LZ_2023}. 

\subsection{Spatial Selection} \label{sect:spat_select}

\begin{figure*}
    \centering
    \includegraphics[width=0.99\linewidth]{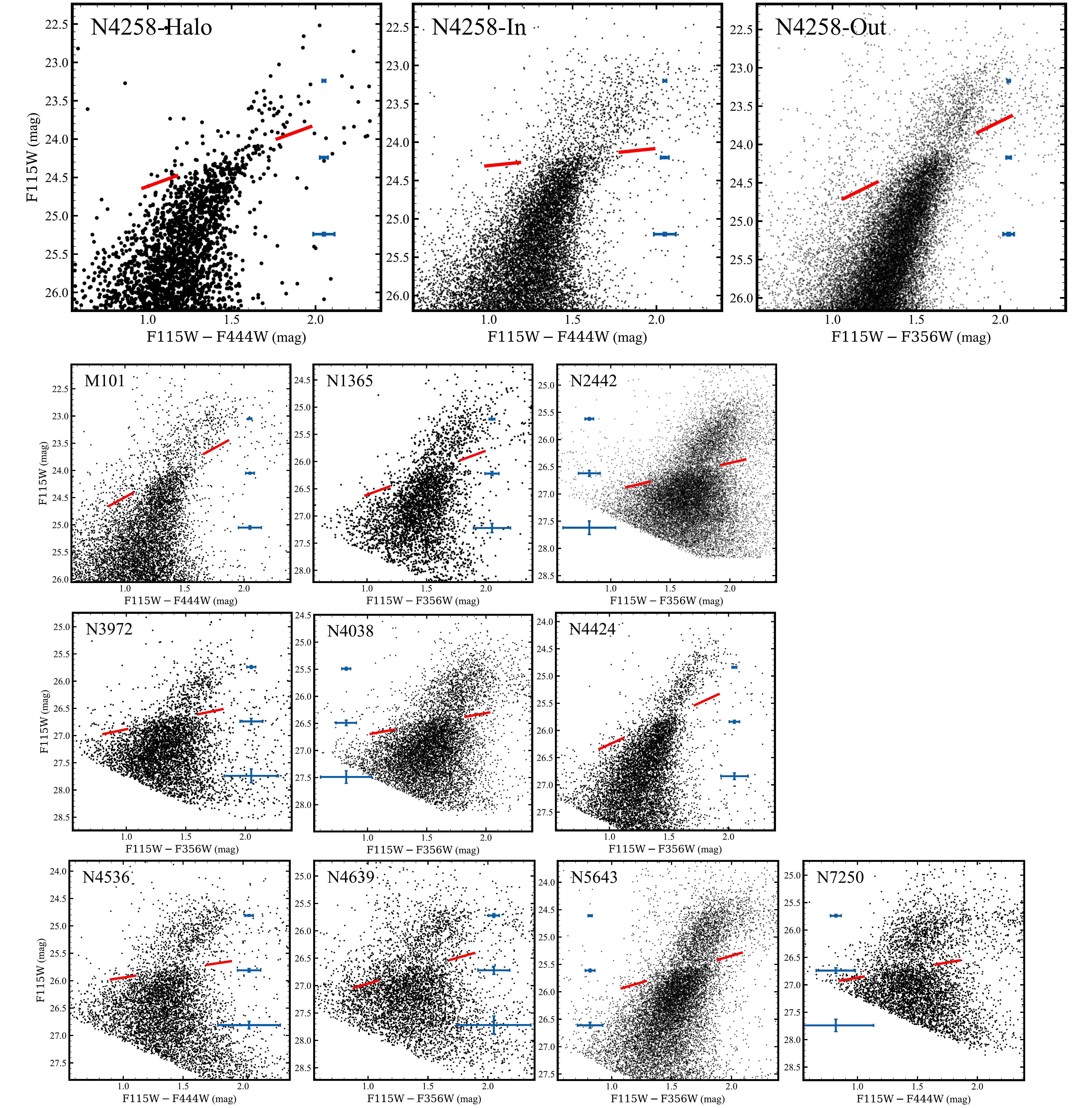}
    \caption{JWST color-magnitude diagrams (CMDs) and best-fit TRGBs. The vertical, magnitude coordinate is in all cases the F115W magnitude. The horizontal coordinate is either the $(\mathrm{F115W}-\mathrm{F356W})$ or the $(\mathrm{F115W}-\mathrm{F444W})$ color index. In each panel, the TRGB is bracketed by a pair of red lines. The three fields imaged of the zero-point calibrator anchor galaxy, NGC~4258, are shown in the top row and the CMDs of the ten SN hosts in the remaining three rows. Representative error bars are plotted at the magnitude of the TRGB and  at one magnitude above and below it (blue capped line segments).}
    \label{fig:trgb_all_cmds}
\end{figure*}

As required by the physical mechanism underlying the TRGB's manifestation on the H-R diagram, it is necessary to target stellar populations that do not coincide with the blue, star-forming regions of galaxies.
The exact prescription for doing so varies in the literature. Some examples include, most commonly, unresolved blue light \citep[see, e.g.][for recent applications to JWST data]{Anand_2024, Hoyt_2024}, unresolved light in the UV and/or mid-infrared \citep{Beaton_2016, Beaton_2019}, information from resolved stellar photometry \citep{Hoyt_2021, Jang_2021, Wu_2023, Scolnic_2023}, or the measured phenomenology of the observed TRGB feature \citep{Hoyt_2023}. The common physical thread that relates these different approaches is the masking of regions of a host galaxy that present with evidence of significant recent star formation.

While the ultimate goal of a TRGB spatial selection is agreed-upon, there remains significant debate over the exact details \citep[see, e.g.,][]{Mager_2024}. It is the view of these authors that the quality of a TRGB measurement is self-evident. That is, the power (significance), clarity (uni-modality), and precision (width) of the TRGB feature is itself the strongest indicator of a good spatial selection for measuring an unbiased TRGB distance. The reader is referred to \citet{Freedman_1989} and \citet{Hoyt_2021} for respective earlier and recent demonstrations of these arguments.

In this study, we follow the methodology laid out in \citet{Hoyt_2024}, wherein published surface photometry was adopted as a starting solution for TRGB sample selection. This initial guess is then perturbed and a solution iteratively converged upon based on the features observed on the CMD. Of particular importance is the clarity of the TRGB feature. Specifically, for each dataset that imaged a mostly smooth background of galaxy light, the LEDA D25 $B$-band isophote \citep{Makarov_2014} or the SGA-2020 $r$-band D26 isophote \citep{Moustakas_2023} was adopted as the shape of the disk light profile, with only the scale adjusted as just described.

The parameters of each adopted profile are presented in Columns (2-6) of \autoref{tab:host_profiles}, starting with the centroid coordinate $(\alpha_0,\delta_0)$, the lengths $a$ and $b$ of the major and minor axes, as well as the position angle $\phi$ of the ellipse, measured in degrees N of E. In the case of the interacting pair of galaxies NGC~4038 and NGC~4039 (the Antennae), a custom profile centered on NGC~4039, the southern member of the pair, was determined by-eye from deep DECaLS $g$-band imaging. We focus on the light of NGC~4039 because only its outer region, and not that of NGC~4038, was captured by the JWST imaging. A smooth profile to parametrize the spatial selection was not adopted for either NGC~1365 or NGC~2442 because the regions of those galaxies that we imaged were dominated by the host galaxies' extended spiral arm structures.

With profiles and initial extents adopted for all fields but NGC~1365 and NGC~2442, we began the iterative spatial selection procedure. The resulting CMD for each field was examined and the ellipse dilated either up or down in size; no changes were made to the \textit{shape} parameters of the adopted profiles. We sought to reduce the presence of blue, young stars to negligible levels, avoid a disproportionate number of intermediate-age AGB relative to RGB stars, maximize the clarity of the TRGB feature, and ensure that we measured TRGB slope values roughly consistent with ingoing expectations. For example, a positive inferred slope was one of the strongest indicators that the spatial selection needed revisiting.

In the cases of NGC~1365 and 2442, a custom spatial selection was adopted instead. In NGC~1365, a gap in the disk left in the wake of its spiral arm was combined with a region of the SE corner that lies just outside the prominent spiral arm. Similarly for NGC~2442, an outermost portion of the imaging was isolated just outside its prominent spiral arm structure. 

In \autoref{fig:spat_select_comp}, we show for all thirteen NIRCAM fields (red rectangles) the final adopted selections, indicated by gold curves outside of which the TRGB measurements are made. One exception is the case of NGC~1365, for which the closed circular region represents the selection was made \textit{interior} to that region. 

Note that the parameters of these unresolved disk light profiles were locked in for all thirteen targets while the program was blinded. However, because the program was designed to simultaneously image the innermost regions of the host galaxies, finding an optimal population for TRGB measurement was one of the more difficult stages of analysis. As a result, for four of the thirteen targeted fields, post-unblinding tests revealed what were inadequate spatial selections. In all cases, the spatial selections that were adopted while the program was still blinded were too strict, leading to an underpopulated TRGB and systematically faint TRGB magnitudes.

The optimal scales of the boundary ellipses were all found to be close to the initial adopted isophote profiles, except for M101 and NGC~7250. The M101 field was embedded in the disk, but, fortunately, the galaxy has very loose spiral arms and, subsequently, many gaps between them within which more pristine stellar populations can be observed. NGC~7250 has a very irregular structure, potentially from a recent interaction, so a deviation from the smooth integrated light profile in this case was not surprising.

Finally, we provide here more information for the special cases of NGC~4258 and M101. For NGC~4258, a different extent of the same elliptical profile was adopted for each of the Inner and Outer Disk fields. For the southern Outer field we adopted a more extended selection, represented by a solid gold curve and solid red footprint in \autoref{fig:spat_select_comp}. For the Inner field we adopted a slightly smaller extent, represented in \autoref{fig:spat_select_comp} by a dotted gold curve and dotted red footprint.
This was done to account for the warp at the edges of the galaxy disk. A similar extended structure can be seen on the opposite end along the major axis. It can be seen that each field's final selection traces well the disk light in that region (left panel of \autoref{fig:spat_select_comp}). For M101, the large angular extent of the galaxy nearly precluded the selection of a population usable for TRGB measurement. However, the outermost edge of our NIRCAM imaging nestles into a gap between M101's loosely wound spiral arms (upper-left thumbnail of \autoref{fig:spat_select_comp}). The selection is further supported by HI imaging of the galaxy, which we present in the Appendix.

Some previous TRGB studies have elected against detailed, case-by-case spatial selection considerations as we have described above, instead choosing to adopt a single, uniform cut on surface brightness for all galaxies in their target sample \citep[e.g.,][]{Anand2022ApJ...932...15A}. However, the breadth of galaxy diversity ensures that a single threshold cutoff adopted to mask disk contamination will prove very unreliable. More specifically, the extents of galaxy disks are known to vary significantly in both physical extent and surface brightness \citep[see, e.g,][]{Mihos2013ApJ...762...82M, Jang_2020}.

\begin{figure*}[t!]
    \centering
    \includegraphics[width=0.9\linewidth]{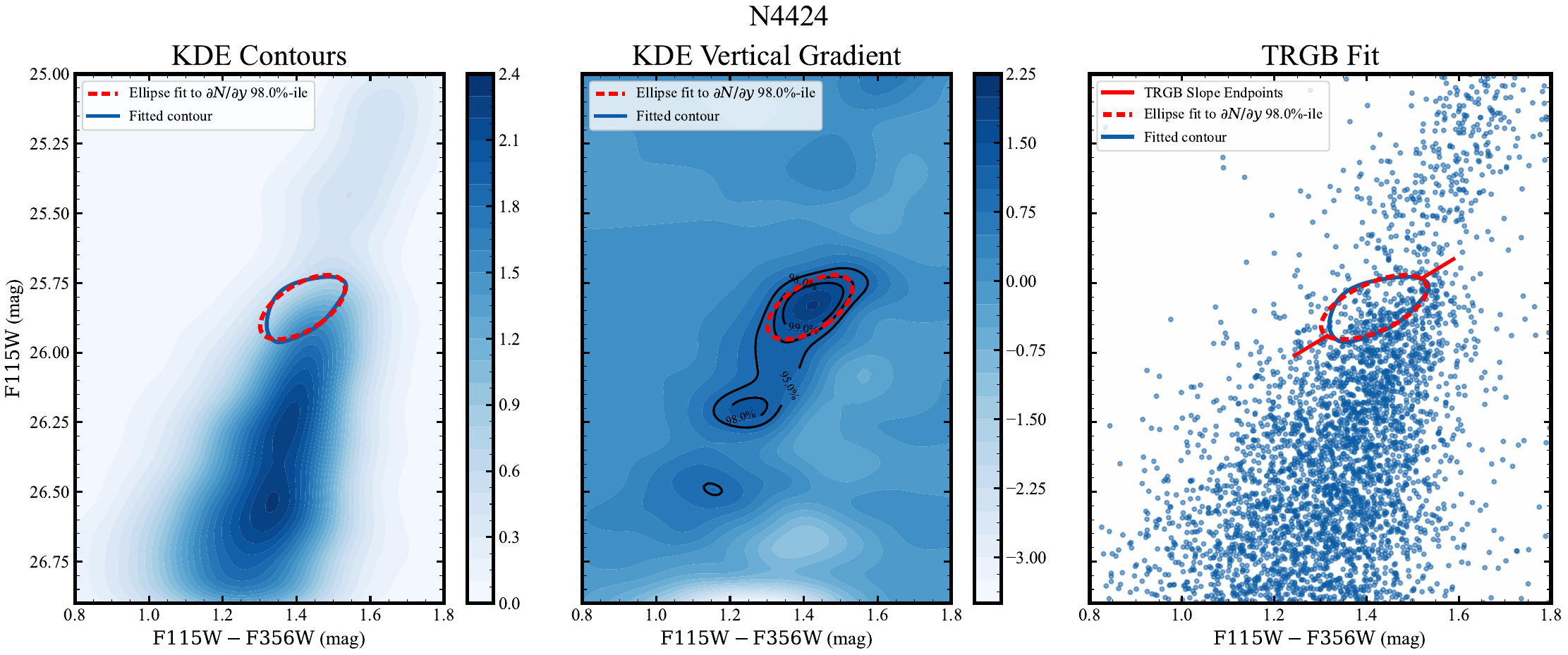}
    \caption{Two-dimensional TRGB measurement for SN host target galaxy NGC~4424. \textit{Left}: Smoothed Hess Diagram (SHD) of the RGB-selected target catalog. \textit{Middle:} Two-dimensional gradient computed along the vertical axis, weighted by the inverse of the horizontal gradient. Positive values indicate a decrease in counts in the upward direction. Three representative thresholds are plotted at 95\%, 98\%, and 99\%. In this case the 98\% contour was adopted, and the best-fit ellipse is plotted (black dashed curve). \textit{Right}: Unbinned CMD (blue scatter points) with the 98\% contour overplotted (blue curve). The best-fit ellipse is plotted (black dashed curve) along with the endpoints that intersect with its major axis. The detections for all target fields are presented as a figure set in the journal's electronic version.}
    \label{fig:4424_2d_edge}
\end{figure*}

\subsection{Color-magnitude Diagrams}
With a well-measured population of RGB stars in hand, we now examine the color magnitude diagrams (CMDs). In \autoref{fig:trgb_all_cmds} we present the cleaned CMDs derived from the regions selected for in the previous section, for all thirteen fields observed in the program. The red giant branch (RGB) is the most prominent feature in all of the CMDs, indicating we have successfully selected for predominantly old stellar populations.

The TRGB estimate for each field is bracketed by a pair of sloped, red lines. Note the linear tilt of the TRGB on the CMD, which is characteristic of its appearance in the NIR (see Section 2). The TRGB is well defined in all cases and unambiguously separated from the intermediate-age AGB stars. The methodology for measuring the TRGB will now be discussed.

\section{TRGB Distances}

In this section, we introduce a new method of measuring the TRGB directly from the two-dimensional CMD. We will use it to derive robust estimates of the tip magnitude and color for each field in our target list.
External HST distances will then be used along with our measurements to constrain an age-averaged, sample-wide estimate of the TRGB color slope.\footnote{Note that HST measurements are only used to constrain the color slope, which is a \textit{relative} parameter and not an absolute one.} That slope is then used to tie each SN host's TRGB magnitude-color coordinate to a fiducial calibration that has its zero point set by tip measurements made in megamaser host galaxy NGC~4258. The zero point measurements were derived using an exactly identical methodology to that used to make the SN host galaxy measurements.

\subsection{TRGB Measurement}
\subsubsection{Background}

\begin{deluxetable*}{lrccrrrrrccr}
\tablecaption{TRGB Measurements\label{tab:trgb_measures}} 
\tablehead{ 
\colhead{Field} &
\colhead{$\sigma_{\mathrm{KDE}}$} &
\colhead{Threshold} &
\colhead{$EBV_{\mathrm{SF11}}$} &
\colhead{$\mathrm{color}$} &
\colhead{$m_{\mathrm{T}}$} &
\colhead{$b$} &
\colhead{$a$} &
\colhead{$N_{fit}$} &
\colhead{Contrast} &
\colhead{Slope} &
\colhead{$\sigma_{tip}$} \\ 
\colhead{} &
\colhead{[mag]} &
\colhead{[\%]} &
\colhead{[mag]} &
\colhead{[mag]} &
\colhead{[mag]} &
\colhead{[mag]} & 
\colhead{[mag]} & 
\colhead{} &
\colhead{} &
\colhead{[mag/mag]} &
\colhead{[mag]}
}
\startdata
M101       & 0.060 & 95.0 & 0.007 & 1.359 & 24.046 & 0.099 & 0.228 & 161  & 2.5 & -1.202 & 0.040 \\
N1365      & 0.100 & 95.0 & 0.018 & 1.473 & 26.200 & 0.148 & 0.228 & 212  & 2.2 & -0.810 & 0.067 \\
N2442      & 0.080 & 90.0 & 0.175 & 1.491 & 26.465 & 0.156 & 0.290 & 1192 & 1.7 & -0.511 & 0.093 \\
N3972      & 0.085 & 96.0 & 0.012 & 1.298 & 26.732 & 0.121 & 0.217 & 232  & 1.9 & -0.457 & 0.062 \\
N4038      & 0.100 & 95.0 & 0.040 & 1.495 & 26.459 & 0.139 & 0.237 & 410  & 1.8 & -0.395 & 0.075 \\
N4258-Halo & 0.070 & 95.0 & 0.014 & 1.460 & 24.226 & 0.106 & 0.237 & 42   & 3.7 & -0.802 & 0.029 \\
N4258-In   & 0.040 & 97.5 & 0.014 & 1.472 & 24.185 & 0.074 & 0.136 & 130  & 3.1 & -0.224 & 0.024 \\
N4258-Out  & 0.055 & 97.0 & 0.014 & 1.554 & 24.155 & 0.104 & 0.162 & 339  & 3.5 & -1.080 & 0.030 \\
N4424      & 0.070 & 98.0 & 0.018 & 1.402 & 25.821 & 0.074 & 0.145 & 151  & 2.9 & -1.006 & 0.026 \\
N4536      & 0.100 & 98.0 & 0.016 & 1.385 & 25.797 & 0.080 & 0.148 & 89   & 1.6 & -0.333 & 0.050 \\
N4639      & 0.060 & 97.0 & 0.022 & 1.366 & 26.701 & 0.098 & 0.209 & 153  & 1.4 & -0.630 & 0.071 \\
N5643      & 0.060 & 98.5 & 0.146 & 1.468 & 25.481 & 0.078 & 0.127 & 193  & 1.2 & -0.648 & 0.066 \\
N7250      & 0.041 & 97.8 & 0.133 & 1.247 & 26.622 & 0.062 & 0.138 & 105  & 1.8 & -0.370 & 0.034
\enddata
\tablecomments{Simultaneous LW imaging for M101, N4258-Halo, N4258-Inner, N4536, N7250 was acquired in the F444W band, with the remainder taken in F356W. $b$ and $a$ are the semi-axes measured normal and tangent to the TRGB ellipse, respectively.}
\end{deluxetable*}

The $I$-band TRGB was first quantitatively defined as the maximal value reached when passing the one-dimensional Sobel edge detection filter $\{-2,0,+2\}$ over the RGB's $I$-band luminosity function (LF) \citep{Lee_1993}. Variations to the \citeauthor{Lee_1993} methodology have surfaced, but are all rooted in the same one-dimensional philosophy, even when the two-dimensional infrared TRGB is involved.
However, as pointed out in \citet{Madore_2018}, \citet{Hoyt2018ApJ...858...12H}, and \citet{Durbin2020ApJ...898...57D}, an accurate measurement of the infrared TRGB's magnitude and color in any single galaxy requires that one simultaneously describe and solve for its finite, two-dimensional structure on the CMD.

Trying to determine an IR TRGB magnitude from the uncorrected, one-dimensional RGB LF \citep[as, e.g.,][]{Wu2014AJ....148....7W} introduces significant, additional uncertainty due to a total loss of the color information needed to accurately localize the TRGB. It should be noted, however, that low-mass galaxies, such as those targeted by \citeauthor{Wu2014AJ....148....7W}, will have narrow RGBs, thereby mitigating to some degree this uncertainty incurred from the loss of color information. However, such an approach would lead to larger and excessive biases in, e.g., more massive SN host galaxies because their larger spread of population characteristics causes the TRGB feature to be more extended on the CMD.
 
This effect was demonstrated in \citet{Hoyt_2024} which studied the color dependence of the F115W TRGB in NGC~4536. That study compared the results when attempting to measure the IR TRGB from a one-dimensional RGB LF marginalized out of a CMD that has either been color corrected or not. When the RGB LF was marginalized out of a CMD that had not been color corrected, they observed an ambiguous, multi-modal structure in the RGB LF near what was ostensibly the TRGB, with two strong peaks separated by $0.3$~mag. Each peak, however, would have produced similar mean TRGB colors according to the \citeauthor{Wu2014AJ....148....7W} methodology which treats the magnitude and color entirely separately, and would lead to significant biases in the distance. We can estimate the size of such a bias in this case by taking roughly half the difference between the two peaks as an estimate of the magnitude uncertainty. And with the TRGB slope value $\simeq -1$~\magSlopeErr{}, this indicates that attempting to identify the TRGB and retroactively associate a mean color based on a color-ignorant measurement from the RGB LF could otherwise lead to systematic distance errors at the level of 0.15 mag.

In \citet{Hoyt_2024}, the color correction was applied via a process known as ``rectification'' of the CMD \citep[][]{Madore_2009}, which rotates the CMD into a coordinate plane on which the TRGB will appear horizontal, or flat, in color. This allows the tip to be measured with the same one-dimensional techniques commonly used with the already flat $I$-band TRGB. We will undertake a different approach in this study, but the philosophy and impact on tip measurement are the same---the two-dimensional tilt must be taken into account during the fitting process to determine a robust tip location.

The solution we adopt in this study presents a step forward for TRGB measurement in terms of both complexity and completeness. We will simultaneously determine the slope and color-magnitude coordinate of the TRGB directly from the two-dimensional CMD. This is novel in the realm of TRGB measurements because color information is not being discarded at any stage of the measurement, as has always been done before. We will see that this approach yields color-corrected F115W TRGB distances that are competitive, in terms of both precision and accuracy, with those derived using the color-insensitive $I$/F814W-band TRGB.

\subsubsection{Two-dimensional Edge Detection}

To measure the tilted IR tip, we build on past edge detection work by expanding the technique along the color axis. As previously mentioned, instead of marginalizing the CMD into a one-dimensional LF and determining a color-ignorant magnitude of the TRGB, we retain all color-magnitude information in the form of a smoothed Hess Diagram (SHD) from which a complete two-dimensional description of the TRGB is to be extracted.

From the SHD, the two-dimensional gradients along the magnitude (vertical) and color (horizontal) axes are computed. The vertical gradient is then weighted by the inverse of the horizontal gradient. This weighting scheme was introduced when we saw in datasets with high photometric SNR that a stray signal coming from the steep discontinuity along the blue edge of the RGB could overpower the tip signal. Despite the majority of the gradient along the blue edge of the RGB being in the horizontal direction, the RGB is so much denser below the tip that even the subdominant vertical component of the RGB blue edge gradient can dominate over the tip signal. This was straightforwardly solved by dividing the vertical gradient by the horizontal gradient, which, of course, downweights gradients in the horizontal direction if they have a significantly larger horizontal component. This approach works because the slope of the F115W TRGB is $\leq -1$~\magSlopeErr{}, i.e., $\tan \phi_{\mathrm{CMD}}$. Additional details are provided in the Appendix.

An ellipse is then fit to the most prominent contour of the weighted vertical gradient (middle panel of \autoref{fig:4424_2d_edge}). The (weighted vertical) gradient computed in all cases revealed an elliptically shaped maximal contour that could be unambiguously identified as the TRGB feature. This feature was more prominent in those CMDs with higher SNR photometry and with cleaner (older) RGB populations. We set for each galaxy a minimum threshold value which would set the outermost boundary of the TRGB. An ellipse was then fit to this maximal TRGB contour and the parameters of that best-fit ellipse provide the formal description of the 2D TRGB. The ellipse angle (equivalent to the TRGB slope) is not used in our analysis and can be described as a nuisance parameter; necessary in the fitting process, but not when computing final distances. 
We will compare the distribution of per-galaxy slopes to our preferred sample-averaged estimate in the following section.

To estimate the uncertainty on a TRGB measurement, the length of the axis perpendicular to the TRGB contour was scaled by the \textit{contrast} of the TRGB ellipse. The \textit{contrast} is defined as the ratio of the number of stars located in the faint half of the TRGB ellipse’s semi-major axis to those above it. So, for a TRGB ellipse with orthogonal half-width $b= 0.10$~mag ellipse plus 40 stars below and 20 stars above its TRGB ellipse major-axis, the uncertainty $\sigma_{TRGB}$ would be equal to 0.05~mag. We show examples of the TRGB error estimation for illustrative limiting cases in the Appendix.

\subsubsection{Fit Results}

The centroid of the ellipse then provides the tip color and magnitude that we use to compute a distance. The gradient threshold that routinely produced stable tip solutions was always at least 90\%, with most values falling between 95\% and 98\%. Higher threshold values were typically found to best capture the tip in higher SNR photometry (e.g., M101, N4258, N5643). 

In \autoref{tab:trgb_measures} we provide the TRGB fitting parameters, including the kernel width used to smooth the Hess Diagram and the minimum threshold value. Also included for each field in the sample is the adopted foreground reddening, the TRGB's extinction-corrected magnitude-color coordinate, the number of stars contained in the TRGB ellipse, the TRGB contrast ratio, the slope of the TRGB (tangent of the ellipse position angle), and the TRGB measurement uncertainty.

All fitting parameters (bandwidth and threshold) were set while the photometric zero points of the program catalogs were still shuffled via random photometric offsets. In response to the discovery of outliers in our post-unblinded internal comparison between each of the Cepheids, TRGB, and JAGB, changes were made to the spatial selection for four targets: N2442, N4039, N5643, N7250. In all cases, we discovered upon closer inspection that the TRGB was underpopulated, a result of being too strict when masking the disk contribution to decontaminate the TRGB---a classic case of having to weigh purity against number statistics in the definition of a fitting population.
It was not surprising that such updates were needed, as we had to carefully isolate the TRGB in the outermost regions of imaging data that were locked onto the Cepheid variables, which themselves reside in the worst possible galaxy location for TRGB. The net shift in distance modulus for these four galaxies was $-0.11$~mag, which resulted in a $+0.8$~km/s/Mpc increase in $H_0$.

We provide in the Appendix additional details on the Tip fitting procedure.
 
\begin{figure}
    \centering
    \includegraphics[width=0.95\linewidth]{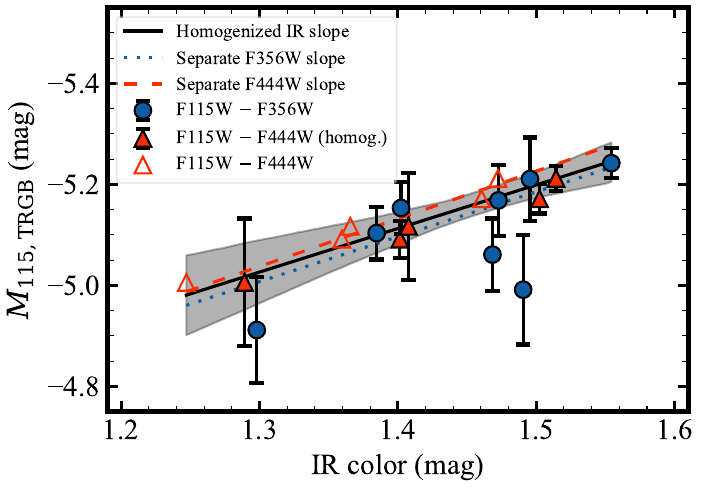}
    \caption{Determinations of the F115W TRGB color dependence. TRGB magnitude-color measurements for the F356W (blue circles) and F444W (open red triangles) \textit{separate} color scales are plotted along with the F444W datapoints that have been shifted onto an \textit{homogenized} IR color scale (filled red triangles). The lines of best fit for the separate F356W (dotted blue line) and F444W (dashed red line) color scales are also plotted, along with the best fit homogenized IR color scale (solid black line) and its 90\% confidence interval (gray band). The apparent F115W TRGB measurements were calibrated to absolute magnitudes using a combined set of external HST distances (SH0ES-22 Cepheids averaged with CCHP TRGB). Note the intercepts of these lines are not equal to the final adopted F115W zero points, which are set directly by measurements in geometric anchor galaxy NGC~4258.}
    \label{fig:sep_and_hom_slope_fits}
\end{figure}

\subsection{Color/Metallicity Slope}

In this subsection, we will use the tip coordinates determined from each target field to calibrate the slope of the F115W TRGB magnitude-color relation. The TRGB color correction can be alternatively and equivalently referred to as ``standardizing'' the best-fit tip magnitudes based on their corresponding colors. To do so, the apparent tip magnitudes $m_i^T$ need to be placed onto an absolute magnitude system.

\subsubsection{TRGB Absolute Magnitudes}
\begin{deluxetable}{llcclrr}
\tablecaption{F115W TRGB Slope Estimates\label{tab:trgb_slopes}} 
\tablehead{ 
\colhead{Color scale} &
\colhead{$c_0$} &
\colhead{$\beta$} &
\colhead{Std.Err.} &
\colhead{$\sigma_{\mathrm{RMS}}$} &
\colhead{$\sum \chi^2$} &
\colhead{$N_{dof}$}
\\
\colhead{} &
\colhead{[mag]} &
\colhead{} &
\colhead{[mag]} &
\colhead{[mag]} &
\colhead{} &
\colhead{}
}
\startdata
F356W        & 1.428 & -0.889  & 0.281            & 0.083   & 7.04 & 6          \\
F444W        & 1.330 & -0.944  & 0.148            & 0.015   & 0.65 & 3          \\
Homogenized  & 1.411 & -0.863  & 0.206            & 0.064   & 7.77 & 10 
\enddata
\tablecomments{TRGB color standardization parameters. From left to right, color scale indicates the IR color that was used to correct the F115W TRGB magnitudes, $c_0$ is the mean color value for the SN Hosts, $\beta$ is the best-fit value of the color slope, determined via least squares minimization of the slope and intercept for the separate color scales (F356W, F444W) and via MCMC sampling for the homogenized scale, with the addition of the F444W color offset term. The standard error on $\beta$ (Std.Err) quantity was determined via bootstrapping for the separate color scales and as the 68\% interval of posterior samples for the homogenized color scale. The $\sigma_{RMS}$ is the dispersion about the best-fit line. $\Sigma \chi^2$ is the sum of the $\chi^2$ statistic for each fit and $N_{dof}$ the number of degrees of freedom.}
\end{deluxetable}

We place the tip measurements of section 4.1 onto an absolute system by adopting an external set of distance constraints published for our target galaxies, i.e.,

\begin{equation}
M^{abs}_i = m^T_i + \mu^{ext}_i
\end{equation}

For the three measurements in NGC~4258, $\mu^{ext}_i$ is straightforwardly set by the megamaser distance $\mu_{0,4258} = 29.397 \pm 0.032$~mag \citep{Reid_2019}. For the ten SN host galaxies we use distances previously determined with HST, computed as the weighted average of the CCHP HST TRGB and SH0ES-22 HST Cepheid distances. The merging of the two distance sets for this purpose is justified. For the six galaxies in common, they agree to 0.001~mag and 0.003~mag in the weighted average and median, respectively, with a total dispersion equal to 0.09~mag.

With absolutely calibrated magnitudes for all targets we can begin calibrating the TRGB magnitude-color relation. Recall first that our data consist of a mix of observations acquired in the F356W and F444W bandpasses. Thus, we consider two different prescriptions for doing the color standardization, one which leaves the two sets of colors as their own \textit{separate} F115W distance scales, and another which \textit{homogenizes} them onto a single IR color scale.

\subsubsection{Separate Color Scales}

For the first approach, which we refer to as \textit{separate} color scales, we determine the slope in each of F356W and F444W separately via a simple weighted least squares fit. The intercept of this best fit relation is not retained (i.e., a nuisance parameter) because the zero point of the F115W TRGB magnitude-color relation will be set directly by the measurements made in the geometric anchor NGC~4258.

The slope on the \colorTHREEFIVE{} color system is found to be $-0.889 \pm 0.281$~\magSlopeErr{}. The RMS dispersion about the best-fit line is 0.078~mag. The sum of the $\chi^2$ is equal to 7.04 with 6 degrees of freedom. 
The slope on the \colorFOURFOUR{} color system is found to be $-0.944 \pm 0.148$~\magSlopeErr{}. The RMS dispersion about the best-fit line is 0.014~mag. The sum of the $\chi^2$ is equal to 0.65 with 3 degrees of freedom. These best-fit parameter estimates are enumerated in the first and second rows of \autoref{tab:trgb_slopes} for the F356W and F444W color scales, respectively.

A visualization of the slope estimation is presented in \autoref{fig:sep_and_hom_slope_fits}. In it, absolute magnitude-color coordinates from each of the thirteen fields are plotted as filled blue circles and open red triangles to represent whether the tip color was formed with F356W or F444W.
The corresponding best-fit slopes are plotted as a blue dotted and red dashed line. 

\begin{figure}
    \centering
    \includegraphics[width=0.95\linewidth]{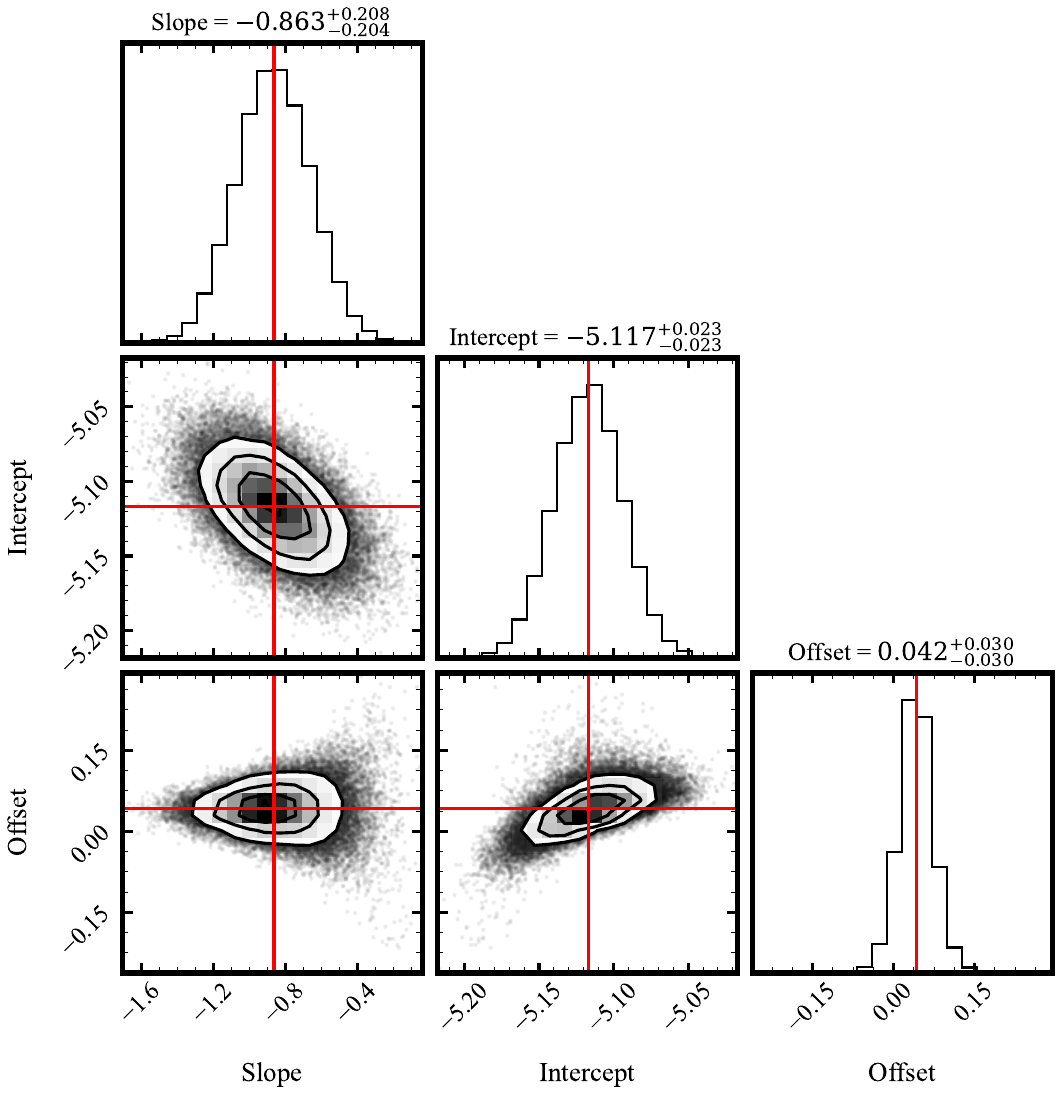}
    \caption{Corner plot for the estimate of the homogenized IR slope, which includes the simultaneous inference of an additive F444W color offset that best aligns it with the F356W color scale. Median sample estimates are marked by red lines and printed above each marginalized histogram, along with the 68\% confidence interval. Generated by the \texttt{corner} package \citep{ForemanMackey_2016} with sampling results from the \texttt{emcee} package \citep{ForemanMackey_2013}.}
    \label{fig:color_offset}
\end{figure}

\subsubsection{Homogenized Color Scale}

The separate F356W and F444W TRGB magnitude-color relations appear as parallel translations of one another. And since the vertical axis of both is the F115W magnitude, the source of the translational offset must be along the horizontal, or color, axis. The cause of this color offset is likely due to the brief departure of the SED from a smooth R-J tail at $\sim 5$\micron{} due to CO molecular absorption that is present in the majority of luminous red giants (see \autoref{fig:trgb_spectrum}). As a result, we can expect that for a constant F115W magnitude, the magnitude in F444W will be slightly fainter than F356W (all Vega magnitudes). We therefore have empirical evidence and physical motivation to support the construction of a single, \textit{homogenized} IR color scale to describe our full set of measurements.

The implementation of this color homogenization is straightforward. We add a single parameter to the color standardization model that quantifies the size of a horizontal color offset used to translate the F444W color scale onto the F356W system. Given the addition of this non-standard third parameter to the model of a two-dimensional, linear relationship, it will prove useful to carry out a more complete Bayesian treatment of the parameter inference. That way, we can inspect the posteriors directly to search for interesting correlations with the new parameter, if any. We sample the posterior using the \texttt{emcee} ensemble sampler \citep{ForemanMackey_2013}. 

In \autoref{fig:color_offset}, we show the resultant corner plot, made using the \texttt{corner} package \citep{ForemanMackey_2016} finding a slope estimate equal to $-0.863 \pm 0.206$~\magSlopeErr{} and color offset term equal to $0.042 \pm 0.03$~mag. We will present additional validations of the IR color homogenization in Section 7. Again, the intercept estimate is a nuisance parameter; it is only needed to provide an accurate constraint on the slope parameter, but is not retained any further. If we were to use this intercept, our F115W distance scale would be pegged directly to the previous HST ones, rendering the rest of this study a tautology. Instead, the zero point of our distance scale will be set independently and internal to the F115W TRGB by the tip measurements made in the megamaser host galaxy NGC~4258. 

\subsubsection{Color Slope Constraints}

In \autoref{tab:trgb_slopes}, the separate color corrections are tabulated in the first two rows and the homogenized color correction in the last row.
In \autoref{fig:sep_and_hom_slope_fits}, the separate color correction approach is represented by the open red triangles and filled blue circles, as well as the accompanying lines. The homogenized approach is represented by all filled datapoints as well as the black line, with its corresponding $90\%$ confidence interval plotted as a gray shaded region.

With a reduced $\chi^2$ of 0.25, the F444W relation appears anomalously tighter than it should be, given the uncertainties, while the opposite is true for F356W with its reduced $\chi^2 = 1.4$. Both cases are likely a result of the small number statistics in each sample. For this reason, we prefer the homogenized result, which exhibits better behaved statistics, e.g., a reduced $\chi^2 \simeq 1$.

In terms of systematics, the first approach has the advantage of not incurring additional uncertainty from the color offset homogenization. The second approach, on the other hand, can leverage the statistics of all three fields of NGC~4258, two of which were taken in F444W and one in F356W. This reduces the overall $H_0$ uncertainty because any uncertainty present at the zero point calibration stage propagates as a 100\% systematic in one's final estimate of $H_0$.

\begin{deluxetable*}{llll|ccccl}
\tablecaption{F115W TRGB Zero Point Calibration and Errors \label{tab:trgb_zp_calib}}
\tablehead{
\colhead{Color system} &
\colhead{Fields} &
\colhead{$M^{TRGB}_{4258}$} &
\colhead{$c_{4258}$} &
\colhead{Measurement} &
\colhead{Photometry} &
\colhead{Extinction} &
\colhead{Color corr.} &
\colhead{Total}
}
\startdata
F356W  &  O  & $-5.242$ & 1.554  & 0.030  & 0.02   & 0.009   & 0.036  & 0.052  \\
F444W  & IH  & $-5.194$ & 1.466  & 0.019  & 0.02   & 0.009   & 0.020  & 0.035 \\
Homog. & IOH & $-5.208$ & 1.549  & 0.016  & 0.02   & 0.009   & 0.023  & 0.036 
\enddata
\tablecomments{TRGB zero point calibration parameters associated with each relative calibration presented in \autoref{tab:trgb_slopes}. A vertical line separates the zero point measurements from the associated error budget terms. Color system indicates which color system is being calibrated. Field lists which in NGC~4258 were used in computing the absolute fiducial with abbreviations defined as: O--Outer disk; I--Inner Disk; H--Halo. $M^{TRGB}_{4258}$ and $c_{4258}$ represent the TRGB magnitude and color estimated from the N4258 data, where the magnitude has been absolutely scaled by the maser distance $\mu_{4258,Maser} = 29.397$~mag. Measurement is the combined Tip measurement error. Photometry is the adopted systematic uncertainty on the photometric zero point. Extinction is the adopted uncertainty on the reddening correction applied to N4258. Color corr. is the uncertainty introduced when extrapolating the 4258 zero point measurements to the SN Host fiducials in \autoref{tab:trgb_slopes}. Total is the quadrature sum of all preceding error terms and, when combined with the maser distance uncertainty, represents the total systematic error on the F115W TRGB distances. The 0.032~mag uncertainty on $\mu_{4258,Maser}$ \citep{Reid_2019} is not included in this table.}
\end{deluxetable*}

Regardless, we will see in the remainder of this section that the small differences between the separate and homogenized color corrected distances to individual galaxies amount to an even smaller net difference in distance modulus less than 0.01~mag. We choose to report both distance scales because of the newness of the methodology.

\subsection{Relative Distances and Uncertainties}

Armed with the newly determined slopes we can now standardize the TRGB magnitude-color coordinates onto a self-consistent distance scale. To do this, we continue with the postulate that there exists a universal, linear relation we can use to predict the intrinsic brightness of the F115W TRGB as a function of its IR color. This is a standard assumption that must necessarily underlie any form of distance measurement (e.g., a universal Cepheid Period-Luminosity relation). The distance modulus between any pair of galaxies $(i,j)$ is then,
\begin{equation}
    \Delta \mu_{ij} = \Delta m_{ij}^{T} + \beta \Delta \mathrm{color}_{ij}
    \label{eq:delta_mu}
\end{equation}
where $\beta$ represents the slope(s) of the TRGB magnitude-color relation(s) determined in the previous section and $\Delta \mu$ the color standardized distance between galaxies $i$ and $j$. $m_T$ and $color$ correspond to the tip coordinates presented in \autoref{tab:trgb_measures}.

To simplify the propagation of slope uncertainties in the error budget we elect to diagonalize the covariances by splitting our distance scale into one made up of two distinct TRGB fiducials, a relative one that is centered on the SN hosts and an absolute one that is centered on the measurements in NGC~4258. The location of the ``relative'' fiducial is set to the mean color of the SN Hosts and that of the ``absolute'' fiducial is set at the midpoint (in both color and color-corrected magnitude) of the NGC~4258 measurements. This decouples the uncorrelated (relative) uncertainties from correlated (absolute) ones and eliminates the need for a full covariance treatment.

The distance of each SN host galaxy from its relative fiducial is then,
\begin{equation}
    \mu_{i, rel} = m^{T}_i + \beta (c_i-c_0) - M_{115}|_{c=c_0}
    \label{eq:fiducial}
\end{equation}
where each $i>0$ is associated with one of the SN host galaxies and $c_0$ is the TRGB color averaged over all SN hosts that belong to each color scale.

The statistical error of a single SN host distance is then,
\begin{equation}
    \sigma_{i,\mu}^2 = \sigma_{i,tip}^2 + \left[\sigma_{\beta} \times (c_i - c_{0}) \right]^2 \label{eq:stat_err}
\end{equation}
where $\sigma_{\beta}$ is the standard error of the slope estimate. There are two additional terms in the uncertainty estimation not explicitly listed in \autoref{eq:stat_err}. The first is an uncertainty associated with the aperture corrections, which we adopt to be 0.02~mag for each field based on internal testing (see Section 3). The second is the uncertainty associated with the corrections for Milky Way foreground dust extinction. We adopt a 16\% uncertainty on the foreground reddening in the cases where the total value exceeds 0.06~mag and otherwise default to an error floor of 0.01~mag for low extinction targets, which corresponds to 0.0086~mag of extinction in the F115W bandpass \citep{Wang_2019}. The adopted quantities are enumerated in the $\sigma_{A_{115}}$ column of \autoref{tab:trgb_dists}.

The estimation of absolute distances and associated uncertainties will now be discussed.

\subsection{Absolute Calibration and Systematic Error Budget}

The zero point of the SN host fiducial $M_{115}|_{c=c_0}$ is set by extrapolating from the absolute fiducial defined by the NGC~4258 TRGB measurements,
\begin{equation}
    M_{115}(c=c_0) = m^{\mathrm{TRGB}}_{4258} + \beta (c_{4258}-c_0) - \mu_{4258,\mathrm{Maser}}
    \label{eq:zp_calib}
\end{equation}
where $m_{4258}$ and $c_{4258}$ are the average magnitude and color coordinates that represent the TRGB in NGC~4258. $\mu_{4258,\mathrm{Maser}}$ is the megamaser distance modulus, equal to $29.397 \pm 0.032$~mag \citep{Reid_2019}.

Constructing the distances in this way allows us to diagonalize the uncertainty matrix so that the zeropoint uncertainty can be subtracted out and a simple diagonal treatment of uncertainties becomes accurate in our statistical inference. The correlated errors can then later be reintroduced into the final quoted error budget on, e.g., $H_0$. 

In \autoref{tab:trgb_zp_calib}, the zero point intercepts derived from NGC~4258 are presented for each of the three color-standardized distance scales, corresponding to the identically indexed rows in \autoref{tab:trgb_slopes}. 
The magnitude-color coordinates $m^{TRGB}_{4258}$ and $c_{4258}$ are computed by averaging the tip measurements for the listed fields, which are drawn from any number of the Outer Disk (abbreviated as O), Inner Disk (abbreviated as I), and Halo (abbreviated as H). The weights used in the averaging are the inverse variances $1/\sigma_{tip}^2$ (see Table 2). 
For F356W, there is only one field in NGC~4258 so the values used are exactly those determined for the N4258-Outer field. For F444W, the Inner and Halo field measurements are first standardized to their mean color and then averaged in magnitude. The difference between the measured magnitudes is 0.040~mag before color correction and 0.029~mag after. The weighted average tip magnitude is 24.203~mag and the standard error, computed as the sum of the inverse variances, is 0.019~mag.
We do the same for the homogenized color scale which includes all three fields in NGC~4258. The standard deviation of the three measurements is 0.071~mag before color correction and 0.030~mag after. The weighted average tip magnitude is 24.187~mag with a standard error equal to 0.016~mag.
The magnitudes are then placed onto an absolute scale using the geometric megamaser distance $\mu_0 = 29.397 \pm 0.032$~mag \citep{Reid_2019}.

The corresponding systematic error budgets are broken down in the right hand side of the table. 
The Measurement column contains the standard errors on the weighted averages computed in the preceding paragraph. 
The Photometry and Extinction columns are adopted identically to their values adopted for the SN hosts. The values are repeated here because, while they are statistical (i.e., uncorrelated) uncertainties on a galaxy to galaxy basis, their contribution becomes \textit{systematic}, or perfectly correlated across the entire distance scale, by their association with the zero point anchor galaxy NGC~4258. 

Finally, the Color corr. term represents the systematic error in the F115W distance scale due to the TRGB being systematically redder in NGC~4258 than in the SN hosts. This quantity is computed by multiplying the slope uncertainty (Std. Err. in \autoref{tab:trgb_slopes}) by the difference in color between the absolute (NGC~4258) and relative (SN host) fiducials or, 

\begin{equation}
  \sigma_{sys,col.~corr.} = \sigma_{\beta} \left(c_{4258} - c_0\right)
\end{equation}

\subsection{Final Distances}

\begin{deluxetable*}{llllccclll}
\tablecaption{Final TRGB Distances and Errors \label{tab:trgb_dists}}
\tablehead{
\colhead{Host} &
\colhead{LW Band} &
\colhead{$\mu_{sep}$} &
\colhead{$\mu_{hom}$} &
\colhead{$\sigma_{tip}$} &
\colhead{$\sigma_{A_{115}}$} &
\colhead{$\sigma_{corr,sep}$} &
\colhead{$\sigma_{corr,hom}$} &
\colhead{$\sigma_{\mu_{sep}}$} &
\colhead{$\sigma_{\mu_{hom}}$}
}
\startdata
M101   & F444W  & 29.140 & 29.151 & 0.040  & 0.009  & 0.004  & 0.002  & 0.042  & 0.042 \\
N1365  & F356W  & 31.370 & 31.366 & 0.067  & 0.009  & 0.013  & 0.013  & 0.069  & 0.069 \\
N2442  & F356W  & 31.651 & 31.646 & 0.093  & 0.022  & 0.018  & 0.016  & 0.098  & 0.097 \\
N3972  & F356W  & 31.746 & 31.747 & 0.062  & 0.009  & 0.036  & 0.023  & 0.074  & 0.068 \\
N4038  & F356W  & 31.649 & 31.645 & 0.075  & 0.009  & 0.019  & 0.017  & 0.078  & 0.078 \\
N4424  & F356W  & 30.928 & 30.926 & 0.026  & 0.009  & 0.007  & 0.002  & 0.030  & 0.030  \\
N4536  & F444W  & 30.914 & 30.923 & 0.050  & 0.009  & 0.008  & 0.003  & 0.052  & 0.052 \\
N4639  & F356W  & 31.775 & 31.774 & 0.071  & 0.009  & 0.017  & 0.009  & 0.074  & 0.073 \\
N5643  & F356W  & 30.647 & 30.643 & 0.066  & 0.019  & 0.011  & 0.012  & 0.071  & 0.071 \\
N7250  & F444W  & 31.609 & 31.629 & 0.034  & 0.017  & 0.012  & 0.025  & 0.041  & 0.047 \\
\enddata
\tablecomments{Quoted uncertainties are statistical and do not include uncertainties on the zero point calibration (see \autoref{tab:trgb_zp_calib}). Subscript $_{sep}$ refers to distances and uncertainties estimated by correcting the TRGB magnitudes via a separate slope calibration for each of the F356W and F444W colors. Subscript $_{hom}$ refers to distances and errors when homogenizing the two sets of colors onto a common scale and using one slope for the color correction. The latter we adopt as the nominal TRGB distance scale.}
\end{deluxetable*}

In \autoref{tab:trgb_dists} we present the final distances estimated to the target galaxies as well as their associated uncertainties. These final distances are computed by combining the values presented in Tables \ref{tab:trgb_measures}, \ref{tab:trgb_slopes}, and \ref{tab:trgb_zp_calib}. In each row is a pair of distances $\mu_{sep}$ and $\mu_{hom}$ depending on which of the treatments of color standardization, separate or homogenized, was used. The TRGB measurement error $\sigma_{tip}$ is repeated from \autoref{tab:trgb_measures}. The statistical uncertainties associated with color correction to the SN host fiducial are listed in the $\sigma_{corr,sep}$ or $\sigma_{corr,hom}$, corresponding to which of the two color systems were used to perform the correction. $\sigma_{A_{115}}$ is the uncertainty on the adopted correction for Galactic dust extinction. Dust attenuation is negligible in either of the F356W or F444W bands. The final (statistical only) distance uncertainties are $\sigma_{\mu_{sep}}$ $\sigma_{\mu_{hom}}$.

These uncertainties can be identically thought of as the \textit{relative} distance uncertainties. The subscript $_{sep}$ and the LW band column associate a single distance with either of the separate F356W or F444W color-corrected distance scales. The subscript $_{hom}$ associates a distance that has been determined using the homogenized IR color correction.

We briefly describe here how the reader can accurately make use of the distances and uncertainties presented here in, e.g. a cosmological parameter inference. The $\sigma_{\mu}$ terms represent a statistical (uncorrelated) uncertainty relative to the SN host fiducial (defined in \autoref{tab:trgb_slopes}). Then, if one is interested in the total uncertainty on any given host distance, the corresponding error from the Total column in \autoref{tab:trgb_zp_calib} would be adopted as the systematic (uncorrelated) error. But when inferring a quantity such as $H_0$ from any single self-consistent set out of these three distance scales, the systematic error need only be reintroduced at the end. That is, the statistical weighting in one's cosmological parameter inference can be computed using $\sigma_\mu$, with the Total zero point uncertainty added at the end as a systematic on $H_0$, in addition to the geometric anchor uncertainty. 

\section{Establishing Distance Concordance}

In this section, we will put into context the new JWST TRGB distances by comparing with published ones derived with HST and either the $I$-band TRGB or Cepheids. These comparisons will reveal stark differences in the quality of published distances estimated to SN hosts, even when using the same imaging datasets. At the same time, we will be able to hone in on what appears to be an increasingly concordant, multi-method, multi-instrument distance scale that is centered on an accurate, unbiased application of the TRGB method. 

We will then combine the $I$-band TRGB distances previously measured by the CCHP using HST with the new ones estimated here using JWST. The two are in excellent agreement for the six targets that overlap between them. This new, composite CCHP TRGB distance scale will then be compared with the SH0ES-22 Cepheid distances, enabling a high-precision, independent check on the distances estimated to 16 SNe~Ia.

\subsection{Comparison with Prior HST TRGB Distances}

\begin{figure*}
    \centering
    \includegraphics[width=0.9\linewidth]{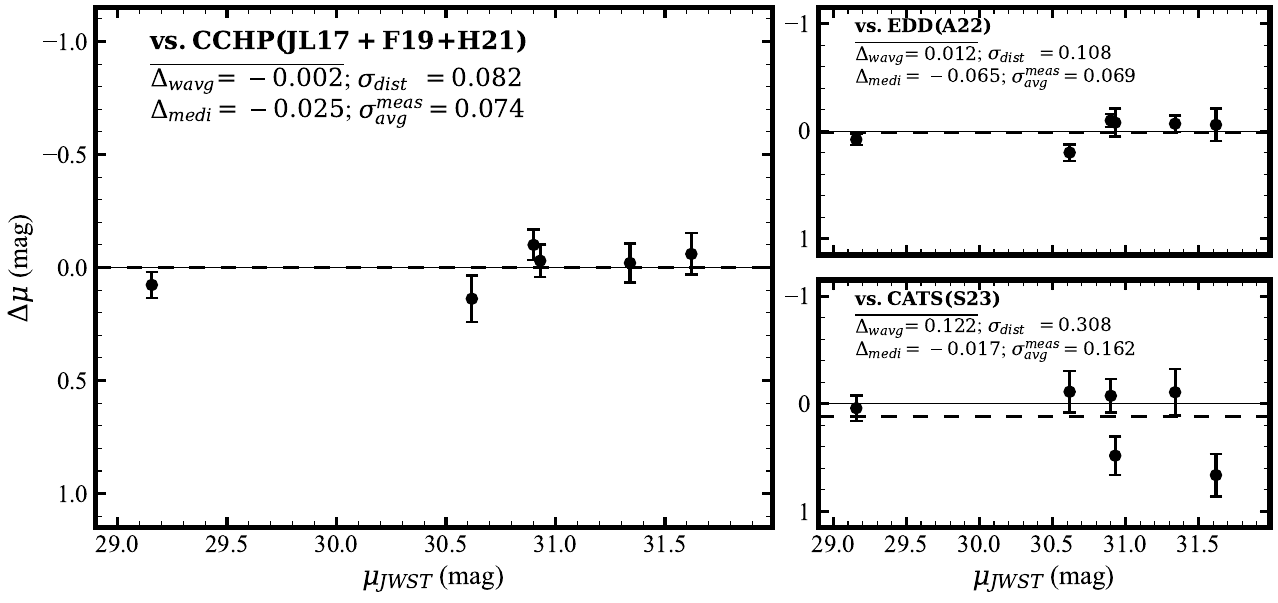}
    \caption{Comparison of new JWST TRGB distances with published HST TRGB distances in the literature. Six of the ten new JWST measurements overlap with the prior HST ones. Positive values on the y-axis correspond to the JWST TRGB distance being farther.}
    \label{fig:trgb_dist_comps}
\end{figure*}

Here, we compare this study's JWST/F115W TRGB distances with three sets that were each derived using the color-insensitive F814W ($I$)-band TRGB with HST. Each distance set represents an independent analysis of the same HST imaging data, of which six targets overlap with this study's sample. The three sources of published distances to which we compare include the previous phase of the CCHP \citep{Freedman_2019, Hoyt_2021}, the Extragalactic Distance Database \citep[EDD or CosmicFlows,][]{Anand2022ApJ...932...15A}, and the Comparative Analysis of TRGBs \citep[CATS,][]{Scolnic_2023}.

We show in \autoref{fig:trgb_dist_comps} the per-galaxy distance comparisons with HST-CCHP, EDD, and CATS, respectively, in the left, top-right, and bottom-right panels. In all three panels, a negative offset indicates the JWST distance from this study is closer than its corresponding estimate in the HST comparison set. Four summary statistics are written in each panel. $\Delta_{wavg}$, $\Delta_{medi}$, and $\sigma_{dist}$ are, respectively, the weighted average, median, and standard deviation of the differences in distance. $\sigma_{avg}^{meas}$ is the average of the reported uncertainties on the subtracted distance estimates. The uncertainty of each single comparison datapoint is estimated as the quadrature sum of the uncertainties on the pair of distances that were subtracted. This means the quantity $\sigma_{dist}/\sigma_{avg}^{meas}$ is very similar to a reduced $\chi^2$. Note the range of the vertical axis was set to a very large 2.1~mag (equivalent to a factor of 2.6 in units of physical distance). This was necessary to prevent points in the CATS comparison from extending beyond the bounds of the plot.

In the remainder of this section, we discuss the individual comparisons. As we will see, the F115W distances provide a high-precision baseline against which the precision of the HST distances can be evaluated. Of course, with a sample size of only six distances, statistical comparisons can become unreliable. However, it must also be emphasized that the \textit{exact} same HST imaging data were used by CCHP, EDD, and CATS in their respective analyses of the TRGB in these galaxies. This causes the three sets of HST distances to be strongly correlated with each other on a per-galaxy basis. Thus, the \textit{differences} seen between each of the three literature distances and this study's JWST measurements are more statistically significant than a naive comparison of the plots would convey.

\subsubsection{Carnegie Chicago Hubble Program (CCHP)}
The CCHP presented nine new TRGB distance measurements derived from their HST imaging program \citep[][F19]{Freedman_2019}. They combined those nine measurements with five others that were derived by \citet[][JL17]{Jang_2017_h0} from archival data. The CCHP result was then updated with improved calibrations of the TRGB zero point \citep{Jang_2021, Hoyt_2023} and augmented by two additional distances derived from new HST data \citep{Hoyt_2021}. These refinements were combined for a determination of $H_0$ in the review and analysis presented in \citet[][F21]{Freedman2021ApJ...919...16F}. We thus use ``HST CCHP'' as an umbrella term for JL17 \citep{Jang_2017_h0}, F19 \citep{Freedman_2019}, F20 \citep{Freedman_2020}, H21 \citep{Hoyt_2021}, F21 \citep{Freedman2021ApJ...919...16F}, and H23 \citep{Hoyt_2023}.

In the left panel of \autoref{fig:trgb_dist_comps} we plot the comparison with HST CCHP. We find excellent agreement between the earlier HST/F814W and new JWST/F115W distances derived by CCHP. The weighted average and median differences are 2 and $-21$~mmag, respectively. The dispersion of 0.082~mag is comparable to the average quoted uncertainty, equal to 0.074~mag, which suggests the distance uncertainties were accurately estimated in both analyses.

\subsubsection{Extragalactic Distance Database (EDD)}
The EDD is a companion to the broader CosmicFlows program, which aims to constrain large scale flows and structures in the nearby universe, employing various methods of distance measurement to do so. For nearby and lower mass galaxies, they use TRGB distances presented in the EDD \citep{Jacobs_09, Anand_2021} that are derived from fitting an analytical model (broken power law with a break at the TRGB) to the 1-D RGB luminosity function \citep{Makarov_2006, Karachentsev_2006, Rizzi_2007}. The latest version of EDD distances to SN host galaxies is presented in \citet{Anand2022ApJ...932...15A} and we compare with those distances here.

The EDD comparison is not qualitatively different from the CCHP one, though the dispersion is slightly larger at 0.108~mag. Compared to an average quoted uncertainty of 0.069~mag, there is evidence that the EDD TRGB methodology may be underestimating their measurement uncertainties. The significance of our quantitative interpretation is limited by the small number of comparison galaxies. Recall, though, that the EDD and CCHP used the exact same HST imaging data, so the dispersion in the EDD comparison being higher than that seen in the comparison with CCHP is more significant than it appears at face value. 

\subsubsection{Comparative Analysis of TRGBs (CATS)}
CATS is the TRGB branch project of the SH0ES collaboration. They adopted an ``unsupervised'' model of TRGB distance measurement. They argue that analysis choices, such as RGB color selection or the width of one's noise suppression kernel, should be held fixed for all TRGB measurements regardless of the characteristics of the underlying data. They also argue that such analysis choices only need to be explored in discrete bins, and further neglect to use artificial star experiments. Their methodology runs contrary to that adopted by the CCHP, which explored analysis choice parameters such as smoothing kernel width and spatial selection over smooth continua for all reasonable values. The CCHP also performed comprehensive artificial star injection experiments to optimize the adopted kernel width directly from the empirical data, as well as to robustly estimate uncertainties associated with the edge detection. See \citet{Hoyt_2021} for a description of the CCHP TRGB methodologies. 

Note that, despite the CCHP's publication of comprehensive data-driven RGB LF injection experiments and resultant inclusion of a smoothing-dependent bias term into their systematic error budget, \citet{Anand_2024} inaccurately claim that this very effect has not been explored, ``recommend[ing] that future studies closely examine the potential systematic effects of adopting different smoothing scales.''

\begin{figure*}
    \centering 
    \includegraphics[width=0.93\linewidth]{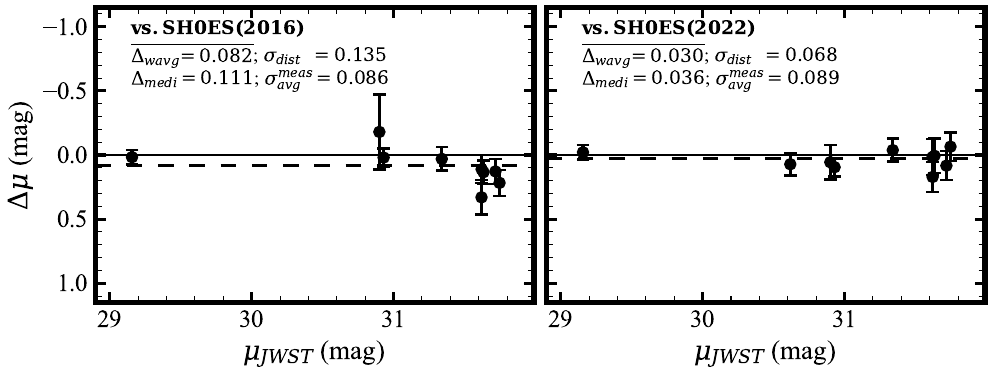}
    \caption{Same as \autoref{fig:trgb_dist_comps} but compared with the SH0ES HST Cepheid distances from 2016 \citep{Riess_2016ApJ...826...56R} and 2022 \citep{Riess_2022ApJ...934L...7R}. 
    Nine and ten of the galaxies for which we present new JWST TRGB distances overlap with the 2016 and 2022 SH0ES samples, respectively.}
    \label{fig:shoes_dist_comps}
\end{figure*}

CATS TRGB magnitudes and errors were taken directly from Table 1 of \citet{Scolnic_2023} which presents the measurements from their baseline analysis. The individual TRGB measurements are all brought onto a common $R=4$ value using their equation (1). If there were multiple reported TRGB magnitudes for a single galaxy, we used the weighted average as representative of the individual measurements. 

The comparison of our TRGB distances with CATS reveals a galaxy-to-galaxy RMS dispersion of 0.308~mag, significantly larger than that seen in our comparison with EDD (0.108~mag) and HST-CCHP (0.082~mag). The weighted average and median offset in distance is 0.122~mag and $-0.017$~mag, respectively. This discrepancy between mean and median suggests that significant outliers are driving the poor precision of the CATS measurements. Indeed, there are two clear outliers in the CATS distance comparison with distance modulus offsets of 0.482~mag and 0.663~mag. These correspond to host galaxies NGC~4536 and NGC~4038/9, respectively. It turns out that, for these two galaxies, CATS misidentified in their baseline analysis the tip of the AGB as the TRGB, which has introduced disconcertingly large biases into their TRGB distance estimates. It is not uncommon to find this same error in past TRGB studies \citep{Saviane_2004, Saviane_2008, Mould_2009, Wu2014AJ....148....7W}. As a result, their $H_0$ estimates are biased toward systematically higher values. We will discuss problems with the CATS analysis in Section 7.4 along with more detailed critiques in the Appendix.

\subsection{Comparison with HST Cepheid Distances}
In this subsection, we compare this study's F115W TRGB distances to two sets of Cepheid distances derived using HST. The two comparison sets used are either the 2016 or 2022 iterations of the SH0ES program \citep{Riess_2016ApJ...826...56R, Riess_2022ApJ...934L...7R}. 
Since we want to compare distances that have been measured directly to the same host galaxies (i.e., just Rungs 1 and 2 of the distance ladder), we adopt the SN-independent distance approximation parameter estimates from each SH0ES paper. Note this quantity is labeled differently in the two papers, as $\mu_{Ceph}$ in R16 and as $\mu_{Host}$ in R22.

Consistent with the TRGB comparisons presented in the previous section, both SH0ES analyses here used the same HST imaging of Cepheids in SN host galaxies. As a result, differences seen between the two distance comparisons are more significant than they appear at face value. The comparison is shown in \autoref{fig:shoes_dist_comps}, with the SH0ES-16 comparison on the left and the SH0ES-22 comparison on the right.

\subsubsection{SH0ES-16}
There is a significant systematic offset and dispersion that exceeds the quoted uncertainties in the comparison with SH0ES-16, suggesting their distances for these ten galaxies were biased and imprecise. The weighted average distance offset is $0.088 \pm 0.032$~mag and the median offset is $0.111 \pm 0.053$~mag, with uncertainties on the statistics estimated via bootstrapping. The RMS dispersion is 0.135~mag, while the average quoted (combined) uncertainty is 0.086~mag, indicating the pre-2022 SH0ES distance uncertainties were on average underestimated by about a factor of 1.5. This is consistent with the \citetalias{Freedman_2019} finding that a Hubble diagram constructed from pre-2022 SH0ES Cepheid distances (no SNe) exhibited a 1.4 times larger dispersion than if it were based on the CCHP TRGB distances.

Of course, summary statistics provide an incomplete description of residual distributions that exhibit a measurable functionality. Indeed, there appears to be a significant step in the distance comparison, with the break falling near $\mu=31.5$~mag (or $d=20$~Mpc). Beyond this distance, the SH0ES distances are all systematically underestimated, by anywhere between 0.15 and 0.35~mag. The clear presence of this distance dependent bias in the previous SH0ES analysis circumstantially supports the hypothesis presented in F25 that a similar distance-dependent bias may still be present in the SH0ES-22 distances, only now it may be hidden in the new, more distant targets still yet to be checked with the TRGB.

Looking at this from another angle, our new F115W TRGB distances were able to unambiguously diagnose the same bias in the pre-2022 SH0ES distances that had already been revealed by prior studies that used the $I$-band TRGB \citep{Jang_2017_h0, Freedman_2019}. This provides more evidence that the F115W TRGB distance scale established in this study is at least as precise and accurate as the CCHP's F814W TRGB distance scale, and significantly more so than the pre-2022 SH0ES Cepheid distance scale.

\begin{figure*}
    \centering
    \includegraphics[width=0.7\linewidth]{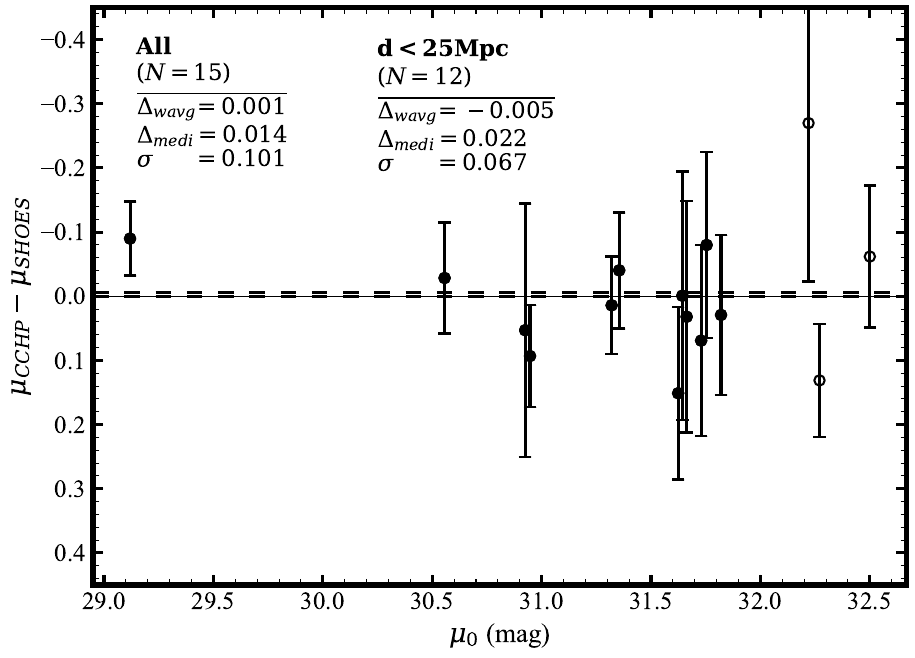}
    \caption{Combined CCHP TRGB distances to fifteen SN host galaxies that overlap with the SH0ES-22 sample. All distances and their uncertainties have been scaled to a common geometric zero point set by the NGC~4258 megamaser system. Six of the CCHP distances are jointly constrained by HST and JWST, four by JWST, and five by HST. The distances to these fifteen galaxies therefore appear to be confirmed to sub-percent accuracy, spanning a range of distances from 7 to 32~Mpc. Note the shorter y-axis limits than in Figures \ref{fig:trgb_dist_comps} and \ref{fig:shoes_dist_comps}.
    }
    \label{fig:final_trgb_shoes_comp}
\end{figure*}

\subsubsection{SH0ES-22}

The dispersion in the distance comparison with SH0ES-22 is the smallest and is also smaller than the average quoted uncertainty in the comparison. Our F115W TRGB distances are in the average (median) $0.030\pm0.023$~mag ($0.036\pm0.030$~mag) farther than the SH0ES-22 distances. The RMS dispersion is 0.068~mag and the average uncertainty 0.089~mag, suggesting slightly overestimated uncertainties in either or both sets of distances. However, this dispersion could be artificially small because these ten SH0ES-22 distances are the same as those used in Section 4.2 to estimate the TRGB's color slope. This could introduce a nonzero correlation between the two sets of distances. However, this would only be the case if there were a correlation of TRGB color with the external HST distances. We explicitly search for and rule out such a correlation in Section 7.2.

The comparison indicates, as has already been covered, that, for these ten galaxies, the SH0ES-22 distances have come into considerably better agreement with CCHP TRGB than were their previous estimates. Indeed, SH0ES' 2022 reanalysis of the same HST data they had used prior resulted in shifts in their distances reported to some of these galaxies as large as 0.3~mag, significantly larger than the size of the Planck-SH0ES Hubble Tension (0.16~mag). 

In the next subsection, we will increase the number of host distances that go into the comparison between SH0ES and CCHP by merging the CCHP's prior HST TRGB distances with the JWST ones presented in this study. 

\subsection{A Multi-method, Multi-wavelength Check on the Extragalactic Distance Scale}

The comparisons in the previous section allowed us to discern for ten SN host galaxies which sets of distances appear both accurate and precise (HST CCHP, SH0ES-22, or this study) and which of them appear less so (CATS, EDD, and pre-2022 SH0ES). The comparisons with CCHP and SH0ES, however, were only approximately ``fair'' at the level of 0.01-0.02~mag because we did not account for differences in which geometric calibrators set the zero point of each distance scale. For example, this study's zero point calibration is set by NGC~4258, but that anchor makes up about one-eighth of the statistical weight in the SH0ES-22 zero point calibration. Indeed, it turns out that half of the 0.03~mag offset between our F115W TRGB distances and SH0ES-22 is due to the differing geometric anchors between the two samples, with SH0ES-22 geometrically anchored to the LMC, Milky Way parallaxes, and NGC~4258, and our distances only to NGC~4258. 

To perform a more exact distance comparison, we need to renormalize all the distances in question onto the same geometric zero point. For this segment, we will only work with the distances from HST CCHP and SH0ES-22, which were found to be the most precise and accurate for these ten galaxies, at least according to the new JWST data. Since this study's distances have been calibrated only to NGC~4258, we need to shift those two HST distance sets from their multi-anchor zero points to one set by just NGC~4258. 

For HST CCHP, their final adopted TRGB zero point was $-4.049 \pm 0.038$~mag, coincidentally within 1~mmag of their separate calibration based on NGC~4258. The central value of the HST CCHP zero point was also exactly equal to that adopted in the initial CCHP TRGB $H_0$ from \citetalias{Freedman_2019}. We therefore can simply use the distances exactly as published in F19 (their Table 3). We augment those with the later \citetalias{Hoyt_2021} distances estimated to NGC~1404 and NGC~5643, which were both set to the same \citetalias{Freedman2021ApJ...919...16F} zero point.

We are then able to combine this study's ten color-standardized F115W TRGB distances with the sixteen $I$-band ones in HST CCHP to construct a new HST+JWST, CCHP-TRGB distance scale. As established at the beginning of this section, this expanded TRGB distance scale exhibits excellent self-consistency both in the average and in terms of uncertainty estimation. For the six galaxies in common between the two samples, we take the weighted average of the two distance estimates. Recall we showed earlier that the two sets of TRGB distances are consistent to 1\% in distance, and that their reported uncertainties have been reasonably and accurately estimated, i.e., the ratio $\sigma_{dist}$ and $\sigma_{avg}^{meas}$ approaches unity. This suggests the weighted average distances for these overlapping galaxies are likely to converge even closer to the true values than either set would have on its own, further tightening the already robust CCHP TRGB distance scale. This new composite CCHP distance scale comprises TRGB distances measured to twenty SN host galaxies (24 SNe~Ia) using either or both of HST/F814W and JWST/F115W. 

Fifteen of the SN host galaxies in the augmented CCHP sample are also in common with the SH0ES-22 sample and they will form the basis of our final distance comparison. We scale the SH0ES-22 distances onto a zero point based only on NGC~4258 by differentially using the corresponding analysis variant from Row 10 of \citetalias{Riess_2022ApJ...934L...7R} Table 5 to systematically shift, and proportionally inflate the uncertainties on, the $\mu_{host}$ distances in their Table 6. The SH0ES value of $H_0$ when using just NGC~4258 as an anchor calibrator was reported as $72.51 \pm 1.54$~km/s/Mpc. Compared to their baseline estimate of $73.04 \pm 1.01$~km/s/Mpc, this translates to a $0.016$~mag lengthening of the $\mu_{Host}$ distances, and a 1.52 times dilation of the associated distance uncertainties. At this level of regularization, the off-diagonal covariance per host is reduced to $<0.01$~mag, ensuring a robust comparison between homogeneously anchored distance scales. With a consistent zero point established for both the CCHP and SH0ES distances, we can now perform a more stringent stress test of the extragalactic distance scale. 

In \autoref{fig:final_trgb_shoes_comp}, we plot the residuals of the new composite TRGB distances against those from SH0ES-22 (on a 4258-only distance scale). We find excellent agreement. This demonstrates that the SH0ES-22 distances to these fifteen SN host galaxies marked a significant improvement over their previously determined, and significantly biased, distances published between 2016 and 2021.

It is crucial to recognize that the HST CCHP TRGB distances published in \citetalias{Freedman_2019} were key to uncovering the biases present in the earlier SH0ES distances (see Section 1). This underscores the need for more TRGB distances in order to check the remainder of the SH0ES Cepheid distance scale, which may still as a whole suffer from a distance-dependent systematic bias. Only 15 of the 37 SN host galaxies in the SH0ES sample have been checked with TRGB. Significantly more time invested into either HST or, preferably, JWST are needed to accomplish this.

\section{The Calibration of SNe~Ia}

The determination of $H_0$ via SNe~Ia is conceptually straightforward \citep[see, e.g.,][]{Sandage_1993}. Time-dependent photometry measurements, or light curves, are acquired for a sample of nearby SNe and used to estimate characteristics of the light curve that, according to some model, can be used to better predict the intrinsic brightness of a SN~Ia at some particular time in some particular band. Correcting the SN photometric measurements using light curve characteristics is by convention referred to as \textit{standardization}, but the process is no different from simply using the color (period) of the IR TRGB (a Cepheid) to better predict its intrinsic brightness, i.e., the use of a more complete empirical description.

Models for standardizing SN magnitudes began with the relation between SN peak magnitude and light curve decline-rate \citep{Phillips_1993}, as well as with SN color \citep{Tripp_1998}. Some modern examples of such corrections include the SALT model's linear stretch $x_1$ and color $c$ parametrization \citep{Guy_2007} or the quadratic stretch $s_{BV}$ employed in SNooPy \citep{Burns_2011}. The aim of an SN cosmology experiment (e.g., CSP or Pantheon) is then to standardize all one's supernovae in a self-consistent way, placing the brightness measurements onto a Hubble diagram that is internally and self-consistently defined by some floating intercept, or zero-point \citep[see, e.g.,][]{Riess_2016ApJ...826...56R}.

Redshift-independent distances derived from external methods such as the TRGB or Cepheids can then be used to place this floating Hubble diagram onto an absolute scale, having themselves been calibrated to a distance derived from trigonometric arguments (e.g., parallaxes, eclipsing binaries, maser cloud proper motions). This chaining of calibrations, sometimes referred to as a distance ladder, allows one to tie a geometrically-derived absolute distance to SNe that probe the mostly smooth Hubble expansion (e.g., for $z>0.02$ $v_{pec}/v_{cosmo} \lesssim 0.04$). We repeat here Eq. 5 from R22, which compactly describes this process,
\begin{equation}
    5 \log H_0 = M_B^0 + 5a_B + 25
\end{equation}
where the $a_B$ term encodes the floating fiducial intercept information. The quantity $M_B^0$ is the absolute magnitude set by the subsample of homogeneously standardized SNe that also have a TRGB or Cepheid distance and can be expanded as, 
\begin{equation}
    M_B^0 = <M_{B,i}> = < m_{B,i} - \mu_i >
    \label{eqn:sn_absmag_calib}
\end{equation}
where $i$ is indexed to each SN with an external distance. Each $m_{B,i}$ is the standardized B magnitude at maximum for each SN and $\mu_i$ is the external, redshift-independent distance estimated with TRGB or Cepheids. This last step is commonly referred to as \textit{calibration} of the SNe~Ia and is what will be discussed in this section. The difference of each SN's  calibrated absolute magnitude from the sample-wide average is referred to as its Hubble residual, or,
\begin{equation}
    \Delta M_{B,i} = m_{B,i} -  <M_{B,i}>
\end{equation}

\subsection{SN Calibration Datasets}

For the TRGB-SN calibration, we consider two sets of TRGB distances $\mu^0_i$ and four published sets of Tripp-standardized SN magnitudes $m_{B,i}$. The four SN magnitude sources consist of two iterations published by each of two collaborations. This particular choice in varying the SN calibration dataset is intended to provide a useful summary insight into how $H_0$ can be impacted by the many choices and selections that underlie large SN compilations, both internally and externally.

The two sets of TRGB distances are:
\begin{enumerate}
    \item This study's new F115W TRGB distances measured to ten SN host galaxies (11 SNe).
    \item The combined set of CCHP TRGB distances measured with either HST or JWST for 20 SN host galaxies (24 SNe). See Section 5.
\end{enumerate}

The four sources of Tripp-standardized $B$-band magnitudes are,
\begin{enumerate}
    \item The first iteration of the Carnegie Supernova Project \citep[CSP-I,][]{Hamuy_2006}, used in the CCHP's previous TRGB determination of $H_0$. Standardized $m_{B'}$ magnitudes are adopted directly from Table 3 of \citet{Freedman_2019}.
    \item The CSP's second iteration, or CSP-II \citep{Phillips_2019} that was used in F25, this program's multi-method determination of $H_0$. Standardized CSP-II magnitudes are calculated from \citetalias{Uddin2023arXiv230801875U} using their Equations 1 and 5 with standardization information from their Table 5 and light curve parameters from their Tables A3 and A4.
    \item The SuperCal compilation of SNe, used in the SH0ES 2016 determination of $H_0$ \citep{Scolnic2015ApJ...815..117S}. Standardized magnitudes are from Table 5 of \citet{Riess_2016ApJ...826...56R}.
    \item The Pantheon+ compilation of SNe used in the SH0ES 2022 determination of $H_0$ \citep{Scolnic_2022ApJ...938..113S}. Standardized magnitudes are from Table 6 in \citet{Riess_2022ApJ...934L...7R}.
\end{enumerate}

With two sets of distances and four sets of SN magnitudes, we will have eight different calibrations to consider, each with varying degrees of correlation due to overlapping SNe. 

The JWST/F115W TRGB distances will be used to calibrate 9 SNe in CSP-I and SuperCal, and 11 in CSP-II and Pantheon+. The difference in size between the two samples is due to the later addition of NGC~5643, host to SN siblings 2013aa and 2017cbv.

The expanded set of 20 CCHP TRGB distances will be used to calibrate 22, 24, 14, and 17 SNe in CSP-I, CSP-II, SuperCal, and Pantheon+, respectively. The difference between the CSP- I and II samples is again the sibling SNe 2013aa and 2017cbv. The additional SNe in Pantheon+ relative to SuperCal also include 2013aa and 2017cbv, as well as SN~2021pit, the recent sibling to SN~2001el that was hosted by NGC~1448. The larger CSP sample sizes are due to some SN host galaxies having a reference TRGB distance, but no SH0ES Cepheid distance, primarily because they are E and S0 galaxies with no Cepheids, but also in two cases where the SNe were later cut out of SH0ES on account of being too reddened. These include SNe~1980N, 1981D, and 2006dd hosted by NGC~1316, SNe~2007on and 2011iv hosted by NGC~1404, SN~1994D hosted by NGC~4526, SN~1989B hosted by M66 (NGC~3627), and SN~1998bu hosted by M96 (NGC~3368).\footnote{Note we do not add the Pantheon+ magnitudes for elliptical galaxies published separately by CATS \citep{Scolnic_2023}. We found $0.03$~mag inconsistencies in the magnitudes of SNe in common between SH0ES and CATS, despite both being attributed to Pantheon+. The source of this internal discrepancy in the magnitudes of calibrator SNe presented in \citet{Riess_2022ApJ...934L...7R}, \citet{Scolnic_2022ApJ...938..113S}, and \citet{Scolnic_2023} is unclear.}

As discussed at the start of this section, all SNe in a single study's published Hubble Diagram will have been standardized to some fiducial definition that is unique to that study's sample of SNe and its adopted standardization model. As such, comparing results across different SN datasets is most accurately carried out in terms of $H_0$, for which these definitional differences cancel out.

\subsection{Estimating $H_0$}
\begin{deluxetable*}{lllrrlll}
\tablecaption{Reference Values of SN Calibrations of $H_0$ \label{tab:sncal_ref}}
\tablehead{
\colhead{SN Source} &
\colhead{$N$} &
\colhead{Dist. Source} &
\colhead{$<M_B>$\tablenotemark{a}} &
\colhead{$\sigma_{M_B}$} &
\colhead{$H_{0}$} & 
\colhead{SN Ref.} &
\colhead{Dist. Ref.}
\\
\hline
\colhead{} &
\colhead{} &
\colhead{} &
\colhead{[mag]} &
\colhead{[mag]} &
\colhead{[km/s/Mpc]} &
\colhead{} &
\colhead{}
}
\startdata
CSP-I      & 18 & CCHP TRGB & $-19.223$ & 0.13 & $69.8\;\: \pm 1.9$  & \citetalias{Freedman_2019} & \citetalias{Freedman_2019,Hoyt_2021} \\
CSP(I+II) & 20\tablenotemark{b} & CCHP TRGB & $-19.165$ & 0.16 & $69.87     \pm 0.69$\tablenotemark{c} & \citetalias{Uddin2023arXiv230801875U} & \citetalias{Freedman_2019,Hoyt_2021}  \\
SuperCal   & 19 & SH0ES Ceph.\tablenotemark{d} & $-19.287$ & 0.13  & $71.99    \pm 1.74$  & \citetalias{Riess_2016ApJ...826...56R} & \citetalias{Riess_2022ApJ...934L...7R} \\
Pantheon+  & 42 & SH0ES Ceph. & $-19.259$ & 0.13 & $73.04    \pm 1.04$ &
\citetalias{Riess_2022ApJ...934L...7R} & \citetalias{Riess_2022ApJ...934L...7R}
\enddata
\tablecomments{This is a compilation from published studies and does not include any new determinations of the SN absolute magnitude. SN source indicates from which study the $m_B$ magnitudes were compiled. $N$ is the number of calibrator SNe used in the original study. Dist. Source specifies which set of distances were used to provide the absolute calibration. $<M_B>$ is the inverse-covariance-weighted average computed here to best reproduce each study's original value. $\sigma_{M_B}$ is the magnitude dispersion of the SN calibration also redetermined here. $H_0$ is the value quoted by the original study, except for SuperCal which was updated from the \citetalias{Riess_2016ApJ...826...56R} to \citetalias{Riess_2022ApJ...934L...7R} distances. SN and Dist. Ref. contain the references for the calibrator SN magnitudes and calibrating distances, respectively.}
\tablenotetext{a}{Each row of $M_B$ values is defined relative to a different fiducial set by the original study's cosmological sample of SNe. As a result $M_B$ values can only be compared within each row, and not across multiple rows.}
\tablenotetext{b}{Table 2 of \citet{Uddin2023arXiv230801875U} quotes 18 TRGB calibrators, but Table A.4 contains 20; we use the latter here.}
\tablenotetext{c}{Only statistical error quoted.}
\tablenotetext{d}{Distances updated to the latest from \citetalias{Riess_2022ApJ...934L...7R}. If the original \citet{Riess_2016ApJ...826...56R} were used, we computed $<M_B> = -19.250$~mag and $\sigma_{M_B} = 0.16$~mag, both in exact agreement with the original study's values.}
\end{deluxetable*}

To estimate $H_0$ in this section, we do not undertake a new analysis of the cosmological, or Hubble Flow, SNe. We will instead update the published SN calibration (the $M_B^0$ quantity) of each study using the specified set of TRGB distances. The change in the SN calibration, defined as $\Delta M_B^0 \equiv M_B^0 (\mathrm{new}) - M_B^0(\mathrm{ref})$, is related to a differential shift in $H_0$ as in,
\begin{equation}
    \Delta H_0 = \frac{H_0 \ln 10}{5} \Delta M_B \label{eq:deltaH0}
\end{equation}
where $\Delta H_0 \equiv H_0 (\mathrm{new}) - H_0 (\mathrm{ref})$ and (ref) refers to the reference value from the original study.

We now compute each source study's reference calibrations as the inverse-covariance-weighted average $M_B^0$ directly from their published tables. The off-diagonal elements of $C_{ij}$, where $i$ indexes to a SN and $j$ to a host galaxy, are populated by the distance uncertainties $\sigma_{\mu_{j}}$ for only those galaxies to have hosted more than one SN. In the case of Pantheon+, additional off-diagonal covariance terms exist in their SN magnitudes because they determine a unique set of light curve parameters for each photometry source of each SN, rather than, e.g., a single set of light curve parameters for each SN determined by simultaneously fitting all of its light curve data at once. We have instead adopted here the $m_B$ magnitudes from \citetalias{Riess_2022ApJ...934L...7R} Table 6, which were computed as a simple average for those SNe that have multiple light curves in Pantheon+.
The reference $M_B^0$ values computed for CSP-I, CSP-II, SuperCal, and Pantheon+, respectively, are $-19.219$, $-19.165$, $-19.250$, and $-19.259$~mag.\footnote{When computing $M_B$ for Pantheon+SH0ES, we use the $\mu_{Host}$ values from their Table 5 to be consistent with the SN-independent nature of the TRGB distances used in this section.}
These can be compared to the values reported in the original studies, or $-19.225$~mag, $-19.171$, $-19.25$, and $-19.253$~mag. The $\sim 5$~mmag of residual disagreement is most likely due to differences in the naive computation of a covariance-weighted average here as opposed to the complete treatment in the original studies in which $M_B$ is marginalized out of a simultaneous $H_0$ fit. Viewing this from the other direction, the good agreement between the simple weighted average values and the marginalized ones emphasizes the degree to which the SN calibration is decoupled from the cosmology sample in measurements of $H_0$, primarily because of the velocity-independent nature of the SN calibration.

Because we are doing a weighted average determination here, we adopt those estimates as the reference $M_B$ values to associate with each study's reported $H_0$, i.e., $(-19.219, 69.80)_{CSP-I}$, $(-19.165, 69.87)_{CSP-II}$, $(-19.250, 73.24)_{SuperCal}$, $(-19.259, 73.04)_{PanPlus}$.\footnote{Note \citetalias{Riess_2022ApJ...934L...7R} do not provide the $H_0$ value from their separate $\mu_{Host}$-SN $H_0$ treatment of the fit in the same range of redshifts, $0.023 < z < 0.15$, as their baseline analysis, only providing a value of 73.30~km/s/Mpc that results from expanding their SN redshift range out to $z < 0.8$. We therefore use their baseline result of 73.04~km/s/Mpc as reference, because it is unclear which of the two changes, the expanded redshift range or the separation of rungs 2 and 3, resulted in the $0.3 \sigma$ shift in $H_0$.}

\begin{deluxetable*}{l|rccc|rcc|ccc}
\tablecaption{Recalibrations of Various SN~Ia Magnitude Samples\label{tab:trgb_sn_cal}}
\tablehead{
\colhead{SN Source} &
\colhead{$N_{SN}$} &
\colhead{$<M_B>$} &
\colhead{$\sigma_{M_B}$} &
\colhead{$H_0$} &
\colhead{$N_{SN}$} &
\colhead{$<M_B>$} &
\colhead{$\sigma_{M_B}$} &
\colhead{$\Delta M_B$} &
\colhead{$H_0$} &
\colhead{$\sigma$ signif.}
\\
\colhead{} &
\colhead{} &
\colhead{[mag]} &
\colhead{[mag]} &
\colhead{[km/s/Mpc]} & 
\colhead{} &
\colhead{[mag]} &
\colhead{[mag]} &
\colhead{[mag]} &
\colhead{[km/s/Mpc]} &
\colhead{}
}
\startdata
CSP-I     & 9  & $-19.267$ & 0.086 & 68.39 & 22 & $-19.225$ & 0.129 & $0.042 \pm 0.016$ & 69.73 & 2.6 \\
CSP(I+II) & 11 & $-19.205$ & 0.139 & 68.58 & 24 & $-19.178$ & 0.160 & $0.027 \pm 0.020$ & 69.45 & 1.4 \\
SuperCal  & 9  & $-19.359$ & 0.055 & 69.60 & 14 & $-19.322$ & 0.095 & $0.037 \pm 0.019$ & 70.83 & 1.9 \\
Pantheon+ & 11 & $-19.360$ & 0.080 & 69.64 & 17 & $-19.301$ & 0.140 & $0.059 \pm 0.019$ & 71.62 & 3.1
\enddata
\tablecomments{Two vertical lines distinguish SN calibrations based on the ten F115W TRGB distances presented in this study (left), from those based on a combined HST+JWST set of twenty-four TRGB distances (middle), and the difference between the two samples in units of $\sigma$ (right).
SN Source indicates the source of standardized magnitudes that are calibrated with the TRGB distances. $N_{SN}$ indicates the number of supernovae in a calibration sample. $\Delta M_B$ is the shift in the Hubble Diagram intercept relative to the original study, equivalently referred to the average magnitude shift in Hubble residuals. $H_0$ is the value estimated by shifting the $H_0$ reported in the reference study by the previous column's corresponding magnitude shift (via \autoref{eq:deltaH0}). $\sigma_{cal}$ is the dispersion of the calibrator Hubble residuals (in magnitudes). $\sigma$ signif. represents the statistical significance of the shift in $H_0$ seen upon expanding the SN sample, estimated by directly bootstrapping the calculation of $\Delta M_B$.}
\end{deluxetable*}

Finally, we update the reference calibration for SuperCal+SH0ES using the latest SH0ES Cepheid measurements from \citetalias{Riess_2022ApJ...934L...7R}. This results in a shift $\Delta M_B = -0.037$~mag, or $\Delta H_0 = -1.25$~km/s/Mpc, bringing the R16 $H_0$ value down to 71.99~km/s/Mpc, i.e., $(-19.287, 71.99)_{SuperCal}$. This intermediate shift seems meaningless at first, but will aid in disentangling the different impact that distance and SN uncertainties have on the measurement of $H_0$. Now, the SuperCal and Pantheon+ reference calibrations will be based on the same set of HST Cepheid distances \citepalias{Riess_2022ApJ...934L...7R}, paralleling the CSP- I and II reference calibrations being based on the same set of HST TRGB distances \citepalias{Freedman_2019, Hoyt_2021}. These reference calibrations are enumerated in \autoref{tab:sncal_ref}.

For the remainder of this section, the impact of the TRGB distances on $H_0$ will be approximated via shifts relative to the reference values. Based on the typical (small) disagreement between our estimate of each study's reference $M_B$ value with the published one, we conclude that the approximate calibrations undertaken in this section are reliable to about 5~mmag, or 0.15~km/s/Mpc. Any shifts measured to be at this level or smaller should be considered insignificant, or consistent with zero.

\subsection{TRGB-SN Calibration of $H_0$}

\subsubsection{JWST sample of SNe}

We first consider the SN calibration based on the ten F115W TRGB distances presented in this study. The resulting average $<M_B>$ values for CSP-I, II, SuperCal, and Pantheon+, respectively, are $-19.267$, $-19.205$, $-19.359$~mag, and $-19.360$~mag, corresponding to updated $H_0$ values of 68.39, 68.58, 69.60, and 69.64~km/s/Mpc. These results are tabulated in the first block of \autoref{tab:trgb_sn_cal}.

There is an encouraging $<0.2$~km/s/Mpc consistency between iterations from the same team, though still a sizable 1~km/s/Mpc systematic offset between the two groups. In the next subsection, we will expand the sample to the larger HST+JWST set of TRGB distances. We will see that the consistency between CSP-I and II is upheld, while the SuperCal and Pantheon+ calibrations diverge significantly.

\subsubsection{Augmented sample of SNe}

We then consider the set of TRGB distances formed by combining this study's JWST/F115W TRGB distances with the CCHP's previous TRGB distances (see Section 5). The resulting average shifts in Hubble residuals for CSP-I, II, SuperCal, and Pantheon+, respectively, are $-19.223$~mag, $-19.178$~mag, $-19.322$~mag, and $-19.301$~mag, corresponding to updated $H_0$ values of 69.73, 69.45, 70.83, and 71.62~km/s/Mpc. These results are tabulated in the second block of \autoref{tab:trgb_sn_cal}.

From this differential SN analysis, it becomes clear that the CSP-standardized Hubble diagram results in consistently smaller $H_0$ values than does either SuperCal or Pantheon+. Recall that the same TRGB distances are being used in all four calibrations here, so trends seen in this section are going to be caused entirely by differences in the SN analysis.

Additionally, the expanded sample of SNe yield systematically higher values of $H_0$ than the JWST-only one. This could be due to some combination of the following: the eleven SNe in the JWST sample disproportionately populate the bright tail of the (standardized) SN~Ia luminosity distribution, or the opposite, wherein the SNe newly included in the expanded sample over-represent the faint tail of SNe. 

We note that our CSP(I+II) $H_0$ value (69.5~km/s/Mpc) differs from that reported in F25 (70.4~km/s/Mpc). This appears to be due to different stretch-correction parameter values reported in \citetalias{Uddin2023arXiv230801875U} and F25. \citetalias{Uddin2023arXiv230801875U} report P1 $-1.09$ $\pm$ 0.10 and P2 $-0.68$ $\pm$ 0.29, while F25 report P1 $-0.91$ $\pm$ 0.10 and P2 $-0.32$ $\pm$ 0.29. Furthermore, the distribution of $s_{BV}$ in the augmented sample of calibrator SNe has a long tail toward smaller values (faster declining SNe), exacerbating the effect this difference in the standardization coefficients has on the Hubble constant. This further emphasizes the need for a larger number of calibrator SNe.

\subsubsection{Comparing the small and large calibration samples}

For the JWST-only sample, the difference between CSP-I and II is $+0.19$~km/s/Mpc, and the difference between SuperCal and Pantheon+ is $+0.04$~km/s/Mpc. This suggests some internal consistency regarding how these 11 SNe are treated in the calibration. For the expanded sample, on the other hand, this trend diverges significantly between the two groups. The difference between CSP-I and II in the expanded sample reverses sign but remains small at $-0.28$~km/s/Mpc, while the difference between SuperCal and Pantheon+ increases to $+0.79$~km/s/Mpc. 

That is, the CSP-I and CSP-II calibrations are very consistent regardless of which set of distances is used, the JWST or the expanded sample. However, the same is not true when comparing SuperCal and Pantheon+ across the two samples. The Pantheon+ trend runs strongly against expectation, wherein the internal self-consistency seen between the SuperCal and Pantheon+ treatments for the smaller JWST-only sample of SNe is lost in the larger sample.  

We can demonstrate this more exactly by restricting the CSP-II and Pantheon+ samples to the exact same 22 and 14 calibrator SNe that overlap with their earlier respective CSP-I and SuperCal iterations. As a result, the CSP(I+II) values of $H_0$ change to 68.93 and 69.58~km/s/Mpc for the small and larger samples, respectively. This worsens the apparent agreement between CSP-I and II for the JWST sample from 0.2 to 0.5 km/s/Mpc, but improves it in the larger sample from 0.3 to 0.1 km/s/Mpc. Doing the same for Pantheon+, the computed $H_0$ values shift to 70.04 and 72.08~km/s/Mpc for the JWST and expanded samples, respectively. Just as in the case of CSP, this worsens the agreement with SuperCal in the JWST sample from exact to 0.4 km/s/Mpc but, contrary to the case of CSP, significantly worsens the agreement in the expanded sample, from 0.8 km/s/Mpc to 1.3 km/s/Mpc.

To begin understanding this discrepancy between the \citet{Scolnic2015ApJ...815..117S} SuperCal and \citet{Scolnic_2022ApJ...938..113S} Pantheon+ magnitudes, we can compare them for all 19 SNe in common between them, and bypass the distance calibration entirely. To do so exactly, we have to first shift the two onto the same fiducial Hubble diagram as set by the $5a_B$ parameter in Equation 12. This difference in the SuperCal and Pantheon+ $m_B$ magnitude systems amounts to 0.007~mag. We discuss this calculation in detail in the Appendix. For these 19 calibrator SNe, SuperCal and Pantheon+ disagree by 0.037~mag in the average, corresponding to 1.28~km/s/Mpc in units of $H_0$. The size of this internal shift in their magnitudes of the same calibrator SNe is larger than the \textit{total} $H_0$ uncertainty of 1.04~km/s/Mpc reported by SH0ES \citep{Riess_2022ApJ...934L...7R}, providing more evidence of underestimated uncertainties in the Pantheon+SH0ES estimate of $H_0$.

\subsubsection{Summary of SN Calibration Findings}
The JWST sample calibrations were generally more consistent between different iterations of the same group and also exhibited smaller dispersions. This could be signaling that the SNe in the smaller JWST-TRGB sample are more ``normal'' in that they cluster near the core of the SN~Ia population distributions, leading to more stable light curve fits and standardization. The instability seen between SuperCal and Pantheon+ in the larger calibration sample may suggest those SNe are predominantly drawn from the wings of the SN~Ia characteristic distributions, where standardization is both less accurate and less precise. If the distribution of any less normal SNe is skewed in the calibration sample, and that skew is not matched by the cosmology sample, then the inference of $H_0$ would become biased.

Interestingly, this was not the case for CSP(I+II) which exhibited similar, larger dispersions for both the JWST and augmented samples (0.14 and 0.16 mag). Furthermore, the shift in $H_0$ between the two samples was also the smallest, with a significance of only $1.4\sigma$. This may be pointing to the other three analyses having over-trained their standardization model to the kinds of SNe more like those present in the JWST-only sample, i.e., the core of the SN~Ia population. On the other hand, the recent CSP(I+II) SN analysis may have overcome this with a more complete training sample than used by the other three analyses \citep[recall that Pantheon+ uses SALT2.4,][]{Betoule_2014}. This could be indicating that despite the statistical dispersion in CSP-II being larger, there may also be less bias toward a particular subpopulation of SNe.

Despite these concerns regarding the treatment of calibrator SNe in determining $H_0$, it is notable that six of the eight calibrations considered in this section suggest $H_0 \leq 69.7$~km/s/Mpc. Furthermore, only one of the eight calibrations suggests a value of $H_0$ larger than 71~km/s/Mpc, and so the TRGB-SN value of $H_0$ appears increasingly at odds with the higher-valued SH0ES estimate of $H_0=73.04 \pm 1.04$~km/s/Mpc. 

When comparing the JWST-only to the expanded CCHP TRGB sample, the Pantheon+ magnitudes yield the most significant shift in $<M_B>$ of $0.059 \pm 0.019$~mag ($3.1\sigma$), corresponding to 2~km/s/Mpc. On the other hand, the CSP-II calibration sees the smallest shift of $0.027 \pm 0.020$~mag ($1.4 \sigma$) equivalent to 0.8~km/s/Mpc. 
The CSP-I and II calibrations agreed in both the JWST and expanded samples, while Pantheon+ shifted into 1.2 km/s/Mpc disagreement with SuperCal when considering exactly the same SNe. While out of the scope of the current study, a more detailed investigation into why Pantheon+ deviates significantly from other SN analyses, including their own earlier SuperCal, is warranted.

\subsection{Comparing TRGB vs. Cepheid Calibrations of the same SNe}

In the previous subsection, we calibrated four sets of SN magnitudes with the same set of CCHP TRGB distances. In this subsection, we address the other side of the $M_B$ equation, now keeping the SN magnitudes the same while changing only the distances. We use the CCHP TRGB and SH0ES Cepheid distances to calibrate the same 17 Pantheon+ SNe common to both datasets. We inspect the resultant $\chi^2$ distributions of the $M_B$ calibration and find that the TRGB significantly reduces the influence that outlier SNe have on the measurement of $H_0$.

For the 17 SNe in common between SH0ES Cepheid and CCHP TRGB, we compute $M_B$ and the corresponding $\chi^2$ statistic from either the Cepheid or TRGB distances, i.e.,
\begin{equation}
    \chi^2 = \frac{(m_{B,i} - \mu_i)^2}{\sigma_{m_{b,i}}^2 + \sigma_{\mu_i}^2}
\end{equation}
where $m_{B,i}$ are the standardized peak SN magnitudes taken directly from Pantheon+ and the $\mu_i$ are either from SH0ES-22 or the combined CCHP TRGB.

The reduced $\chi^2$ based on the SH0ES Cepheid distances is 1.53 while that based on the CCHP TRGB ones is 1.07. The weighted standard deviation in the $M_B$ calibration is equal to 0.147~mag and 0.121~mag when based on the SH0ES and CCHP distances, respectively.

\begin{figure}
    \centering
    \includegraphics[width=0.9\linewidth]{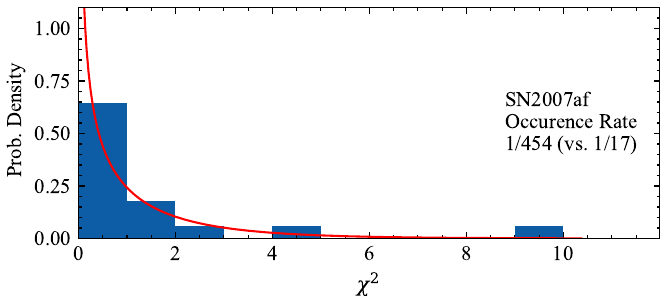}
    \includegraphics[width=0.9\linewidth]{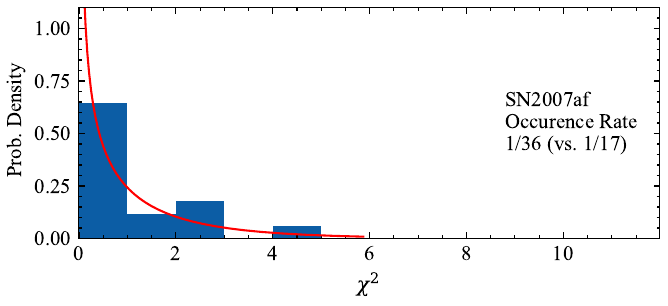}
    \caption{$\chi^2$ distribution of the $M_B$ calibration based on SH0ES Cepheid (top panel) and TRGB (bottom panel) distances. The $\chi^2$ statistic is computed about a simple weighted average of the Pantheon+ standardized magnitudes for 17 calibrator SNe. The outlier SN~2007af is highlighted along with the expected frequency of its $\chi^2$ value of 9.4 occurring in the computation of $<M_B>$. \textit{top:} SH0ES Cepheid distances are used to calibrate the Pantheon+ SN magnitudes. Here, just as in the full SH0ES sample of 42 SNe, SN~2007af is a $>3\sigma$ outlier, beating the odds of occurring in a sample of this size by 50 times. \textit{bottom panel:} CCHP TRGB distances are used to calibrate the Pantheon+ SN magnitudes. In this case, the $\chi^2$ value of SN~2007af equal to 4.8 has occurred at only twice the expected frequency---1 in 36 for a sample of 17.}
    \label{fig:shoesVScchp_chi2}
\end{figure}

Most of the SH0ES $\chi^2$ is contributed by the outlier SN~2007af. Dropping SN~2007af shifts $M_B$ brighter by 0.029~mag, corresponding to a concomitant drop in $H_0$ equal to $-1.02$~km/s/Mpc and a decrease in the reduced $\chi^2$ from 1.53 to 1.01. Doing the same with the calibration based on CCHP TRGB shifts $M_B$ by only 0.004~mag, or $-0.15$~km/s/Mpc, dropping its reduced $\chi^2$ from 1.07 to 0.82.

The story does not significantly change if we consider instead the full SH0ES-22 sample of 42 calibrator SNe. With SN~2007af included in the Pantheon+SH0ES calibration, the unweighted mean calculation of $<M_B>$ is exactly $-0.01$~mag brighter than the weighted mean, and the median is $-0.01$~mag brighter yet. The weighted and unweighted standard deviations are 0.129~mag and 0.134~mag, respectively. After dropping just SN~2007af from the calibration, the three averages come into near perfect 0.003~mag agreement of each other. Similarly, the weighted and unweighted standard deviations drop significantly to 0.115~mag and 0.127~mag, respectively. In the end, dropping 07af brightens $M_B$ by 0.015~mag, corresponding to a 0.5~km/s/Mpc drop in $H_0$ to 72.5~km/s/Mpc. 

This emphasizes a notable drawback inherent to Cepheid distance measurement: inhomogeneous uncertainties. The quoted uncertainties in the SH0ES Cepheid distances range from 0.039 to 0.203~mag, translating to an order of magnitude dynamic range in statistical weight. This is why certain SNe (like 2007af) can have an influence on the value of $H_0$ that is disproportionately large relative to the numerical fraction they make up. However, the CCHP TRGB provides homogeneous and likely more accurately estimated uncertainties, on account of the nature of the TRGB as a standard candle.
That is, older stellar populations are more stably present from galaxy to galaxy, while the measurement of a sufficiently large number of Cepheids depends on galaxy inclination, star formation history, morphology, source crowding, and more. This is borne out further in the absolute calibration of SNe for which the use of TRGB distances regularly results in a smaller dispersion in the $M_B$ calibration than do SH0ES Cepheid distances.

We presented in this section evidence that TRGB likely provides a more accurate constraint on the SN absolute magnitude than do the SH0ES Cepheid distances. 
The impact that these seemingly less reliable Cepheid distance uncertainties have on $H_0$ is further amplified by what appears to be an outlying analysis of the calibrator SNe in Pantheon+ relative to both CSP and their own earlier SuperCal analysis. Indeed, it may prove to be that a sizable portion of the Hubble Tension is sourced by some combination of these two phenomena---unstable HST Cepheid distance uncertainties and an anomalous Pantheon+ analysis of the calibrator SNe.

\section{Discussion of the New F115W TRGB Distance Scale}

We will now discuss key aspects of the F115W TRGB distance scale established in this study, including a broader contextualization of the color standardization, internal checks on self-consistency both in the slope and zero point calibrations, comparisons with other TRGB studies, and new implications for the spectroscopic standardization of SNe~Ia.

\subsection{The IR TRGB as a standardizable candle}

In this subsection, we place into context the IR TRGB color standardization employed in this study by comparing with the standardization equations used with other distance indicators, namely Cepheid variables and SNe~Ia.

The equation for making a color-standardized TRGB distance measurement with two bandpasses like those used in this study can be written as,
\begin{equation}
    \mu + M_{\lambda_1,corr} = m_{\lambda_1} + \beta^{m_{\lambda_1}}_{c_{\lambda_1,\lambda_2}} c_{\lambda_1, \lambda_2}
\end{equation}

where $m_{\lambda_1}$ and $c_{\lambda_1, \lambda_2}$ are the TRGB magnitude and color determined from a two-dimensional fit to the CMD, labeled $m_T$ and Color in \autoref{tab:trgb_measures}.
$\beta^{m_{\lambda_1}}_{c_{\lambda_1,\lambda_2}}$ is the slope of the absolute magnitude-color relation that describes the bandpass combinations in question and correspond to the $\beta$ column of \autoref{tab:trgb_slopes}. Recall that, if a bandpass besides the color-insensitive ones ($I$/F814W/F090W) is used, $m$ and $c$ must be constrained simultaneously with the observed slope in an inference performed on the two-dimensional CMD of a single observed stellar population, and \textit{not} after marginalization over color. This is because the observed TRGB is tilted and cannot be accurately described without the full color \textit{and} magnitude information. 

Meanwhile, the Cepheid PL relation can be defined as,
\begin{equation}
  \mu + M_{\lambda,1} = \alpha \log P + \beta \left(m_{\lambda,1} - -m_{\lambda,2}\right)_0 + \gamma [O/H]
\end{equation}
where the Cepheid magnitude at a given wavelength $\lambda_1$ is a function of the logarithm of the period ($P$) with coefficient $\alpha$, a color term with coefficient $\beta$, and a term with coefficient $\gamma$, that allows for a metallicity dependence (where $[O/H]$ represents the logarithmic oxygen-to-hydrogen ratio for HII regions in the vicinity of the Cepheids, relative to the solar value).

\begin{figure*}
    \centering
    \includegraphics[width=0.7\linewidth]{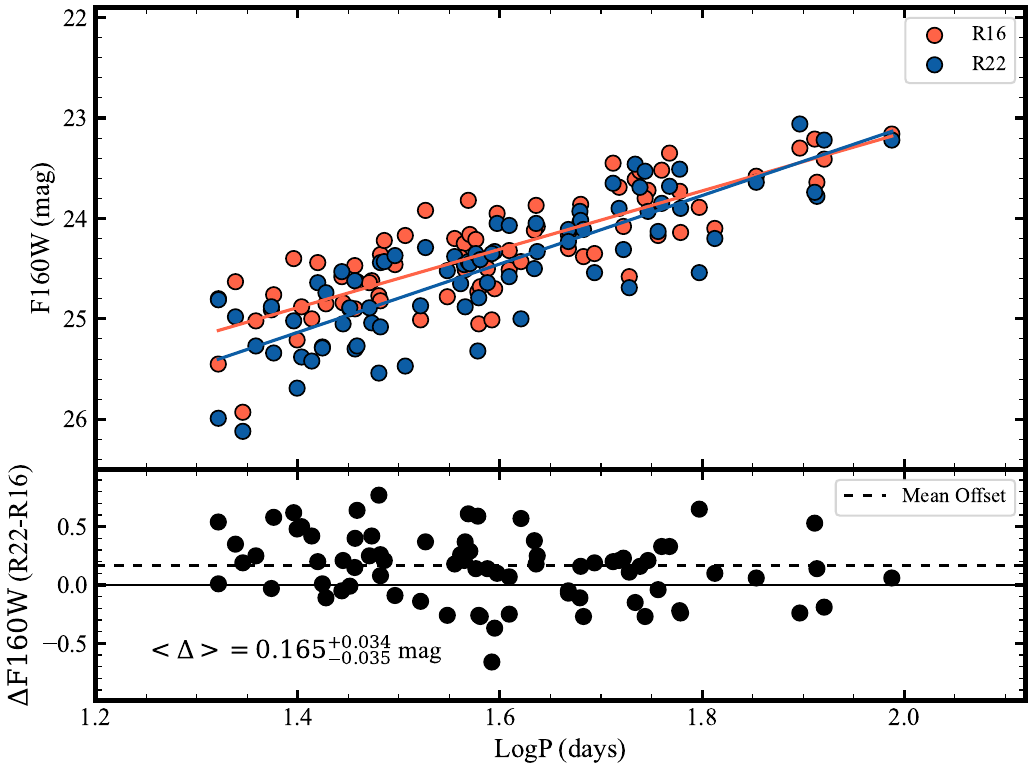}
    \caption{Drift in SH0ES Cepheid flux measurements for host galaxy NGC~5584 exceed quoted uncertainties by several times. \textit{top:} The F160W PL relations from the 2016 (orange markers) and 2022 (blue markers) SH0ES analyses are plotted. The two best-fit slopes are also plotted (colors matched to the respective markers).
    \textit{bottom:} The point-wise difference of the F160W magnitudes from the top panel. The average shift to their Cepheid magnitudes from 2016 to 2022 was $0.165 \pm 0.035$~mag (error on the mean determined via bootstrapping), which is over three times the total distance uncertainty of 0.053~mag that SH0ES quotes for this host galaxy and equal to the size of the full Planck-SH0ES Hubble Tension (0.16~mag).}
    \label{fig:n5584_hband}
\end{figure*}

In the case of SNe~Ia, the standardization equation takes on various forms of the kind,
\begin{align}
  \mu + M_{corr,max} = m_{max} + \alpha \times (stretch) - \beta \times (color) \nonumber \\ + \delta (host~property)
\end{align}
where $M_{corr,max}$ is the standardized absolute magnitude of the SN at the time it reaches its brightest. $m$ is the same but for the apparent magnitude and before any standardization. $\alpha$ parametrizes the correction based on the width (referred to as decline-rate or stretch) of the SN. $\beta$ corrects for the color of the SN (sometimes reparametrized as an extinction, though it remains an open question whether photometric light curves contain the information needed to disentangle intrinsic SN color from external extinction). $\delta$ broadly refers to residual, nonlinear biases (or steps) that have been observed in the distributions of $\alpha$- and $\beta$- corrected SN distances that appear correlated with local and global properties of their host galaxies. This is most commonly parametrized by the total stellar mass of the host galaxy. However, it can also, and arguably more accurately, be parameterized by the local specific star formation rate \citep[see, e.g.,][]{Rigault_2015, Rigault_2020}.

It immediately appears from the above equations that, despite some claims that the IR TRGB is too difficult or uncertain to use as a distance indicator due to its color dependence \citep[e.g.,][]{Riess_2024}, it is in principle a much simpler correction than either of the two other leading distance indicators used in measuring $H_0$. Why, then, is the term ``standard'' candle typically ascribed to a Cepheid, while ``standardizable'' is used to describe either an SN~Ia and the IR TRGB? Indeed, all three rely on color measurements as part of their standardizing equations, and TRGB is the only one for which color is the \textit{only} standardization term required. For Cepheids, the standardization precision actually worsens because, unlike TRGB with its smooth, tight, linear relation between color and metallicity, Cepheids possess no such empirical probe for their metal content (besides infrared color curves formed from observations at 4.5\micron{}). In fact, in the following calculations, we will demonstrate that distances derived using the Cepheid PL relation (PLR) are likelier to suffer from systematic bias due to uncertainties in the standardization coefficients, e.g., the slope of the relation.

The typical slope of the Cepheid PLR ranges between $-2.5$ to $-3.4$~mag/dex depending on the observed bandpass. Meanwhile, the slope of the F115W TRGB magnitude-color relation ranges between $-0.7$ and $-1.0$ \magSlopeErr{}. The range of TRGB colors we observed here was between 1.3 and 1.6~mag, for a maximal total range in apparent brightness at the TRGB equal to $0.3 \times -1.0 \simeq 0.3$~mag, while the typical range of Cepheid periods used in extragalactic distance measurement range from 10d ($\log P = 1$) and 100d ($\log P=2$), for a maximal dynamic magnitude range of about 3~mag. Thus, the Cepheid PL slope needs to be constrained an order of magnitude more tightly than does the slope of the F115W TRGB to reach comparable precision as standardizable candles. 

Two unique aspects of the Cepheid distance scale can then amplify this effect into a sizable source of potentially underestimated systematic bias in one's final estimate of $H_0$. First, the average period of Cepheids in SN hosts (40d or $\log P = 1.5$) is 0.5~dex longer on average than that of the average in the geometric anchor galaxies (10d, or a $\log P = 1$). Second, SH0ES use the fluxes of their Cepheids measured in SN hosts to constrain the slope. However, those flux measurements have proven unstable at a level that far exceeds the quoted uncertainties. This is predominantly a result of the extreme levels of  source crowding present in the HST IR imaging that underpins the SH0ES distance scale; in most cases, the Cepheid contributes less than half of the total flux contained in an aperture centered on it. 

We demonstrate this explicitly in \autoref{fig:n5584_hband}. The $F160W$ magnitudes and periods published by SH0ES for the same 78 Cepheids in SN host galaxy NGC~5584 are plotted in the top panel. The per-Cepheid magnitude difference is then plotted in the bottom panel. The net shift in magnitude is $0.165 \pm 0.035$~mag, with the uncertainty on the average having been estimated via bootstrapping. SH0ES reported in both 2016 and 2022 a distance uncertainty of just 0.05~mag to this SN host galaxy. That is, the Cepheid magnitude measurements alone shifted by more than $3\sigma$ of the quoted SH0ES uncertainties. Furthermore, the impact of this unstable photometry on the estimated slope is severe. The 2016 Cepheid magnitudes yield a PL slope equal to $-2.905 \pm 0.196$~mag/dex while the 2022 measurements yield $-3.405 \pm 0.214$. Note that SH0ES determined these Cepheid magnitudes from the \textit{exact same} HST imaging data. The average period is $<\log P> = 1.59$, compared to the average value in the geometric calibrators, or $<\log P >= 1.15$. The difference between the two slopes is 0.5~mag/dex, which corresponds to a 0.22~mag systematic error in the distance to any SN host galaxy (which have similar period distributions). This emphasizes that the difficult-to-measure Cepheids in SN host galaxies must not be included in SH0ES' determination of the PL slope. If a similar fraction of TRGB

Indeed, we can see this explicitly borne out in the literature by comparing how the SH0ES Cepheid distances reported in \citet{Riess_2022ApJ...934L...7R} changed in \citet[R24][]{Riess_2024} for the same galaxies due to a change in the Cepheid PL slope. The slope in R22 was reported to be $-3.30$~mag/dex and was changed to $-3.25$mag/dex in R24. This seemingly small shift in the slope of the Cepheid PLR actually resulted in an average shift to the SH0ES Cepheid distance moduli equal to 0.025~mag (A. Riess, priv. comm.)---this is equivalent to 0.86~km/s/Mpc in units of $H_0$, or 85\% of the entire uncertainty of 1.04 km/s/Mpc quoted by SH0ES. On the contrary to produce an identical systematic shift in our F115W distances, the F115W slope would have to be shifted by 0.2~\magSlopeErr{}, which is already pushing the outer edges of our TRGB slope confidence intervals and, more importantly, is exactly and fully accounted for already in our error budget, contrary to the case of the PL slope and the SH0ES distance scale.

Looking forward, this propagated uncertainty in the TRGB color correction will decrease further with the addition of more geometric calibrators beyond NGC~4258. Indeed, the SMC and LMC TRGB are expected to cover the full color range we have observed in our F115W TRGB measurements. According to the BaSTI isochrone suite, RGB populations with $[Fe/H] \simeq -1.1$~dex and $[Fe/H] \simeq -0.6$~dex would be observed with \colorTHREEFIVE{} TRGB colors equal to $\sim 1.25$~mag and $\sim 1.45$~mag, respectively. Together with the TRGB color of 1.55 mag in NGC~4258, the geometric calibration of the IR TRGB would span the entire range of TRGB colors observed in the SN hosts (see \autoref{tab:trgb_measures}). The systematic uncertainty currently incurred from extrapolating the red TRGB measurements to the bluer SN host measurements would be reduced to near zero. It is not physically possible for the Cepheids to ever achieve a similar reduction in the PL slope systematic because new long period Cepheids will not suddenly be discovered in the Magellanic Clouds, and because we are embedded in the disk of the Milky Way which severely limits the completeness of galactic Cepheid samples.

We conclude with the remark that any ``candle'' is inherently standard-\textit{izable} in the limit of empirical measurements that reach infinitely high SNR. In fact, the express \textit{goal} of distance scale studies is the identification of new, or the refinement of existing, axes along which to improve distance standardization and tighten the dispersion about the Hubble diagram. This is not novel or unique to the distance scale, of course. Though buried in archaic astronomical jargon, the existence of a ``standardizable'' candle does not suggest that a distance indicator is of poor precision or accuracy but simply that \textit{all} science is predicated on the refinement of models in response to higher precision empirical measurements. 

\subsection{F115W Color Correction}

A novel aspect of the distances presented in this study is the use of the tip color to correct the corresponding magnitude measurements onto a standard, absolute system. We estimated the amount that this added parameter impacts $H_0$ as the difference between the TRGB colors of the SN host and geometric anchor fiducials ($\Delta color = 0.13$~mag) multiplied by the slope uncertainty ($\sigma_{\beta} = 0.148$~\magSlopeErr{}) corresponding to a 0.019~mag, or a 0.6 km/s/Mpc systematic uncertainty in $H_0$. For comparison, this is equal to just over half the 0.032~mag uncertainty on the geometric distance to NGC~4258, meaning the color correction is a sub-dominant systematic in the F115W TRGB distance scale. And this would shrink greatly with the addition of more geometric calibrators such as the Magellanic Clouds (MCs). From APOGEE abundances of MC red giants from \citet{Nidever_2020}, a typical TRGB star in the SMC and LMC can be expected to have $[Fe/H]$ equal to $-1.1$ and $-0.6$~dexl resepectively. And according to the BaSTI isochrone suite, these metallicites correspond to $F115W-F356W$ colors of 1.25 and 1.45~mag. Along with the TRGB color of 1.55~mag we measured in NGC~4258, this trio of geometric calibrators would fully cover the TRGB colors observed in our SN host galaxy sample, which ranges from 1.29 to 1.50~mag (after homogenization to the F356W color scale).

Despite contributing a small amount to the total error budget, the IR TRGB has never been used in $H_0$ experiments, so it is important that we still explicitly and thoroughly demonstrate the robustness of our color standardization for the sake of proving the accuracy of the method itself.

\subsubsection{IR color homogenization}

\begin{figure}
    \centering
    \includegraphics[width=0.98\linewidth]{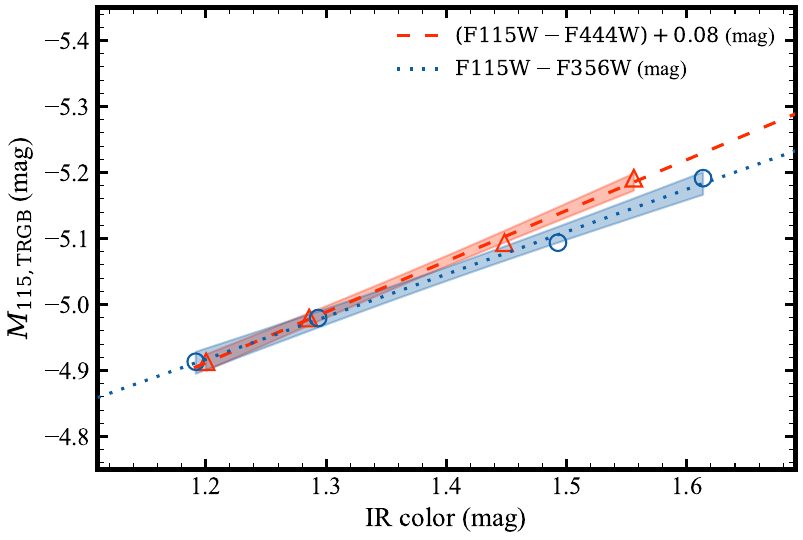}
    \caption{Test of the color homogenization offset in the BaSTI isochrone suite. Plotted for each of F356W (blue open circles) and F444W (red open triangles) are four TRGB magnitude-color pairs in the BaSTI isochrone suite that span the same TRGB color range observed in our targets, corresponding to metallicities $[Fe/H] = \{-1.30, -1.05, -0.60, -0.30\}$~dex and averaged over ages between 4 and 13~Gyr in 1~Gyr steps, for an effective age of 8.5~Gyr.
    Also shown is the best-fit line for each set of TRGB points (blue dotted and red dashed lines) along with their the $2\sigma$ standard error interval (blue and red shaded regions).
    The BaSTI TRGB F115W-F444W relation has been shifted by the same horizontal offset that was used for the empirical IR color homogenization. The agreement of the shifted F444W relation with the F356W one is excellent, providing an additional check on the IR color homogenization.
    }
    \label{fig:basti_slope_overplot}
\end{figure}

As described in Section 4.2, we undertook two different approaches to color-correcting the observed F115W TRGB magnitudes. In one case, we considered separately the color formed between F115W and either of F356W or F444W. In the other case, we computed a homogenized IR color based on shifting the \colorFOURFOUR{} colors onto the same system as \colorTHREEFIVE{}. The latter was adopted as the preferred distance solution both here (Sections 5 and 6) and in the program's cosmology analysis \citep{Freedman2024arXiv240806153F} because of the more robust number statistics attained by leveraging all the F115W measurements. This IR color homogenization was motivated by the fact that the empirical F115W vs. \colorTHREEFIVE{} or \colorFOURFOUR{} relations appear in our measurements as parallel translations of one another and that both F356W and F444W fall along the tail of a TRGB star's SED (which peaks around $1.5 \micron{}$). 

A significant constraint on the IR color offset term comes from the three TRGB measurements made in NGC~4258. Because the Outer Disk field (observed in F356W) also came out to have the brightest tip measurement relative to the Halo and Inner Disk measurements (observed in F444W), the inferred value of the IR color offset term implicitly suggested that the Outer Disk TRGB is \textit{intrinsically} redder by about 0.04~mag than that in the Inner Disk field. This is surprising given the Halo and Inner Disk fields agreed to within 0.01~mag in TRGB color, so one would expect the Inner and Outer Disk fields to be even closer in color. Indeed, a potentially more accurate description here is that a statistical fluctuation led to a slightly brighter TRGB measurement in the Outer Disk. That fluctuation is then being over-interpreted in the minimization as its TRGB being redder relative to the other two. This alternative explanation is physically motivated because the Outer and Inner Disk fields are located in similar regions of the host galaxy and should be expected to be closer in TRGB color than either is to the tip in the Halo field. Instead, if we simply take the difference in the TRGB colors of the two ``Disk'' fields (though, recall that we mask the majority of the disk in our spatial selection) as the color offset term, we find 0.08~mag.

This larger value for the color homogenization offset term is also corroborated by isochrone predictions. In \autoref{fig:basti_slope_overplot}, we plot the four brightest timestamps reached by BaSTI isochrones corresponding to metallicities $[Fe/H] = \{-1.30, -1.05, -0.60, -0.30\}$~dex and averaged over ages between 4 and 13~Gyr in 1~Gyr steps, for an effective age equal to 8.5~Gyr. The F115W vs. \colorTHREEFIVE{} tips are plotted as blue open circles with the line of best fit and its $2\sigma$ confidence interval plotted as a blue dotted line and blue shaded region. The same information is plotted as red open triangles, a red dashed line, and red shaded region for F115W vs. \colorFOURFOUR{} after adding to it the alternative value of 0.08~mag for the color homogenization term. After the shift, the two relations are in excellent agreement, providing an additional check on the accuracy of the adopted IR color homogenization. The isochrones appear to also predict a slight additional color slope to the offset, i.e., a dilation as a function of color, but we refrain from over-interpreting the model predictions at this time and leave the isochrones as a zeroth order check on the empirical homogenization that we have adopted. Additional JWST IR data will help to constrain the IR TRGB color-color relations to higher precision and potentially constrain above the noise higher order terms in the transformation.

In practice, changing the color offset from 0.042 to the larger 0.080~mag shifts those homogenized distances estimated to hosts observed in F444W farther by 0.011~mag in the average, and those observed in F356W closer by $-0.021$~mag in the average, for a net shift of just $-0.008$~mag toward a shorter F115W TRGB distance scale. This corresponds to just 0.25~km/s/Mpc in units of $H_0$, reiterating that the color homogenization uncertainty contributes very little to the final error budget.

\subsubsection{Search for Circularity in the Slope Estimation and Final Distances}

\begin{figure}
    \centering
    \includegraphics[width=0.92\linewidth]{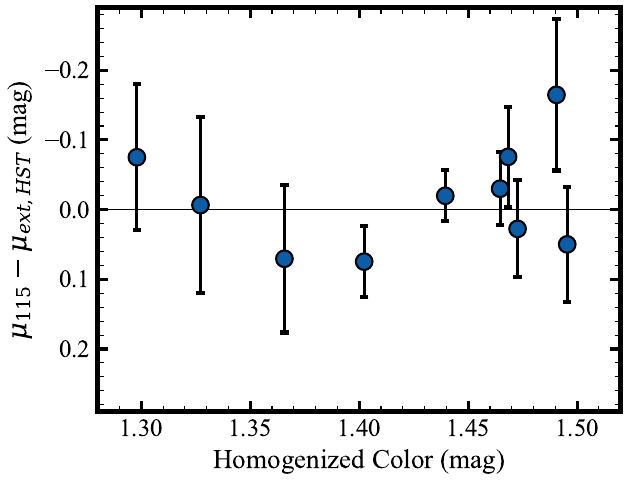}
    \caption{Plotted for each SN host galaxy as a function of TRGB color is the difference between the external HST distances adopted for estimating the F115W color slope (section 4.2) and this study's final measured distances. No correlation is seen here, indicating this study's F115W TRGB distances are not being pre-determined by the external ones that were used to calibrate the color standardization.}
    \label{fig:color_dist_correlation}
\end{figure}

As just discussed, the dependence of the F115W TRGB magnitude on its IR color was accounted for in our analysis in either of two ways, with the F356W or F444W measurements being treated as two separate color scales, or all TRGB measurements being placed onto a single, homogenized IR color scale. In either case, in order to estimate the slope of the TRGB absolute magnitude-color relation, we used an externally determined and published set of distances to anchor the TRGB apparent magnitudes (presented in Table 2) onto an absolute system. For the ten SN host measurements we adopted a set of published HST distances, computed as the weighted average of Cepheid (SH0ES-22) and TRGB (CCHP) distances. And for the three independent tips measured in NGC~4258 we used the trigonometric megamaser distance.

Importantly, only the \textit{slope} of the TRGB's magnitude-color relation was estimated in this way, and we did not retain any of the intercept information. This was intended to keep the final JWST distance measurements independent of the HST ones that were used to derive the TRGB color dependence, i.e., to ensure that the HST distances do not pre-determine our new JWST distances. However, a circularity could still arise if there were to exist a correlation of the HST distance residuals with IR TRGB color. This would bias an estimate of the TRGB slope toward a value that would statistically best replicate the anchoring HST distances. Such a correlation, however, is highly unlikely given that HST Cepheid distances are totally unrelated to those determined with the JWST TRGB and that the HST TRGB distances were determined \textit{without} any color correction. However unlikely, we still look to explicitly rule out such a conspiracy left for us by nature. 

We show such a test in \autoref{fig:color_dist_correlation}, we plot as a function of TRGB color the difference between our new JWST distances and the external HST ones. As can be seen, there is no correlation or functionality seen in the residuals between the final JWST distances and the external HST ones as a function of color. This confirms that our final F115W distances are indeed decoupled from, and independent of, the HST distances that were used to constrain the TRGB's color dependence.

\subsubsection{Age Effects}

\begin{figure}
    \centering
    \includegraphics[width=0.91\linewidth]{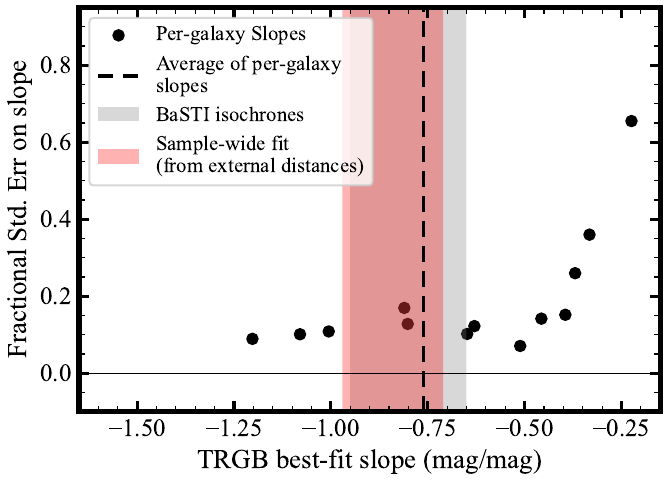}
    \caption{TRGB slope values determined from each galaxy in the sample. Plotted along the y-axis is the uncertainty associated with each slope determination. The sample-wide slope determined in Section 4.2 is plotted (red band), along with the range of isochrone predictions for reasonable values of age ($4-12$~Gyr) and metallicity $-1.30 < [Fe/H] < -0.30$~dex.}
    \label{fig:intra_gal_slopes}
\end{figure}

As introduced in Section 2.3, age serves as a second parameter along with the metallicity in determining the brightness and color of the TRGB. In the case of F115W, the IR bandpass used in this study, age variation can result in magnitude variations up to 0.09~mag in size, which places already a strict and conservative upper limit on the impact of age on F115W TRGB distance measurement. This is reduced further for numerous reasons. For the metallicity/color range of our target TRGBs, the color slope due to age is close to parallel to that resulting from metallicity. Furthermore, our TRGB distance measurements do \textit{not} use the slope observed in any one target field to perform the color standardization. Instead, we rely on the mean value theorem to effectively smooth over the spread in age per target stellar population.

We demonstrate this in \autoref{fig:intra_gal_slopes}. We plot on the $x$-axis the slope that was marginalized out of each 2D TRGB fit and on the $y$-axis the standard error on each slope constraint. For comparison, we plot as a red band the sample wide slope from Section 4 and as a gray band the range of slopes predicted by the BaSTI isochrone suite for any age $>4$~Gyr and between metallicites $-1.30 < [Fe/H] < -0.30 $~dex. It is evident that the slopes observed in each target population contain significant scatter, yet still cluster around both the sample-wide and isochrone-predicted bands. This would seem to support our hypothesis that variations in TRGB magnitude due to age (and SFH) are largely smoothed over given a large enough sample of observed composite populations.

\begin{deluxetable*}{lcll}
\tablecaption{TRGB Slope Calibrations in the F115W or $J$ bands \label{tab:slope_lit_comp}} 
\tablehead{ 
\colhead{Source Study} &
\colhead{$\beta$} &
\colhead{Error} &
\colhead{Calibration Sample}
\\
\colhead{} &
\colhead{[mag/mag]} &
\colhead{[mag/mag]} &
\colhead{}
}
\startdata
F115W vs. \colorTHREEFIVE &       &                       & \\ \hline
\citet{Newman_2024}       & -0.98 & $_{-0.17}^{+0.31}$    & 4 Low Mass Galaxies \\
This Study                & -0.89 & 0.28                  & 8 $L_*$ galaxies            \\
\hline \hline
F115W vs. \colorFOURFOUR  &       &                       & \\ \hline
\cite{Hoyt_2024}          & -0.99 & 0.16                  & NGC~4536                 \\
\citet{Newman_2024}       & -0.78 & $_{-0.45}^{+0.30}$    & 4 Low Mass Galaxies \\
This Study                & -0.94 & 0.15                  & 4 $L_*$ galaxies          \\ \hline \hline
$J$ vs. $(J-K_s)$         &       &                       & \\ \hline
\citet{Valenti_2004}      & -1.15 & 0.17\tablenotemark{a} & 24 MW Globular Clusters       \\
\cite{Serenelli_2017}     & -0.81 & 0.04\tablenotemark{a} & Theory+Empirical BCs \\
\citet{Madore_2018}       & -0.85 & 0.12                  & IC 1613                  
\enddata
\tablecomments{Published calibrations of the TRGB slope observed in the JWST/F115W or 2MASS/$J$ bands ($\lambda_{eff} \simeq 1.2~\micron{}$) as a function of infrared colors based on a second band ($K$, F356W, or F444W) that lies in the Rayleigh-Jeans regime of a TRGB star's SED (see Section 2).}
\tablenotetext{a}{Standard error not reported, so published dispersion about best-fit line is adopted here.}
\end{deluxetable*}

\subsubsection{Comparison with Literature}

In this section and in \autoref{tab:slope_lit_comp}, we place our estimates of the F115W color dependence in the context of the published literature. Note, however, that JWST has only recently launched so the literature is still limited in breadth. As such, we supplement the table of slope estimates for JWST bands F115W, F356W, and F444W with slopes estimated in the ground-based $J$ and $K/K_s$ bands.
As previously noted by \citet{Hoyt_2024}, these bands will behave similarly for TRGB stars. The F115W and $J$ filters are separated by only 90~nm in central wavelength, while the $K/K_s$, F356W, and F444W all lie in the low energy, Rayleigh-Jeans tail of both Vega and a TRGB star.
This was independently corroborated by the JWST RGB isochrone study of \citet{McQuinn_2019}. In their Figure 5, they show that the VEGA magnitude of the TRGB asymptotes to a near constant value for all bands as red as or redder than $K$ (2.2 micron).

In the first sub-block of \autoref{tab:slope_lit_comp}, we compare this study's estimate of the F115W vs. \colorTHREEFIVE{} TRGB slope with that published in \citet{Newman_2024}. \citeauthor{Newman_2024} used cospatial HST and JWST imaging of four lower mass Local Group galaxies and estimated a slope equal to $-0.98^{+0.31}_{-0.17}$~\magSlopeErr{}. In section 4.2, we estimated $-0.889 \pm 0.281$~\magSlopeErr{}. The two estimates are in excellent agreement. 

The second sub-block of \autoref{tab:slope_lit_comp} presents a similar comparison but for F115W vs. \colorFOURFOUR{}. The \citeauthor{Newman_2024} result was again similarly derived, although they appear to have excluded via their color cut some of the bluest RGB population that was by contrast included in their F356W estimate. They found $-0.78^{+0.30}_{-0.45}$~\magSlopeErr{}. In our previous analysis of the TRGB slope \citep{Hoyt_2024}, we found $-0.99 \pm 0.16$ \magSlopeErr{}. In this study, we found $-0.944 \pm 0.148$~\magSlopeErr{}. All three estimates are again in very good agreement.

In the third and final sub-block, we tabulate published calibrations of the ground-based $J$ vs. $(J-K_s)$ TRGB magnitude-color relation, which, as we argued at the start of this section, is likely a sufficiently analogous pair of bandpasses to those used in this study. \citet{Valenti_2004} found $-1.15 \pm 0.17$~\magSlopeErr{} from a sample of 24 globular clusters with (spectroscopically determined) metallicities ranging from $-2.2$ to $-0.4$~dex. \citet{Serenelli_2017} in their comprehensive study on modeling the IR TRGB reported a slope equal to $-0.81$~\magSlopeErr{}. \citet{Madore_2018} found using ground based imaging of the Local Group galaxy IC~1613 a value of $-0.85 \pm 0.12$~\magSlopeErr{}. The slope estimates are all in good  agreement with each other, as well as with the estimates in the F115W bandpass.

Note that \citet{Valenti_2004} quote $K$ in their photometry but also report that they calibrated their instrumental $K$ magnitudes to 2MASS magnitudes, suggesting that what they refer to as $K$ (or $K'$, depending on which of the two was used on the ESO/2.2m IRAC2 camera) magnitudes were likely transformed to the 2MASS $K_s$ system. The effect is minor in any case, as $(J-K)$ color terms in the $K_{long}$ to $K_{short}$ transformation are typically of order 0.02~\magSlopeErr{} or less \citep{Carpenter_2001}. Compared to the TRGB slope of order unity and uncertainties of order 25\%, this 2\% uncertainty due to which system their $K$ magnitudes are on makes little to no difference to the slope comparisons we are making here.

\subsection{Consistency Checks in the Zeropoint Anchor}

In this section, we assess the stability of the F115W TRGB, as well as the robustness of our zero point calibration. We do this with the high SNR data we have acquired in three locations of the geometric anchor galaxy NGC~4258. Two of the imaging datasets paired F115W with F444W (Inner Disk and Halo) and one with F356W (Outer Disk). This is the only galaxy for which we have multiple fields that also cover all the program bandpasses.

The \textit{full range} of the F115W TRGB measurements in NGC~4258, which are each located throughout the galaxy and separated on the sky by several arcmin, was equal to 0.071~mag. And, after applying the homogenized color correction, this shrunk further to just 0.058~mag. The corresponding standard deviations are 0.029~mag before and 0.024~mag after the color correction. These statistics are in excellent accordance with our reported measurement uncertainties, equal to 0.024, 0.029, and 0.030~mag for the three fields. This emphasizes that the 2-D tip fitting methodology introduced in this study provides magnitude-color ordinates, and corresponding uncertainties, that are sufficiently precise to be used in TRGB distance standardization and tighten TRGB distance precision even over very small gradients of $0.03$~mag in TRGB color. Furthermore, the agreement between the measurement errors and the dispersion underscores the \textit{accuracy} of the methodology, defined in terms of accurate error estimation.

Similarly, we can consider just the measurements from the two F444W datasets to circumvent the color homogenization step and test only the color correction. The difference in TRGB magnitude between the Halo and Inner Disk fields decreases from 0.021~mag to 0.014~mag, after correcting for the 0.01~mag difference in color, further underscoring that our IR TRGB methodology, in terms of both the measurement on the CMD and the application of the color standardization is sufficiently precise and accurate that it can refine distance measurements even at the 0.01~mag level.
 
A corollary of the above consistency check is that the F115W TRGB is repeatable and stable to better than 0.025~mag (1\% in distance) based on multiple independent measurements made in different regions of the same galaxy (NGC~4258 in this case).
This is over an order magnitude smaller than the extremely large (exceeding 1~mag or 40\% in distance in some cases) location-dependent variations in TRGB distance in the same galaxy reported by \citet{Scolnic_2023} in their CATS analysis. We can therefore conclude that their claims of high amplitude TRGB variations within a single galaxy do not reflect the true TRGB intrinsic dispersion, but rather are artificially introduced by their RGB sample selection methodology. We will now discuss this in detail by using as an example a pair of SN hosts where the CATS methodology for finding the TRGB failed.
 
\subsection{Detailed Comparison with CATS}

\begin{figure*}
    \centering
    \includegraphics[width=0.88\linewidth]{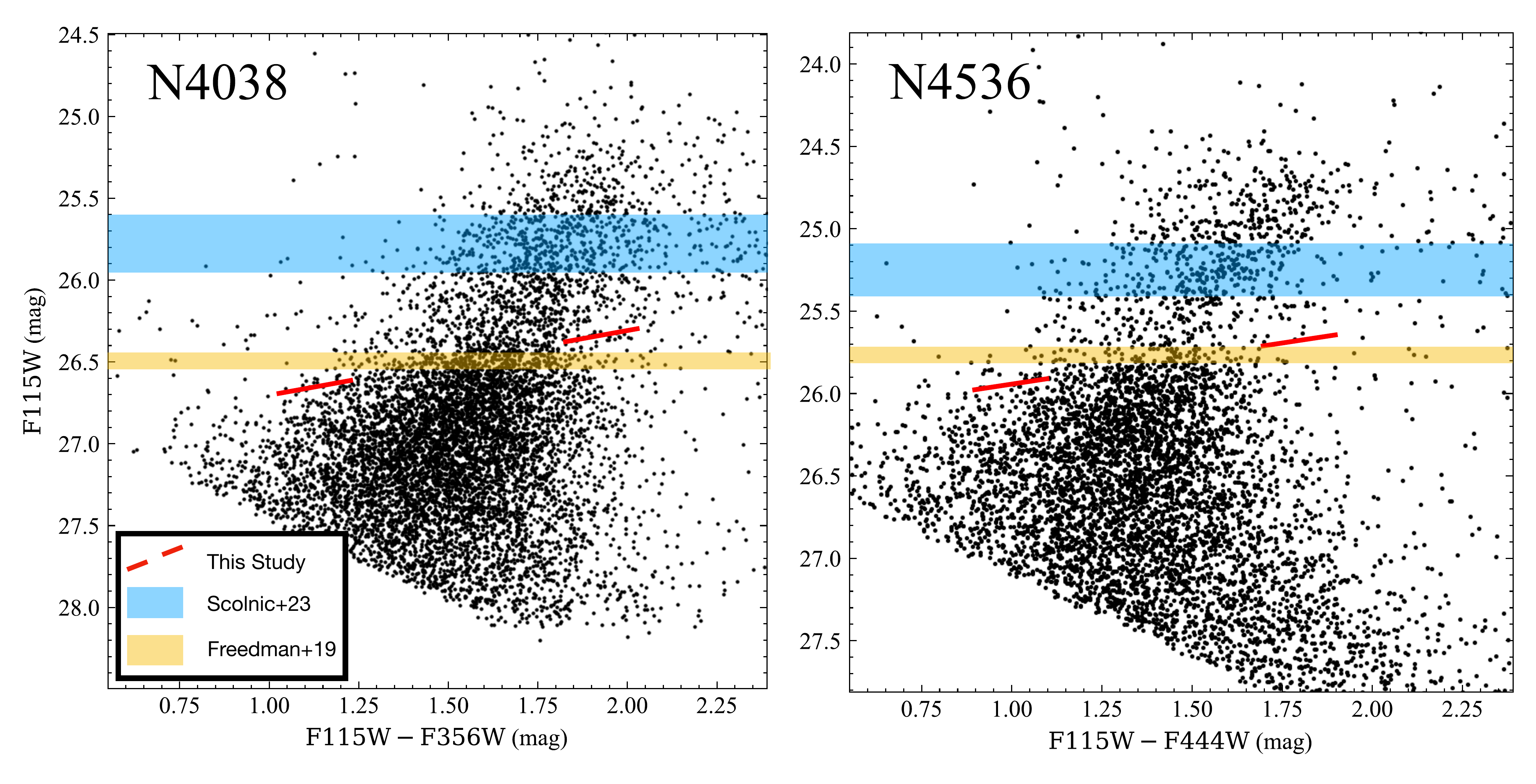}
    \caption{Systematic biases present in CATS TRGB measurements \citep{Scolnic_2023} as visualized on this study's new high signal-to-noise JWST data. F115W vs. \colorTHREEFIVE{} CMDs are shown (black dots) for galaxies 
    NGC~4038/9 (left panel) and NGC~4536 (right panel). This study's F115W tip measurements (sloped red lines with gap at TRGB) are repeated from \autoref{fig:trgb_all_cmds}. Overplotted are the weighted average F814W TRGB magnitudes and errors reported from the ``baseline'' CATS analysis \citep[horizontal, blue bands,][]{Scolnic_2023} and those reported by the CCHP \citep[horizontal, gold/yellow bands,][]{Freedman_2019}. Both sets of F814W TRGB measurements were derived from the same HST imaging datasets (not pictured). The JWST data confirm that the CATS analysis has confused the AGB for the TRGB in these SN host galaxies, leading to distance modulus biases of 0.45~mag and 0.68~mag, corresponding to 20\% and 33\% in units of physical distance.}
    \label{fig:cats_cchp_compare}
\end{figure*}

In Section 5.1, we compared the new F115W TRGB distances from this study to the six F814W TRGB distances from CATS that overlapped between the two samples. The residuals against the CATS distances exhibited significantly larger dispersion (0.31~mag) than did either of the other two F814W TRGB comparisons (0.08 and 0.11 mag for CCHP and EDD, respectively). This indicated in a summary way that the CATS TRGB measurements are imprecise relative to other published sets of TRGB measurements. 

As a result, the majority of the analysis ``variants'' reported by CATS resulted in abnormally large values of the SN calibration dispersion (triple the expected value) and high values of $H_0$ because of biases in their TRGB distance measurements; in some cases, these biases exceed 0.4~mag (or 20\% in distance). This is a result of several flawed assumptions that underpin their analysis, which subsequently biases their global estimate of $H_0$.

Here, we explore in greater detail the two significant outliers in the CATS comparison that are associated with the SN host galaxies NGC~4038/9 and NGC~4536. We do this by projecting their published HST F814W TRGB measurements onto our new JWST CMDs. 

As was previously described in Section 5, we compute distance moduli from the CATS baseline analysis by exactly following their published prescription so as to accurately represent their findings. We first ``standardize'' the TRGB magnitudes reported in their baseline analysis to $R=4$, take the weighted average, then compute the distance relative to their $R=4$ zero point calibration based in NGC~4258 (their equation 3). This distance is then projected onto our F115W CMD by scaling our measured F115W magnitude by the difference between this study's distance and that computed from the baseline CATS analysis.

We will see that the TRGB measurements reported by CATS in their baseline analysis for these two galaxies are each $>3\sigma$ discrepant with independent evidence coming from some combination of: this study's TRGB distances, SH0ES-22 Cepheid distances, and CATS' own TRGB measurements to fellow cluster members.
\subsubsection{NGC~4038/9}

The Antennae Galaxies, NGC~4038 and 4039, are a pair of interacting galaxies, the distance to which was debated quite heavily in the past. \citet{Saviane_2004, Saviane_2008} claimed to have measured the TRGB around F814W$ = 26.7$~mag, for a $\mu_0=30.6$~mag or $d=13$~Mpc. Then, \citet{Schweizer_2008} pointed out those authors had likely confused the tip of the AGB for the TRGB. \citet{Jang_2015} presented an improved analysis of the Antennae TRGB, wherein the star-forming regions associated with the tidal tail were explicitly masked, and a clear TRGB signal was revealed from this HST dataset. That measurement was recalibrated onto the CCHP zero point in F19 yielding a value $\mu_0 = 31.68 \pm 0.05$~mag or $d=23$~Mpc. We will show here unambiguously that CATS has mistaken the tip of the AGB for the TRGB.

CATS reported in their baseline analysis two TRGB measurements for this galaxy equal to $30.893 \pm 0.192$~mag and $31.543 \pm 0.580$~mag with corresponding ``contrast'' parameter values 3.8 and 2.3, respectively. 
The weighted average distance modulus, computed after shifting both values to $R=4$, is $\mu_0 = 30.957 \pm 0.182$~mag. Note that the CATS ``contrast'' parameter is a misnomer. It is calculated from a 0.5~mag split of a 1~magnitude interval, while the actual TRGB signal has a width of $\lesssim 0.1$~mag. We discuss this in the Appendix and emphasize that their definition should not be confused with ours, which is based on actual tip stars. 

In the left panel of \autoref{fig:cats_cchp_compare}, we plot as black dots the JWST CMD for stars contained in our halo selection for NGC~4038/9, which is not embedded in the star-forming tidal tail, contrary to the HST imaging data commonly used to measure the TRGB distance to this galaxy pair. The F115W TRGB determined in this study is bracketed by a pair of sloped red lines. The TRGB distance published by the CCHP \citep{Freedman_2019} is plotted along with its $1\sigma$ interval as a gold band. The same but estimated by CATS in their baseline analysis is shown as a blue band. Note that CATS report two TRGB distances to this one galaxy in their baseline analysis, each separated by 0.65~mag (or 30\%) in distance. 

The CATS measurement is biased 0.7~mag brighter (closer) than both the CCHP F814W and F115W measurements (which are themselves in excellent agreement). Notice how the CATS measurement falls along the locus of the JAGB, confirming they have in their analysis confused the TRGB with a signal produced by AGB stars. The source of this error can be understood upon inspection of their spatial selection for this target, which retains significant contamination from intermediate-age stellar populations associated with tidal debris. In the Appendix, we discuss this in greater detail along with other methodological weaknesses in the CATS analysis.

We conclude with an additional, external confirmation that the CATS distance to this galaxy is biased. The SH0ES-22 distance reported for this galaxy based on Cepheid variables is $\mu_{Ceph} = 31.615 \pm 0.117$~mag. In this study, we found $\mu_{TRGB} = 31.645 \pm 0.078$~mag. Due to the very good agreement, we take the average of the two for a tighter constraint on the true distance, or $\mu_{0} = 31.636 \pm 0.065$~mag. Compared to S23's value of $\mu_{TRGB} = 30.957 \pm 0.182$~mag, the offset is $-0.679 \pm 0.194$~mag, indicating that the TRGB distance to this galaxy preferred by CATS is $3.5 \sigma$ discrepant with a totally independent constraint that is jointly set by the F115W TRGB and SH0ES-22 Cepheids.

\subsubsection{NGC~4536}

Similar to the case of N4038/9, the CATS analysis of the TRGB in NGC~4536, a member galaxy of the Virgo cluster, appears to be in error. CATS report three TRGB measurements in their baseline analysis of NGC~4536. Again, we follow their reported methodology here exactly as published in order to accurately represent their findings, i.e., by ``standardizing'' all of the measurements to an $R=4$, taking the weighted average, and calibrating the distance using their $R=4$ measurement in geometric anchor NGC~4258 with their Equation (3). The individual distances come out to be $30.229 \pm 0.200$~mag, $30.949 \pm 0.364$~mag, and $31.219 \pm 0.676$~mag with corresponding ``Contrast'' parameter values of 4.0, 3.3, and 2.3. The weighted average, $R=4$ distance modulus is $\mu_0 = 30.447 \pm 0.169$~mag.

In the right panel of \autoref{fig:cats_cchp_compare}, we show the JWST CMD for sources contained in our halo selection for NGC~4536, which in fact overlaps to some degree with the same HST imaging used by both CATS and CCHP \citep[see][]{Hoyt_2024}. Again, this study's F115W TRGB is bracketed by a pair of sloped red lines, the CCHP F814W TRGB and its $1\sigma$ interval as a gold band. The F814W TRGB from the CATS baseline analysis is plotted as a blue band.

In this case, the CATS measurement is biased 0.45~mag brighter (closer) than both the CCHP F814W and F115W measurements, which are again in very good agreement. The CATS measurement again falls on the AGB suggesting they have in this case also inaccurately confused a discontinuity detected near the ``tip of the AGB'' for the TRGB signal. The error here is again reminiscent of past biases in published TRGB measurements that have already been resolved. \citet{Mould_2009} similarly triggered off sparse tips of the AGB in their WFPC2 photometry of SN host galaxies M66 and M96.

Being a member of a well-defined galaxy cluster of finite depth (about 0.15~mag in distance), we can use the distances to cluster members as an additional check on the CATS measurement to this galaxy. CATS report a single TRGB distance to fellow Virgo member NGC~4526 equal to $\mu_0 = 31.01\pm0.043$~mag. Compared to their measurement in NGC~4536 to which they assign the highest weight
equal to $\mu_0 = 30.229\pm0.200$~mag, this corresponds to a $3.1\sigma$ internal tension present in CATS' own TRGB measurements. Put another way, CATS' measurements are arguing that NGC~4536 does not belong to the Virgo cluster and in fact lies at the same distance as the Leo I group, and they place that claim at a $>3 \sigma$ significance. Again, the much likelier explanation here is the existence of serious methodological flaws and biases in the CATS analysis, some of which we expand upon in the Appendix.

\subsection{On The Distance Between Twin SNe 2011by and 2011fe}

Recent studies have demonstrated that well-calibrated optical spectrophotometry of SNe~Ia can be used to significantly increase both the precision and accuracy of their use as distance indicators. Specifically, variations in flux-wavelength space ($R \sim 300$) were found to encode the information needed to accurately predict the intrinsic brightness of an SN~Ia \citep{Fakhouri_2015, Boone_2021_dist, Stein_2022}. These spectrophotometric approaches to standardization produced typical values for the dispersion in Hubble residuals (a gauge of one's object-to-object distance precision) around 0.08~mag, a significant improvement over the typical 0.15~mag observed when SNe are standardized via the observed properties of their wide-bandpass photometric light curves. Part of this drop in the total dispersion is a direct consequence of a significant decrease in the correlation of the model residuals with host properties \citep[see, e.g.,][for in-depth discussions of this effect]{Hamuy_2000, Neill_2009, Sullivan_2010, Rigault_2015, Rigault_2020}. For example, \citet{Boone_2021_train} demonstrated that the host mass step decreased from $0.092 \pm 0.027$~mag when standardizing with SALT2 to just $0.036 \pm 0.025$~mag when standardizing with their spectral model. The reduction of such population biases is likely to prove key in reducing the systematic error floor in SN cosmology, and even moreso in the case of $H_0$. As we saw in the previous section, the distance ladder determination of $H_0$ is still at the mercy of the small number of calibrator SNe (few tens) compared the number in cosmology samples (several hundreds to few thousands).

However, some concerns have been raised regarding spectrophotometric standardization. \citet{Foley_2020} reported a significant discrepancy in Hubble residuals between twin SNe 2011by and 2011fe (hosted by galaxies NGC~3972 and M101, respectively). Specifically, they report a $5 \sigma$ discrepancy in the luminosities of these twin SNe which in magnitudes is equal to $\Delta M = 0.335 \pm 0.063$~mag. However, their calculation of the statistical significance has several weaknesses. First, they only used the twin spectra to constrain reddening, and not a relative distance, then simply applied that reddening constraint to $V_{max}$ values derived from light curves. They do, however, briefly mention a gray differential offset they determined directly from the spectra. This resulted in a slightly smaller magnitude difference of 0.303~mag. Second, their differential reddening determination only used bluer wavelengths $\lambda < 5600$\AA, while all of the published twinning analyses made use of the full visual spectral range $3300 < \lambda < 8600$\AA. Third, the phases used in their ``phase-matched'' differential analysis differed by as much as 1.7~d, introducing considerable additional uncertainty due to SN time evolution. Fourth, they ignore the contribution of intrinsic dispersion when calculating the significance of the luminosity discrepancy. Finally, and the point we will focus most on here, their determination relies on the use of SH0ES HST Cepheid distances to provide the absolute flux scale, since peculiar velocities are too large in these nearby SNe for them to be placed on a Hubble Diagram from redshift measurements alone.

\citet{Foley_2020} measured and adopted HST Cepheid distances to the hosts of these SNe that were determined following the same methodologies of SH0ES. They reported $\mu_{3972} = 31.594 \pm 0.071$~mag and $\mu_{101} = 29.135 \pm 0.047$~mag for a $\Delta \mu = 2.459 \pm 0.085$~mag. As discussed throughout this text, several of the SH0ES distances that predated R22 were later found to be biased by up to 0.3~mag. Indeed, these two galaxies actually systematically shifted in their distances by 0.05~mag in R22, being reported as $\mu_{Host} = 29.179 \pm 0.041$~mag and $\mu_{Host} = 31.635 \pm 0.089$~mag.\footnote{Interestingly, despite the significant 0.05~mag systematic shift to both galaxies' HST Cepheid distances, the $\Delta \mu$ changed by just 0.002~mag, equal to $2.457 \pm 0.090$~mag. And while it appears at face value that each of these distances shifted within less than one sigma of a statistical fluctuation the shifts are far more significant because SH0ES used the exact same HST data to determine both sets of distances. This provides additional evidence that uncertainties have been underestimated in the SH0ES Cepheid distance measurements.}

As it turns out, we have newly determined F115W TRGB distances to the hosts of both of these SNe. For M101 we determined $\mu_{sep,M101} = 29.140 \pm 0.042$ and $\mu_{hom,M101} = 29.151 \pm 0.042$~mag for the separate and homogenized IR color scales respectively. For N3972, we found $\mu_{sep,N3972} = 31.746 \pm 0.074$ and $\mu_{hom,N3972} = 31.747 \pm 0.068$~mag. These correspond to $\Delta \mu =-2.606\pm 0.085$~mag and $\Delta \mu =-2.596\pm 0.080$~mag, from which we estimate a simple representative midpoint value equal to $\Delta \mu_{TRGB} = -2.60 \pm 0.08$~mag. This TRGB estimate of the separation between these two SNe is $0.14$~mag larger than the SH0ES Cepheid constraint adopted by \citeauthor{Foley_2020} equal to $\Delta \mu_{Ceph}-2.46 \pm 0.085$~mag and immediately reduces by 42\% the ``Twinning discrepancy'' they report from 0.335 to 0.195~mag. Just using the differential TRGB distance, without correcting any of the other problems with the \citeauthor{Foley_2020} computation of evidence against twinning, accounted for just over two-fifths of their evidence. This may be another sign that points to uncertainties in the SH0ES HST Cepheid distance measurements having been underestimated.

We conclude here with a more reliable estimate of the magnitude difference between SNe 2011fe and 2011by and its significance that rectifies some of the other flaws described at the start of this section. First, we take the actual spectroscopic magnitude difference reported by \citeauthor{Foley_2020}, or $\Delta M = 0.303 \pm 0.006$~mag, rather than the comparison they do with $V$-band photometry. We then shift that to the new TRGB distance determined here, resulting in $\Delta M = 0.16 \pm 0.08$~mag. Finally, the Twinning dispersion is reported in \citet{Fakhouri_2015} to be 0.072~mag, nearly identical to the 0.073~mag unexplained dispersion reported by \citet{Boone_2021_dist}. Being a pairwise comparison between two objects, this adds $ 0.072 \times \sqrt{2}$~mag dispersion into the comparison, for a total $\Delta M^{11fe,11by}_{TRGB}  = 0.16 \pm 0.11$~mag, corresponding to just a $1.5 \sigma$ fluctuation. For comparison, the magnitude difference derived from their Cepheid distances would equal $\Delta M^{11fe,11by}_{Ceph}  = 0.30 \pm 0.11$~mag, or a $2.6 \sigma$ fluctuation. In either case, it is clear that a more rigorous accounting of the distance measurement errors and the intrinsic SN distance dispersion indicates the luminosity difference between twin SNe 2011by and 2011fe is not as significant as claimed by \citet{Foley_2020} and is in fact consistent with a statistical fluctuation.

\section{The Future of the TRGB with JWST}

The TRGB method has been employed extensively from the ground and with HST, providing distance measurements to hundreds of galaxies via observations taken in the $I$ or F814W bandpasses through which the TRGB brightness exhibits a minimal dependence on color (metallicity and age). The advantage of being able to use the TRGB in the F090W, a JWST/NIRCAM bandpass that behaves similarly to those two, is clear. It was therefore incredibly exciting when JWST's wavefront sensing solution reached twice the precision of mission requirements, ensuring a diffraction-limited PSF to wavelengths as short as 1\micron{}. This made F090W imaging significantly more sensitive than pre-launch expectations, dramatically increasing the bandpass' utility for measuring TRGB distances in its color-insensitive form with which the field is most comfortable. 

However, it must be stated that a key finding of our analysis here was that the color-corrected F115W TRGB seems to perform equally as well as the F814W TRGB in terms of precision and accuracy. This begs the question: can we do \textit{even better} and/or probe \textit{even further} with the color-corrected TRGB in F115W than we can with F090W? As a first step in the direction of addressing this question, we enumerate here and throughout this section some key advantages of the F115W TRGB relative to its classical application in the $I$ (or an equivalent) band.

\begin{enumerate}
    \item Observed in JWST/F115W, the TRGB is $-0.5$ and $-0.8$~mag (AB mag) brighter than it is in JWST/F090W and HST/F814W, respectively.
    \item The IR TRGB does not require the use of a color selection, since it is explicitly included in the two-dimensional fit.
    \item In the IR, C-rich and Extreme AGB stars are pulled redward and out of the O-rich AGB and away from the bright RGB, as a result reducing AGB contamination at and under the TRGB.
    \item The ratio of absolute to selective absorption in F115W is 1.8 and 1.6 times smaller than it is for F814W and F090W, respectively, significantly reducing extinction uncertainties.
\end{enumerate}

\subsection{Superior Sensitivity in the near-infrared}

\begin{figure}
    \centering
    \includegraphics[width=0.95\linewidth]{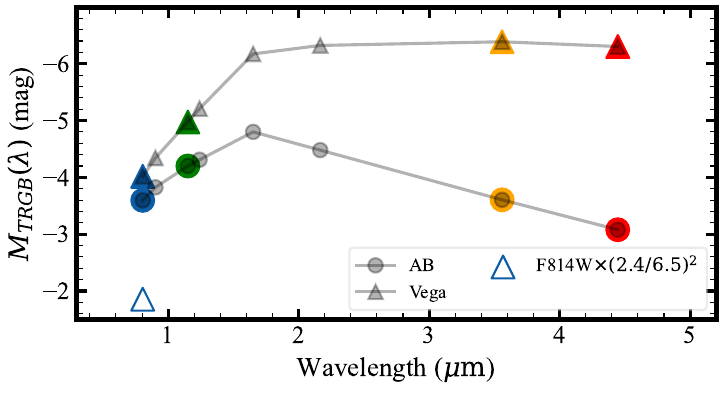}
    \caption{Absolute Magnitude of the TRGB as a function of wavelength. $M_{TRGB} (\lambda)$ is shown for all wide bandpass filters available to JWST/NIRCAM as well as for HST/ACS/F814W. Both the AB (circles) and VEGAMAG systems (triangles) are plotted. F814W, F115W, F356W, and F444W are outlined and matched in color with their corresponding transmission curve in \autoref{fig:trgb_spectrum}. All magnitudes were shifted vertically to set $M_{814} (VEGA) = -4$~mag. For the HST/F814W band, an additional pseudo-magnitude is plotted (open triangle) after being scaled down by the approximate difference in sensitivity between JWST and HST.
    }
    \label{fig:trgb_snr}
\end{figure}

One of the most exciting components in the lead-up to JWST's successful launch has been the possibility of observing TRGB stars in the NIR where they are brightest. We demonstrate this quantitatively in \autoref{fig:trgb_snr} in which synthesized photometry of the model spectrum shown previously in \autoref{fig:trgb_spectrum} is plotted for all JWST NIRCAM bands plus the ACS/WFC/F814W bandpass. Magnitudes were synthesized in both the VEGAMAG and AB systems using \textit{sncosmo} \citep{Barbary2016ascl.soft11017B}. Four points are colored to match the same four filters with transmission curves plotted in \autoref{fig:trgb_spectrum}. Again, a second datapoint for the HST F814W synthetic magnitude (VEGAMAG) is shown to illustrate the difference in mirror size between HST and JWST (open blue circle).

The F115W TRGB is $-0.5$~mag brighter than the F090W TRGB and $-0.8$ brighter than the F814W TRGB (both in AB mags), making F115W TRGB programs able to probe much deeper and more efficiently than F090W ones. This is corroborated from the published work of \citet{Li_2024}, who were unable to make a F090W TRGB measurement to the furthest galaxy in their sample ($\mu=33.1$~mag) due to insufficient SNR (=3) at the TRGB. Based on that and their exposure time being 2000~sec, the limiting distance of the F090W TRGB would be around 40~Mpc ($\mu = 33.0$~mag) for a reasonably deep 10~ks exposure, still a marked gain over HST's limit around 30~Mpc ($\mu = 32.5$~mag). However, because of its greater sensitivity both intrinsically and on JWST's detectors, F115W looks capable of pushing the threshold of JWST TRGB distances into the 60~Mpc range, again in a reasonable 10~ks exposure time. This particular regime is critical reaching just far enough into the smooth Hubble flow to make viable a measurement of $H_0$ directly from TRGB distances, \textit{without} the need to use SNe or another secondary indicator.

\subsection{Direct use of color information}

\begin{figure}
    \centering
    \includegraphics[width=0.9\linewidth]{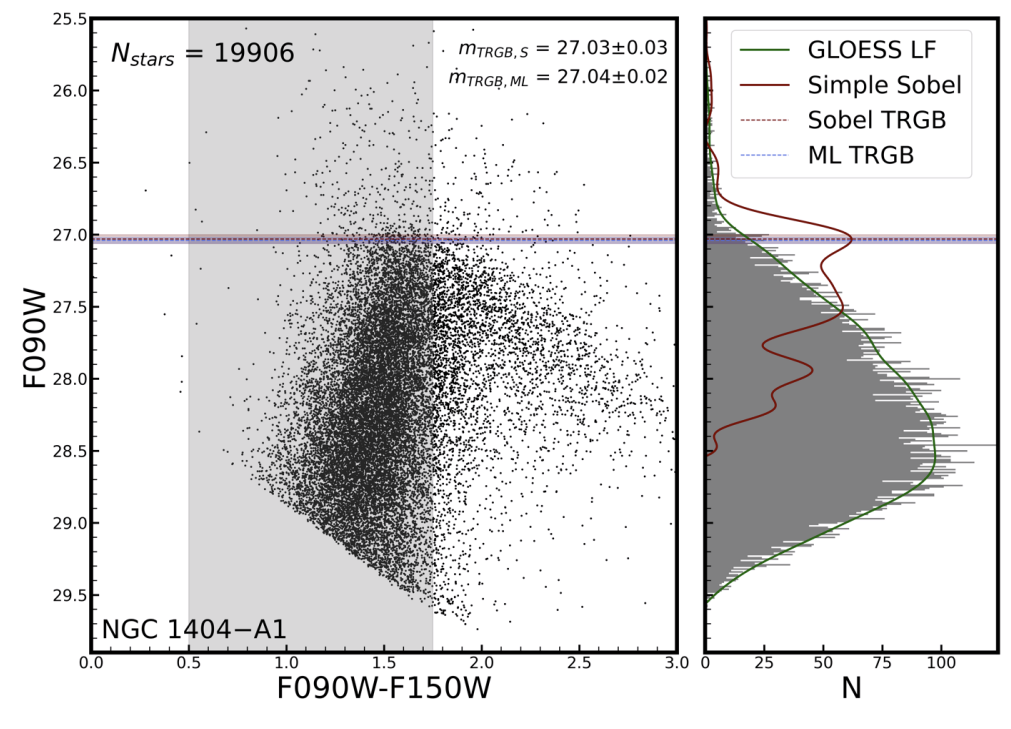}
    \caption{JWST F090W, F150W CMD of NGC~1404. Replicated from \citet{Anand_2024_tully}.}
    \label{fig:anand_fornax_cmd}
\end{figure}

Recall that we did not employ a color selection of tip stars in our analysis of the F115W tip. This is common practice in the $I$-band TRGB to try and select for lower metallicity RGB stars and maintain the $[M/H] \lesssim -0.5$~dex standard candle regime. However, this adds another ``author choice'' into the analysis of the TRGB, which is completely removed from the NIR TRGB because the color information is used directly in the tip measurement. 

We can see an example of the uncertainty that such a selection introduces into $I$-band TRGB measurement using recent data from JWST. \citet{Anand_2024_tully} presented F090W vs. F150W CMDs and TRGB distances to several Fornax Cluster elliptical galaxies. The edge detection response has noticeable structure below what appears to be the tip, which may be a result of this nonlinear $I$-band color dependence. Furthermore, there is not an obvious best choice for where one should place their color selection in order to exclude the metal-rich TRGB stars.\footnote{\citet{Anand_2024_tully} do not do this, but the variation of the F090W magnitude as a function of the color cut could (and should) be determined from the data by, e.g., squeezing the red end of the color cut bluer until some convergence criterion is reached. Recall that our F115W methodology newly introduced here does not rely on color cuts at all.}

\subsection{Reduced contamination from non-RGB stellar populations}
When the TRGB is identified in a CMD in the optical, or where the redder of the two bands is on the magnitude axis, photometric incompleteness slices into the real structure of the CMD, potentially masking information that could be used to improve TRGB measurement accuracy. However, this is not a problem in the near infrared when one's second bandpass used to form the color baseline runs further into the red than the magnitude at which one wants to measure the TRGB. In this scenario, the detection of the most metal-poor (bluest) TRGB stars immediately ensures the detection \textit{all} red giants brighter (oxygen-rich AGBs) and redder (metal-rich TRGBs, carbon-rich AGBs and extreme AGBs).

Furthermore, contamination from these very red AGB stars (predominantly Carbon-rich and dust-enshrouded extreme AGBs) at the tip feature on the CMD is reduced to almost none, due to the AGB being pulled further away from the TRGB both in magnitude (O-AGBs) and color (C-AGBs) when viewed in the infrared. This cleaner distinction between AGB and RGB is precisely why we were able to use our new JWST F115W CMDs to directly confirm where CATS had mistaken the tip of the AGB for the TRGB in at least two of their TRGB measurements. (see \autoref{fig:cats_cchp_compare}). 
Coming at this from a different direction, recall that one of the other distance indicators studied as part of our JWST program, the JAGB, uses these very carbon-rich AGB stars as distance indicators. That is, the very presence of a distinct horizontal J-band branch of C-rich AGBs on the CMD (the JAGB) is a self-evident confirmation that there must also be less AGB stars masquerading underneath the TRGB. 

Note this stipulation fails for regions of color-magnitude space in which the C-rich AGBs are not separated from their O-rich counterparts, such as the HST/WFC3/IR F110W and F160W bandpasses. This has led to past misidentifications of the tip of the AGB for the TRGB even in IR CMDs \citep{Wu2014AJ....148....7W}.

\subsection{Reduced effects of interstellar extinction}

A simple but powerful advantage of the IR TRGB is the reduced effect of both foreground and \textit{in situ} dust extinction on distance measurements. The selective extinction in the F115W bandpass is 1.6 times smaller than it is for F090W and 1.8 times smaller than F814W \citep{Wang_2019}. In this study's sample, there are three low latitude galaxies (NGC~2442, NGC~5643, and NGC~7250) hosting four SNe for which the reduction in the contribution of foreground extinction to the error budget is important. Additionally, several of our RGB spatial selections likely contain more contributions from an outer disk rather than a stellar halo population (M101, the two NGC~4258 disk fields, NGC~2442, and NGC~5643). 

Encouragingly, the measurements in NGC~4258 allow us to quantify the impact that \textit{in situ} dust extinction could have on our F115W TRGB distances, if any. The RGB populations selected from the outermost regions of both the Outer and Inner Disk fields yield TRGB magnitudes (24.155 and 24.185~mag) that are both brighter than the tip measured in the NGC~4258 Halo field (24.226~mag). That is, if there is any \textit{in situ} dust extinction along the TRGB sightlines, it is not measurable above the statistical noise.

\subsection{A Unique Role for each of the F090W and F115W in TRGB applications with JWST}

The findings of \citeauthor{Newman_2024} emphasize that significantly larger sample sizes are needed to sufficiently populate the elongated IR TRGB than the comparatively narrow and sharp $I$-band TRGB. By contrast, the findings of this study and those of \citet{Anand_2024_tully} indicate that, given sufficient number statistics and mixed stellar populations, the linear magnitude-metallicity-color relation of the IR TRGB may be easier to define on a CMD than the F090W/F814W TRGB. Furthermore, \citet{Bellazzini_2004} reported that the TRGB's $J$-band magnitude (similar to F115W) exhibits the smallest sensitivity to age (for any metallicity considered) when compared to its $I$, $H$, and $K$ band magnitudes.

This brings us to an interesting answer to the question of which single band is best for applications of TRGB with JWST: that there probably is \textit{not} one single best band that fits all applications. It seems that an optimal application of the TRGB could be the use of F090W observations for the measurement of TRGB distances to lower mass stellar populations which may not be able to fully populate the two-dimensional IR TRGB. On the other hand, F115W may be ideal for more massive targets such as SN host galaxies. These predominantly high mass calibrator host galaxies likely lead to an effective smoothing over second-order variations in the IR TRGB that might skew an attempted measurement in a smaller stellar population that has a more narrowly defined population in terms of age and metallicity. 

We therefore posit an optimal choice of filters for use of the TRGB with JWST/NIRCAM: $I$-band/F814W/F090W TRGB can still ensure distance precision in the $0.05-0.06$~mag to any galaxy, including low mass ones, while the F115W TRGB is potentially more optimal for measuring distances to high mass targets such as SN hosts that reach even lower in precision, due to the clearer, elongated form the IR tip takes on the CMD. The strength of $I$-band lies in its insensitivity to metallicity/color (and age), which ensures a denser RGB and sharper TRGB on the CMD even in small number samples. But this very strength may simultaneously set a floor on how its precision in higher mass galaxies for which number statistics are not an issue. That is, the $I$-band tip's quadratic, nonlinear form and the compression of the RGB isochrones in the low metallicity regime may set an irreducible floor that is larger than can be achieved at F115W where the TRGB is more easily standardizable across most metallicities, i.e., in the limit of infinite number statistics, a standardized F115W TRGB distance scale may perform even better (2\% in distance) than the already precise $I$-band (3\% in distance).
Note this is likely only true for TRGB in F115W, $J$-band, or other equivalents, since the TRGB color slope steepens considerably for redder wavelengths, nearly doubling in the F150W/$H$ band. 

Keep in mind that the benefits of the TRGB in different bandpasses discussed here are geared toward attaining small, yet nonzero, gains in TRGB distance precision from the current 3-4\% level to potentially 2\% with F115W in massive host galaxies. And recall that, as we showed in Section 5, the $I$-band and F115W TRGB agree extremely well with each other to better than 1\% in the average and 3.5\% in point-to-point RMS.

\section{Summary and Conclusions}

In this study, we have established a new, near-infrared TRGB distance scale based on JWST observations in the F115W, F356W, and F444W bandpasses as part of the CCHP's latest $H_0$ program. A new methodology for triangulating the location and shape of the infrared TRGB was introduced. This new analysis technique leverages all information on the CMD (empirical H-R diagram) for an unbiased determination of the TRGB's magnitude-color coordinate. The TRGB measurements were standardized onto a universal distance scale by correcting for the known color/metallicity dependence of the IR TRGB.

The new distances were found to agree with previous CCHP estimates based on the color-insensitive $I$-band TRGB to better than 1\% in the average and 4\% in RMS. Furthermore, they were able to reveal significant biases in the (pre-2022) SH0ES Cepheid distances and the CATS TRGB distances. This demonstrates that our approach to determining color-standardized F115W distances has achieved a precision and accuracy comparable to distances derived using the $I$-band TRGB.

Of the four sets of SN magnitudes considered, only the use of Pantheon+ resulted in an $H_0$ value greater than 71~km/s/Mpc. Similarly we found that the SN subsample selection bias posited by \citet{Riess_2024_sample} is most statistically significant only if the Pantheon+ SN magnitudes are used ($3.1 \sigma$).
Notably, the weakest evidence for a bias was seen when using CSP-(I+II), which is the same SN dataset used in F25, which presents our program's primary $H_0$ analysis, ultimately rejecting the claim that our sample of 11 JWST calibrator SNe could be significantly skewed with respect to a larger sample of 24 SNe. This suggests the Pantheon+ treatment of the calibrator SNe is anomalous relative to other analyses. Considering the improvement of the SH0ES distances in 2022 that brought their distances into near 1\% agreement with the CCHP's, it appears that the Pantheon+ SN analysis, and not the Cepheid distances, may now be the primary reason behind the ongoing disagreement between the CCHP and SH0ES estimates of $H_0$, sometimes referred to as ``a tension within the Hubble tension.''

A pair of discussion sections provided comprehensive validation of this study's new methodolology, as well as implications of the new results, and finally a look forward to the future of the TRGB with JWST. We summarize those sections here. 

\begin{enumerate}
    \item Validation of the color standardization methodologies introduced in this study, including a growing consensus in the literature that the color slope of the F115W TRGB lies between $-0.8$ and $-1.0$~\magSlopeErr{}.
    \item Multiple JWST pointings of one host galaxy (NGC~4258), each separated from the other by at least 10~kpc, confirmed the TRGB is repeatable to $<0.025$~mag, or 1\% in distance.
    \item Significant biases ($>0.4$~mag or $>20\%$ in distance) were identified in the CATS TRGB analysis based on joint evidence from the CCHP TRGB, SH0ES Cepheids, and cluster/group membership.
    \item The $5\sigma$ distance discrepancy between twin SNe~2011by and 2011fe claimed by \citet{Foley_2020} was relaxed to $<3 \sigma$ by our new JWST TRGB distances, and decreases further to just a $1.6\sigma$ disagreement once the \citeauthor{Foley_2020} pairwise distance comparison is corrected to properly take into account contributions from intrinsic dispersion.
    \item Advantages of the color-corrected F115W TRGB compared to its $I$/F814W/F090W color-insensitive counterpart were highlighted.
    \item A new and potentially optimal way to pair the TRGB with JWST was posited. The F090W (or $I$-band) TRGB may be better suited for application to low mass target galaxies, while F115W may perform better in higher mass galaxies with stellar populations that are more mixed on average.
\end{enumerate}

In summary, this study has established a robust, near-infrared TRGB distance scale using JWST, introducing a new methodology for precise and unbiased TRGB measurements. Our findings demonstrate the reliability of color-corrected F115W TRGB distances, revealing systematic biases in previous analyses and clarifying the primary sources of uncertainty in H0 measurements. As we continue refining the TRGB as an independent calibrator, JWST’s unparalleled sensitivity and resolution will be instrumental in expanding the sample of TRGB-calibrated supernovae and further reducing systematic uncertainties. By leveraging these advances, future JWST observations will play a decisive role in resolving the Hubble tension and providing a definitive measure of the expansion rate of the universe.

\begin{acknowledgments}
TJH is indebted to Greg Aldering for insightful discussions regarding, and detailed comments on, this manuscript. TJH acknowledges Saul Perlmutter for his support of this research. TJH's contributions to this work were supported in part by the U.S. Department of Energy, Office of Science, Office of High Energy Physics under Contract No. DE-AC02-05CH11231. We acknowledge the usage of the HyperLeda database (\url{http://leda.univ-lyon1.fr}).

The Legacy Surveys consist of three individual and complementary projects: the Dark Energy Camera Legacy Survey (DECaLS; Proposal ID \#2014B-0404; PIs: David Schlegel and Arjun Dey), the Beijing-Arizona Sky Survey (BASS; NOAO Prop. ID \#2015A-0801; PIs: Zhou Xu and Xiaohui Fan), and the Mayall z-band Legacy Survey (MzLS; Prop. ID \#2016A-0453; PI: Arjun Dey). DECaLS, BASS and MzLS together include data obtained, respectively, at the Blanco telescope, Cerro Tololo Inter-American Observatory, NSF’s NOIRLab; the Bok telescope, Steward Observatory, University of Arizona; and the Mayall telescope, Kitt Peak National Observatory, NOIRLab. Pipeline processing and analyses of the data were supported by NOIRLab and the Lawrence Berkeley National Laboratory (LBNL). The Legacy Surveys project is honored to be permitted to conduct astronomical research on Iolkam Du’ag (Kitt Peak), a mountain with particular significance to the Tohono O’odham Nation.
The Legacy Surveys imaging of the DESI footprint is supported by the Director, Office of Science, Office of High Energy Physics of the U.S. Department of Energy under Contract No. DE-AC02-05CH1123, by the National Energy Research Scientific Computing Center, a DOE Office of Science User Facility under the same contract; and by the U.S. National Science Foundation, Division of Astronomical Sciences under Contract No. AST-0950945 to NOAO.
The Siena Galaxy Atlas was made possible by funding support from the U.S. Department of Energy, Office of Science, Office of High Energy Physics under Award Number DE-SC0020086 and from the National Science Foundation under grant AST-1616414.

This research is based in part on observations made with the NASA/ESA Hubble Space Telescope obtained from the Space Telescope Science Institute, which is operated by the Association of Universities for Research in Astronomy, Inc., under NASA contract NAS 5–26555. 
This work is also based in part on observations made with the NASA/ESA/CSA James Webb Space Telescope as part of program GO-1995. 
The data were retrieved from the Mikulski Archive for Space Telescopes at the Space Telescope Science Institute, which is operated by the Association of Universities for Research in Astronomy, Inc., under NASA contract NAS 5-03127.
A.J.L. was supported by the Future Investigators in NASA Earth and Space Science and Technology (FINESST) award 80NSSC22K1602.
\end{acknowledgments}
\facilities{JWST(NIRCam)}

\software{DOLPHOT \citep{Dolphin_2000}, Astropy \citep{astropy:2013, astropy:2018, astropy:2022}, jwst \citep{Bushouse_JWST_Calibration_Pipeline_2023}, Jupyter Notebook \citep{Kluyver2016jupyter}
}
\vfill
\clearpage 

\appendix

\section{Photometric Catalog Source Selections}

\begin{figure*}
    \centering
    \includegraphics[width=0.9\linewidth]{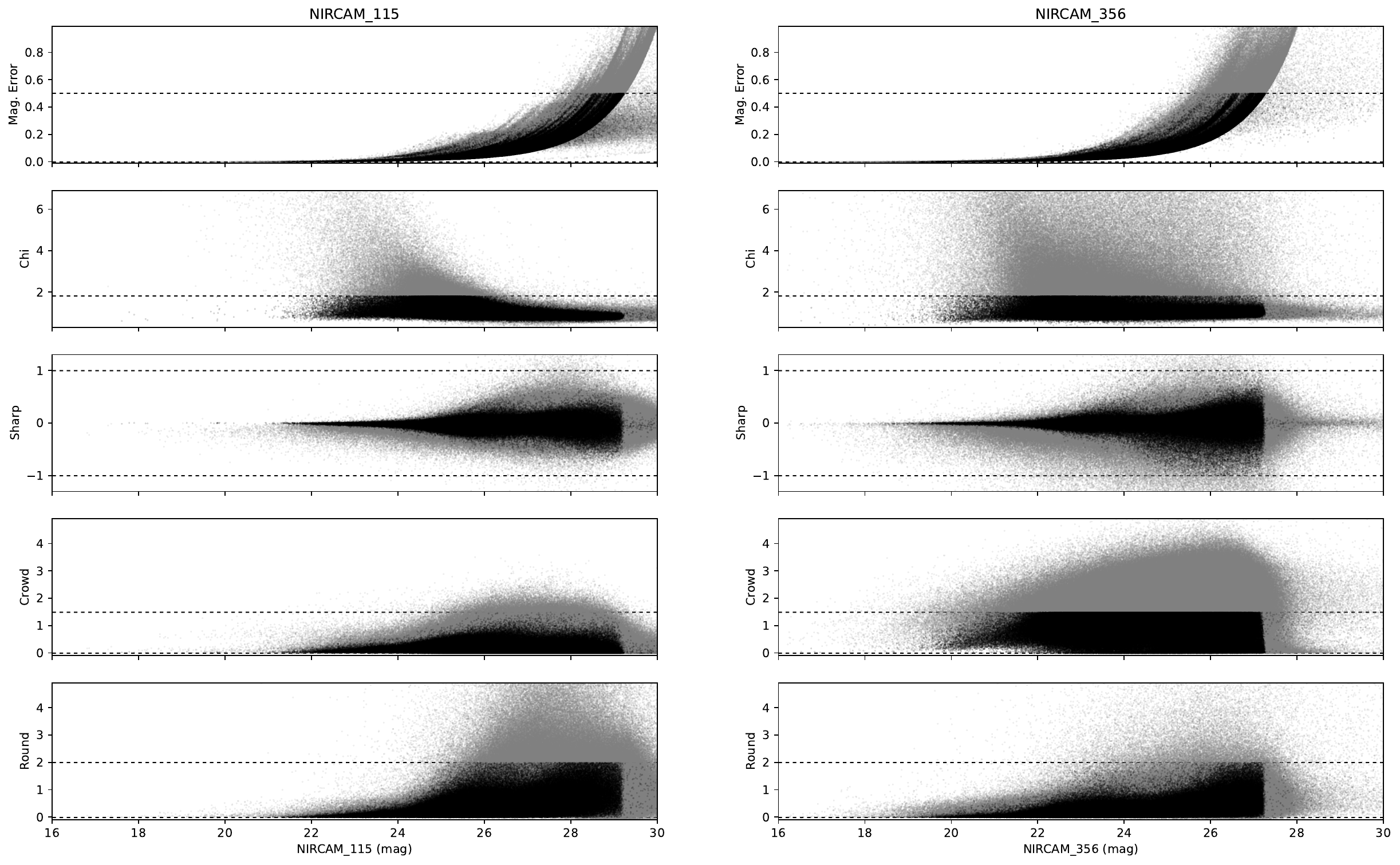}
    \caption{Photometric parameter estimates from DOLPHOT used in point source selection plotted as a function of magnitude for the NGC~4038/9 imaging dataset. Plotted from top to bottom is the magnitude error, chi, sharpness, crowding, and roundness. The left column is for the F115W (short wavelength, SW) and the right for the F356W (long wavelength, LW) band.}
    \label{fig:dq_clean_4038}
\end{figure*}

\begin{figure*}
    \centering
    \includegraphics[width=0.9\linewidth]{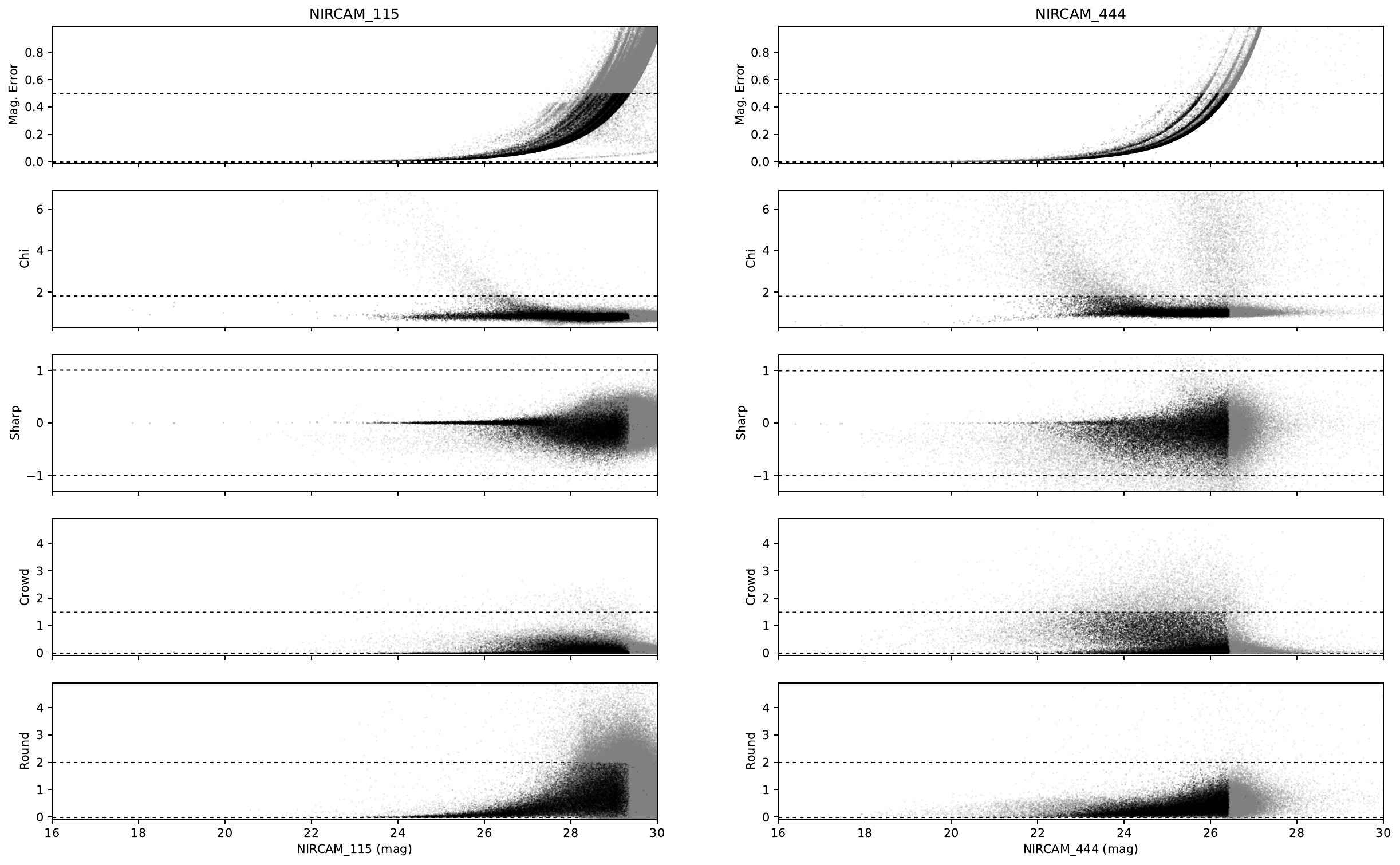}
    \caption{Same as \autoref{fig:dq_clean_4038} but for the NGC~4258 Halo field. Instead of F356W, the LW magnitudes here are F444W.}
    \label{fig:dq_clean_4258}
\end{figure*}

Conservative point source selection criteria were imposed on the photometric catalogs. We present here two examples that represent the extremes of observed source density/crowding. NGC~4038/9 is the most crowded field in our program and NGC~4258-Halo the least. The point source selection for each is shown in \autoref{fig:dq_clean_4038} and \ref{fig:dq_clean_4258}. 

Conservative values of each morphological parameter were adopted uniformly across the sample. The selection cuts made were: (1) $\sigma_{mag} < 0.5 $,  (2) $ \chi <1.9 $,  (3) $ | \mathrm{Sharp} | < 1 $, (4) $ \mathrm{Crowd} < 2 $, (5) $ \mathrm{Round} < 2 $. \\

\clearpage
\section{M101 Spatial Selection Validation}

\begin{figure*}
    \centering
    \includegraphics[width=0.8\textwidth]{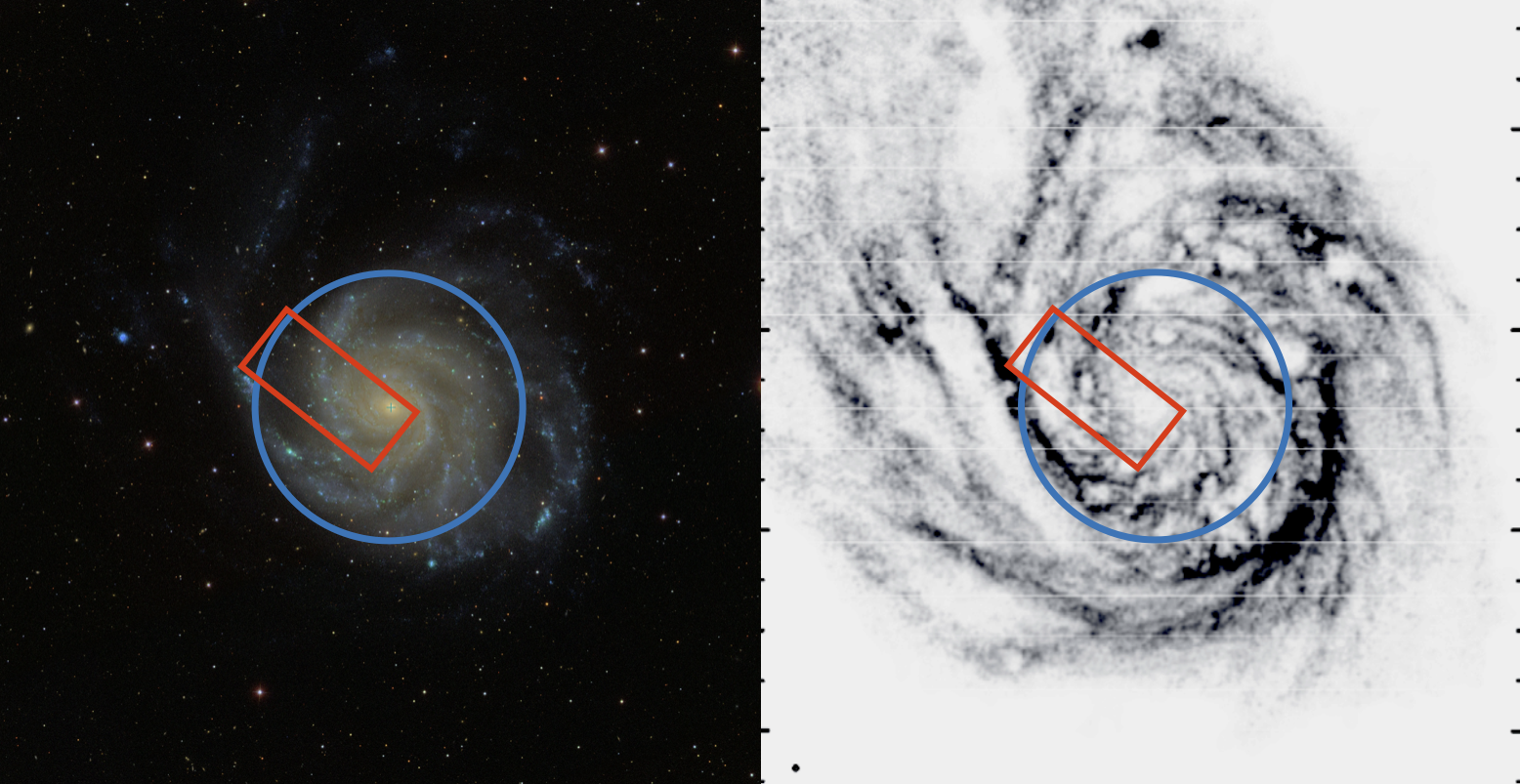}
    \caption{M101 TRGB Spatial Selection. \textit{Left}: SDSS \textit{gri} image (source: \url{https://simbad.u-strasbg.fr/simbad/sim-id?Ident=M+101}). \textit{Right}: HI map from \citet{Walter_2008}. Plotted in both panels is the JWST/NIRCAM field (red rectangle) and the adopted boundary outside which the TRGB is measured (blue circle). North is up, East is left.}
    \label{fig:m101}
\end{figure*}
The spatial extent of M101 on the sky is too large for NIRCAM to have sampled in one pointing the Cepheids in its inner regions, as well as an older RGB population its outer regions. Using the same criteria for the other SN Hosts, we converged on a very limited spatial selection, making up about 6\% of the total sampled NIRCAM area and just 3\% of all detected sources. Because of this significant constriction in sample size, we cross-compared with neutral hydrogen maps constructed of M101 \citep{Walter_2008}. It turns out that our selection nestles right into a gap
of M101's wispy star-forming tendrils, leading to concomitantly undetectable levels of HI (and by proxy dust).

\section{Expanded Discussion on TRGB Fitting Procedure}

We present here additional details on the tip fitting procedure newly introduced in Section 4.1. 

\subsection{Two Dimensional Solution}
As introduced in the main text, attempts to measure the IR TRGB from the one dimensional RGB LF can lead to biases due to the loss of information along the color axis. Simply put, the tip solution is not separable along both the magnitude and color axes (due to the sloped TRGB) and thus must be solved for simultaneously for an unbiased magnitude-color estimate, i.e., 
\begin{equation}
    T(m,c) \neq M(m)C(c)
\end{equation}
where $T(x,y)$ refers to a function that completely describes the 2D TRGB. $M(m)$ and $C(c)$ would be the solutions that could predict the frequency of tip stars as a function of magnitude and color of the TRGB \textit{if} $T$ were separable (but it is not).

\begin{figure*}
    \centering
    \includegraphics[width=0.9\linewidth]{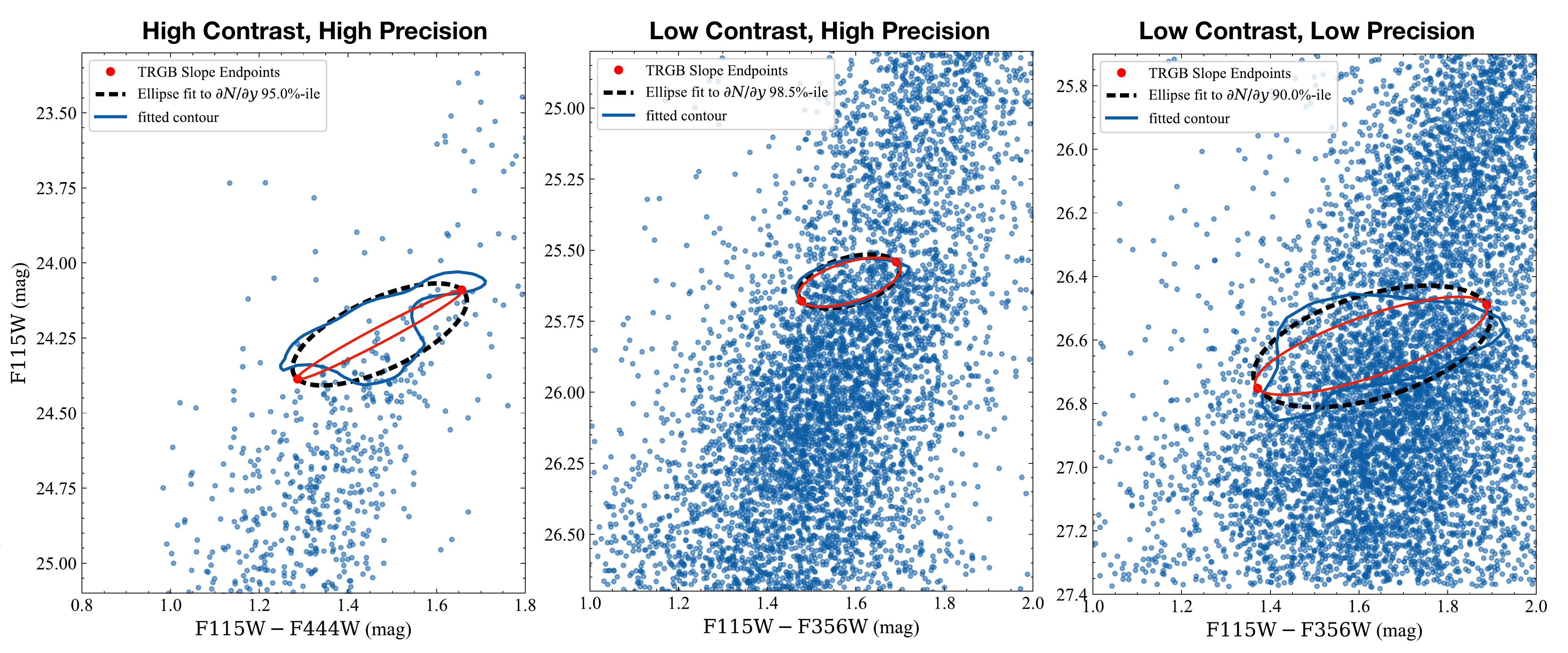}
    \caption{Three representative examples to demonstrate the relationship between TRGB ellipse parameters, contrast, and the measurement uncertainty. For N4258-Halo, N5643, and N2442, the adopted TRGB contour (solid blue curve), best-fit TRGB ellipse (dashed black curve), and the final uncertainty ellipse (solid red curve) are plotted over the color magnitude diagram (blue dots). The contrast---used to scale the ellipse down to an estimated error ellipse---is the ratio between the semi-minor axis of the solid red and dashed black curves. The three examples are for high contrast, high precision (N4258-Halo, left panel), high precision, low contrast (N5643, middle panel), and low precision, low contrast (N2442, right panel). The estimated error ellipse does a more than adequate job capturing the region outside of which one can confidently say the TRGB is very likely not located.
    }
    \label{fig:contrast}
\end{figure*}

\subsection{Tip fitting}
Bandwidths typically had to set larger values for less populated RGBs, to fill in gaps in the Hess Diagram. Optimal threshold values were set between the 95th and 98th percentile, but for N2442, the lowest precision measurement in the sample, which was set to 90\%. All choices in the TRGB fitting parameters were optimized for and locked in while the photometric catalogs were blinded.

During the blinded process of setting these parameter values, we sought to produce a fit that resulted in the most reasonable value of the TRGB slope based on approximate ingoing knowledge of its value (somewhere between -0.4 and -1.0). We found this criterion to be, perhaps unsurprisingly, tightly coupled to the TRGB fit that appeared the best to the eye. For instance, in the high-SNR datasets (M101, N1365, N4258, N4424, N5643), over-smoothing the Hess diagram would result in TRGB slopes that were far too steep due to blurring the TRGB signal with that coming off the blue edge of the RGB. In some of the lower-SNR datasets (N2442, N3972, N7250), over-smoothing did the opposite, and led to flattened slopes or even reversed it to a positive value. Thus, by aiming for reasonable slope values, unphysical solutions could be identified, which imposed clear bounds on the smoothing and threshold parameters. With parameter bounds set by the slope criterion, we could then iterate and reach a converged best-fit TRGB in each galaxy. 

Nevertheless, we found that the TRGB distances were largely insensitive to the analysis choices such as the bandwidth and minimum threshold, so long as they were kept within reasonable values. This is due to different analysis choices simply sliding the centroid of the TRGB ellipse along and parallel to the TRGB feature itself, therefore preserving the TRGB's magnitude-color relation and minimizing the impact on inferred distance.

\subsection{Tip Contrast and Error Estimation}

In \autoref{fig:contrast}, we show three cases of tip ellipse fits (black dashed lines), along with the corresponding error ellipse (red curve) estimated in our methodology. Each case shows how our methodology behaves in different scenarios. 

First, we show a case of high contrast and high precision based on the NGC~4258 Halo field. Due to the lower number statistics, the initial error ellipse is large $b=0.106$, $a= 0.237$. However, because of the very high contrast, again defined as the ratio of tip stars below the bisecting semi-major axis relative to the other half, the uncertainty comes out to just 0.029~mag.

In the middle panel we show the same for NGC~5643 as a low contrast, high precision case. Here, the initial tip is tightly constrained ($b=0.078$~mag) due to large number statistics, but the contrast is equal to 1.2, and so the final error ellipse does not shrink much below the initial ellipse shape, resulting in a tip uncertainty equal to 0.066~mag.

In the right most panel, we show the same for NGC~2442, the lowest precision tip measurement in our sample as a case of low contrast and low precision. The initial tip localization is not tightly constrained ($b = 0.156$) despite the large numbers of stars, indicating perhaps the presence significant contamination from non-standard candle red giants. The ellipse is then slightly tightened on account of the contrast value of 1.7, resulting in a final tip error of 0.093~mag, the largest in our set of measurements.

We comment briefly on some observations here. There is some degeneracy between the size of the initial tip localization ellipse $b$ and the contrast, which in fact provides some self-regulation that in turn stabilizes the tip uncertainty estimation. For example, if the ellipse for NGC~5643 were the size of that in either of the two, the contrast ratio would be significantly larger, and the final error would likely come out very similar to what we currently estimate from the small initial localization ellipse. 

Finally, and perhaps most encouragingly, we can see that more datasets like the NGC~4258 halo one will reduce tip measurement errors to near negligible levels. If that CMD were populated by $10\times$ more stars, the tip measurement uncertainty would likely shrink to sub-percent levels since the initial tip ellipse would be much smaller, i.e., imagine a small initial ellipse such as that seen in NGC~5643, but with the contrast seen in the tip of NGC~4258 Halo. This bodes well for programs that target TRGB fields from the outset, rather than trying to squeeze such measurements out of the edges of primarily disk pointings as we did in this study. That is, the F115W TRGB can only get better, to be even more precise than it has already been shown to be.

\section{Expanded Discussion of Flaws in the CATS Methodologies}

In the main text, we raised several concerns regarding the CATS TRGB analysis. In this appendix we will expand upon some of these in an attempt to understand the cause of the large TRGB distance biases present in their analysis.

At its core, the CATS methodology is flawed in the sense that it enforces the exact same analysis choices on all TRGB imaging datasets at once. This is not a valid assumption given the significant variation of the properties of the underlying imaging datasets. The photometric catalogs underlying TRGB $H_0$ span at least an order of magnitude in photometric SNR and are derived from galaxies of all types, from giant ellipticals to early-type spirals, and located in various different regions, some on the border of disk and halo, others purely in the halo, and one in the star forming tidal debris of a galaxy merger (NGC~4038/9). 

\subsection{Spatial Clipping Algorithm}

S23 construct a spatial clipping algorithm based on the \textit{fractional} density of blue stars relative to the peak density observed in each field. They discretize this into three bins, a 20\% reduction, 10\%, and 5\%, in order of increasing strictness of the spatial clip. This algorithm, however, \textit{cannot} be uniformly applicable to a sample of galaxies that lie at vastly different distances and with varying disk inclinations, wherein the projected column density of blue stars in the inclined disk will peak more sharply than the more spheroidally-distributed halo red stars. 

This problem becomes particularly apparent in their analysis of NGC~4038/9. The analyzed field is pointed directly into the star-forming tidal debris of the Antennae. By inspecting their published CMDs, an improvement in contrast at the \textit{true} TRGB can be seen when going from the the 20\% to the 5\% cut. However, as a result of their decision to only consider three discrete values for all their analysis ``choices,'' the authors terminate at 5\%, their strictest spatial clip, even when the imaging of galaxies such as NGC~4038/9 demand stricter clipping. That field had an abnormally large number of blue stars due to sampling the tidal tail of the Antennae, and therefore a relative fraction of that peak density will sample a far younger population of stars than, say, the same \textit{relative} clipping applied to a pure stellar halo field.

NGC~4038/9 is a notable case in which the CATS methodology has failed, and is one of the two cases we discussed in Section 7.
From their Fig. 1 and the supplementary figures provided at \url{https://github.com/JiaxiWu1018/CATS-H0/blob/main/clipping_map/NGC4038_5p.png}, it is immediately clear that their spatial clipping routine failed to clip the star forming regions if the tidal debris and their associated younger stellar populations.
This reveals a critical weakness in the CATS spatial clipping algorithm, which applies the same \textit{fractional} clipping of blue star contribution to the CMD regardless of the underlying astrophysical environment. That is, their algorithm is tuned to clip the same fraction of blue stars from each field, regardless of whether that field contained a large \textit{total} number of blue stars or not.
The algorithm will therefore fail increasingly worse in fields that are populated with dense star-forming regions, because a 95\% reduction in blue star fraction for a field \textit{dominated} by blue stars will leave a significant amount of young star contamination as compared to a 95\% reduction of blue stars for a halo-dominated field that already has a small starting number of young stars.

\subsection{Contrast Parameter}

CATS argue that their $R$ parameter represents a form of ``contrast'' of the TRGB, but that is a misnomer if one considers how it is actually computed in a 1~mag interval that encompasses the AGB, and sometimes even excludes the RGB itself in cases where their algorithm has confused a tip of the AGB signal with the tip of the RGB. Indeed, their definition of ``contrast'' is in fact simply the ratio of RGB to AGB stars, which is commonly used as a proxy of the underlying stellar population, with larger RGB fractions signifying an older population. In fact, the very idea of a ``contrast'' computed for the 1D luminosity function used in $I$/F814W/F090W TRGB measurements does not make sense in this context, nor does it provide unique information that the edge detector response function does not. Indeed the first derivative (or some weighted variation) of the RGB LF is itself already a more precise and direct measurement of contrast. The use of some scalar-valued, summary estimate of the tip contrast only makes sense when the problem is increased to at least two dimensions, such as the appropriately named contrast parameter determined as part of our two-dimensional F115W TRGB analysis.

As an additional wrinkle in their contrast parameter, recall that CATS do not carry out artificial star experiments. Such tests are crucial to account for the different levels of incompleteness seen on either side of the boundary about which they compute the AGB/RGB ratio (again, they misattribute this ratio to a gauge of tip contrast). Incompleteness is not symmetric and therefore their contrast parameter will always attribute artificially larger values to brighter discontinuities such as their misdetections on the AGB. This skews the TRGB measurements in their entire sample to systematically closer distances and concomitantly higher values $H_0$, particularly because the SN host measurements are not as deep as the data in the geometric anchor, leading to a steeper incompleteness gradient that biases the contrast more for those targets.

\subsection{Photometric Reductions}

In the case of NGC 4536, all of the aforementioned issues are at play. However, there is also a problem with the photometry that leaks into the other issues. Their photometric reductions of NGC~4536 appear to be unstable. Upon inspection of S23 Figure 2, the mean color of the N4536 RGB appears to fall around an F606-F814W color of 1.5~mag. However, if one downloads the photometry directly from their GitHub, the RGB would be almost entirely excluded by CATS' adopted color cut. It is unclear which of these two concerningly different sets of photometry was used in their TRGB distance measurement.

As we showed in Section 7.4, the correct TRGB measurement to this galaxy is their middle value around 27th mag. Thus, we hypothesize that the erroneous photometric reductions that were adopted by S23 for this galaxy suppressed the real TRGB signal and introduced many false peaks that were picked up by their unsupervised analysis.

From further inspection of S23, their Figure 2, it appears that similar photometric reduction issues recurred for NGC~4526 and NGC~4424, but not enough to blur the true TRGB signal as severely as in the case of NGC~4536 just discussed here.

\subsection{Multiple tips per Host Galaxy}
The assumption in the CATS analysis that there can exist multiple real TRGBs in a single host discrepant by over 1~mag is an unphysical assumption, because the intrinsic dispersion of the TRGB is not expected to exceed 0.1~mag. Allowing for multiple signals within that range (convolved with the photometric error function) is reasonable, but allowing for over 1~mag of distance variance violates the most fundamental basis underlying the measurement of distances, that standard candles indeed occur at a \textit{standard} intrinsic brightness that approaches a constant to within some physical/irreducible dispersion. Instead, the correct way to interpret multiple tips that are separated by amounts that exceed by orders of magnitude the expected intrinsic dispersion of the TRGB is that either or both the color and spatial selection for an old metal-poor RGB population has failed and needs to be refined.

\section{Translating between SuperCal and Pantheon+ SN magnitudes}

\begin{figure}
    \centering
    \includegraphics[width=0.45\linewidth]{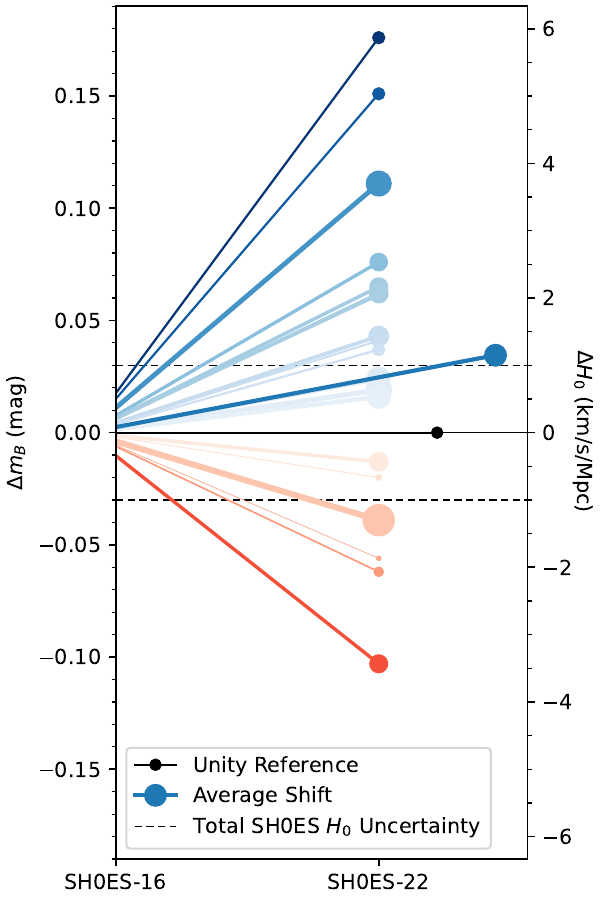}
    \caption{Shift in apparent SN magnitude between SuperCal \citep{Scolnic2015ApJ...815..117S} and Pantheon+ \citep{Scolnic_2022ApJ...938..113S}. Each line represents one of the 19 SNe in common between the two samples. Each line capped by a dot represents the magnitude shift of each calibrator SN. Each line is colored and shaded according to whether the SN's magnitude became fainter (deeper hues of blue) or brighter (deeper hues of red). Both the size of each point and thickness of each line is linearly scaled to the ratio of weights of that SN between R16 and R22 (thicker line and larger point represents increased weight in 2022). Also shown is a line representing a zero magnitude shift and a weight ratio of unity (black line and marker). The net shift between SH0ES-16 and SH0ES-22 (solid blue line) extends slightly further on the x-axis than the individual SNe and has arbitrary thickness and dot size, for clarity. This net shift in apparent SN magnitude is equal to 0.037~mag, or 1.28~km/s/Mpc, which is larger than the total uncertainty currently quoted in SH0ES-22. The total range of the change in SN magnitudes is 0.3~mag, or 10~km/s/Mpc, significantly larger than the size of the Hubble Tension. This comparison is only between the \textit{apparent} standardized SN magnitudes, and does not make use of the SH0ES distances in any way. The difference in Hubble diagram intercepts $5a_B$ has been removed to bring the two sets of values onto a consistent standardization fiducial.}
    \label{fig:snmag_supercal_panplus}
\end{figure}

Here we present the regularization of the SuperCal and Pantheon+ $B$-band magnitudes, as well as a direct comparison between the two sets of calibrator magnitudes. One set was used in the SH0ES-16 calibration of $H_0$ \citep[from SuperCal,][]{Scolnic2015ApJ...815..117S} and the other in the SH0ES-22 calibration \citep[from Pantheon+,][]{Scolnic_2022ApJ...938..113S}.

As previously described, an $H_0$ measurement consists of calibrating a fiducial or the floating intercept of a Hubble Flow diagram. Thus, to compare SN magnitudes between two $H_0$ experiments, the difference in this floating intercept needs to be subtracted out. The following simplified description of the Hubble expansion is sufficient for this purpose, which we repeat here from Eq. 5 in R22,
\begin{equation}
    5 \log H_0 = M_B^0 + 5a_B + 25
\end{equation}
where the $a_B$ term encodes the fiducial definition set by the Hubble Diagram SNe. To bring the SH0ES-16/Supercal $m_B$ magnitudes onto the same fiducial as SH0ES-22/Pantheon+, we add $-5(a_B^{R22}-a_B^{R16}) = -0.007$~mag to the SH0ES-16 $m_B$ magnitudes. As a quick sanity check on this regularization, the $H_0$ value between the baseline R16 and R22 analyses changed by less than 6~mmag (0.2 km/s/Mpc) and the $M_B,max=-19.25$~mag reported in R16 is within 3~mmag that reported in R22. Note where the truncation at two decimals places was done in the original study, not here. This confirms the R16 and R22 definitions of $m_B$ set by the Hubble Flow sample must be nearly identical, if not exactly so after accounting for the shift in $a_B$.

\bibliography{bib_gaia,bib_trgb,bib_ogle,bib_cephs,bib_trgb_metal,bib_main,bib_ext_data,bib_unionfactory,bib_csp,bib_cchp,bib_sn_lowz,bib_pantheon,bib_h0,bib_jl17,bib_software,bib_edd,bib_arxiv,bib_proposals}{}
\bibliographystyle{aasjournal}



\end{document}